\documentclass{aa}

\newcommand{\myemaila}{eistrup@virginia.edu}

\usepackage[fleqn]{amsmath}
\usepackage{hyperref}
\usepackage{amssymb}
\usepackage{url}
\usepackage{gensymb}
\usepackage{graphics}
\usepackage{subfigure}
\usepackage{wasysym}
\usepackage{textcomp}
\usepackage{multirow}
\usepackage[version=3]{mhchem}
\usepackage{txfonts}

\hypersetup{draft}
\begin{document}

\title{Cometary compositions compared with protoplanetary disk midplane chemical evolution}
\subtitle{An emerging chemical evolution taxonomy for comets}
\author{Christian Eistrup\inst{1,2,\thanks{Virginia Initiative on Cosmic Origins (VICO) Fellow}}, Catherine Walsh\inst{3} \and Ewine F.~van Dishoeck\inst{1,4}}

\institute{Leiden Observatory, Leiden University, P.~O.~Box 9531, 2300~RA Leiden, The Netherlands  \label{1} \and
Department of Astronomy, University of Virginia, 530 McCormick Road, 22903 Charlottesville, VA, USA  \\\email{\myemaila} \label{1} \and
School of Physics and Astronomy, University of Leeds, Leeds, LS2 9JT, UK \label{3} \and 
Max-Planck-Institut f\"{u}r Extraterrestriche Physik, Giessenbackstrasse 1, 85748 Garching, Germany  \label{4}
}

\date{Received $\cdots$ / Accepted $\cdots$} 

\titlerunning{Cometary compositions compared with disk midplane chemical evolution}
\authorrunning{Christian Eistrup et al.}

\abstract
{Comets are planetesimals left over from the formation of planets in the solar system. With a growing number of observed molecular abundances in many comets, and an improved understanding of chemical evolution in protoplanetary disk midplanes, comparisons can be made between models and observations that could potentially constrain the formation histories of comets.}
{Our aim is to carry out the first statistical comparison between cometary volatile ice abundances and modelled evolving abundances in a protoplanetary disk midplane.}
{A $\chi^{2}$-method was used to determine maximum likelihood surfaces for 14 different comets that formed at a given time (up to 8 Myr) and place (out to beyond the CO iceline) in the pre-solar nebula midplane. This was done using observed volatile abundances for the 14 comets and the evolution of volatile abundances from chemical modelling of disk midplanes. Two assumptions for the chemical modelling starting conditions (cloud inheritance or chemical reset), as well as two different sets of cometary molecules (parent species, with or without sulphur species) were investigated.}
{Considering all parent species (ten molecules) in the reset scenario, $\chi^{2}$ likelihood surfaces show a characteristic trail in the parameter space with high likelihood of formation around 30 AU at early times and 12 AU at later times for ten comets. This trail roughly traces the vicinity of the CO iceline in time.} 
{A statistical comparison between observed and modelled chemical abundances in comets and comet-forming regions could be a powerful tool for constraining cometary formation histories. The formation histories for all comets were constrained to the vicinity of the CO iceline, assuming that the chemistry was partially reset early in the pre-solar nebula. This is found, both when considering carbon-, oxygen-, and sulphur-bearing molecules (ten in total), and when only considering carbon- and oxygen-bearing molecules (seven in total). Since these 14 comets did not previously fall into the same taxonomical categories together, this chemical constraint may be proposed as an alternative taxonomy for comets. Based on the most likely time for each of these comets to have formed during the disk chemical evolution, a formation time classification for the 14 comets is suggested.}


\maketitle

\section{Introduction}
\label{introduction}

When the solar system formed 4.6 billion years ago, the planets formed their cores from solid material in the pre-solar nebula. In the outer, colder regions of this nebula volatile molecules, such as \ce{H2O}, \ce{CO2}, and CO were frozen out as ices on the surfaces of grains, and later larger bodies. Some of these bodies merged to form the planetary cores, and eventually the Jovian planets, but some of this solid material remained unused by planets and is still present as comets in our solar system today .

Comets are made up of partly refractory dust and partly volatile ices. These ices reside deep inside the comets, and they are thought to be pristine samples of the material that was present in the pre-solar nebula \citep[see review by][]{charnley11}. Comets are thus interesting because of what they can tell us about the chemical composition in the icy outer pre-solar nebula 4.6 billion years ago, but also because comets are known to have impacted on the Earth after having been dynamically scattered towards the Sun from the outer solar system. The volatile and organic material they carry on them has thus added to the chemical make-up of the Earth. Furthermore, understanding the origin of water and life on Earth may be traced back to comets.

Comets have been observed from the ground and from space for decades in various wavelength regimes \citep[see e.g.][]{biver1999,biver2014,bockelee-morvan2000,bockelee-morvan2015}. Several efforts have gone into detecting molecular species in the comae of comets and using these to classify them \citep{ahearn1995,fink2009,charnley11,cochran2012,leroy2015}. At least two classification groups have been proposed for cometary compositions: `typical' and `depleted', where depleted refers to a depletion in organic carbon-chain molecules, compared with the typical compositions \citep[see e.g.][]{cochran2012}. Hundreds of comets have been analysed for composition. The majority of these (75-91\%) fall under the Typical category, as found by \citet{cochran2012}. They also find good agreement with the other studies \citep[e.g.][]{ahearn1995,fink2009} in which comets are typical and which are depleted.

However, relating observed cometary species to the actual cometary compositions is still a challenge. This is because some parent species, sublimating from the comet, get dissociated into chemical daughter species, such as radicals (for example, \ce{NH3} gets dissociated into \ce{NH2}, and HCN gets dissociated into CN) when moving from the surface to the coma of a comet. Tracing which daughter species originate from which parent species and how the daughter species abundance in the coma translates to parent species' abundances near the surface is tricky, as pointed out by \citet{leroy2015}, for example.

Recently, ESA's \emph{Rosetta} mission visited comet 67P and orbited the comet for two years with an armada of instruments, providing unprecedented details about the comet. The ROSINA instrument onboard the mission has been particularly powerful for determining chemical composition. The comet showed very different amounts of produced species from the summer to the winter hemispheres \citep{leroy2015}, and hence it is difficult to say which amounts of which species are representative of the bulk composition since temperature plays a role. It is, in turn, difficult to classify this comet's composition according to the typical-and-depleted-scheme. Because the summer and winter hemispheres of comet 67P show very different, and, so far, inexplicable chemical characteristics, the two hemispheres will be treated as separate comets in the analysis in this work \citep[this separation is also highlighted by][]{leroy2015}. Considering the two hemispheres as separate brings the number of comets considered for the analysis in this work to 15, but with the actual number of analysed cometary bodies remaining at 14 (as noted in the Abstract).

In this work, a quantitative comparison between the observed cometary abundances and the protoplanetary disk midplane chemical evolution models from \citet{eistrup2018} will be made. Molecular abundances, observed mainly from remote sensing with infrared (IR) and millimetre (mm) facilities, for each of the 15 comets presented in Tables 2, 3 and 5 in \citet{leroy2015} \citep[which are based on][]{bockelee-morvan2004,charnley11} will be compared statistically in time and space with volatile abundances from the models 
\citep[observed cometary abundances originally reported in][]{altwegg1994,biver1999,biver2006,biver2007,biver2008conf,bockelee-morvan1995,bockelee-morvan2014,brooke1996,colangeli1999,combes1988,dellorusso2001,dellorusso2007,dellorusso2008,dellorusso2009,dellorusso2011,despois2005conf,disanti2003,disanti2007codepl73p,disanti2007comet9p,eberhardt1994,eberhardt1999,gibb2003,gibb2007,kawakita2013,krankowski1986,lis1997,magee-sauer2008,mcphate1996conf,mumma1996,mumma2000,mumma2003,mumma2005,mumma2011,charnley11,ootsubo2012,paganini2014,radeva2013,rubin2011,villanueva2006,weaver1994,weaver1999,weaver2011,woodney1997}. 

The aim is to test if there is a statistical connection between current cometary abundances, and where and when such abundances were found in the pre-solar nebula, thereby possibly tracing the formation histories of the comets. Based on this test, a possible `chemical evolution'-taxonomy of comets may be established if multiple comets have similar formation histories. 

\begin{table*}[]
\caption{Molecules detected in each comets included in this work}           
\centering                          
\begin{tabular}{lcccccccccccc}              
\hline\hline 
&&&&&&&&&&&\\
Comet                                                        & \ce{H2O}        &  CO            & \ce{CH4}            &  \ce{CO2}            & \ce{C2H6}       & \ce{CH3OH} & \ce{H2CO}     & \ce{O2}         &   \ce{SO2}          & \ce{H2S} & OCS \\
&&&&&&&&&&&\\
\hline
&&&&&&&&&&&\\
1P/Halley                                                   &  \checkmark   &   \checkmark &    \checkmark      &   \checkmark & \checkmark     &   \checkmark    &   \checkmark &  \checkmark  &                           & \checkmark    &     \\
\hline
&&&&&&&&&&&\\
C/1995 O1 (Hale-Bopp)                             &  \checkmark   & \checkmark   &    \checkmark   &    \checkmark   &  \checkmark     &    \checkmark  &   \checkmark&                          & \checkmark      & \checkmark    &  \checkmark   \\
\hline
&&&&&&&&&&&\\
C/1996 B2 (Hyakutake)                             &  \checkmark   &  \checkmark  &    \checkmark   &                          &  \checkmark      &    \checkmark   &   \checkmark&                          &                        & \checkmark   &  \checkmark   \\
\hline
&&&&&&&&&&&\\
C/2001 A2 (LINEAR)                                 &  \checkmark   &  \checkmark  &    \checkmark    &                         &   \checkmark      &   \checkmark   &   \checkmark&                          &                           &  \checkmark &     \\
\hline
&&&&&&&&&&&\\
C/2012 F6 (Lemmon)                               &  \checkmark   &   \checkmark &      \checkmark  &                            &                          &  \checkmark   &   \checkmark&                          &                           &                           &     \\
\hline
&&&&&&&&&&&\\
C/2013 R1 (Lovejoy)                                &   \checkmark  &  \checkmark  &                           &                           &                         &  \checkmark   &   \checkmark &                          &                           &                           &     \\
\hline
&&&&&&&&&&&\\
103P/Hartley 2                                         &    \checkmark &  \checkmark  &                           &   \checkmark      &   \checkmark   &  \checkmark   &   \checkmark&                          &                           &                           &     \\
\hline
&&&&&&&&&&&\\
73P/Schwassmann-Wachmann 3           &   \checkmark  &                        &                           &                          &  \checkmark     &   \checkmark  &   \checkmark&                          &                           & \checkmark  &     \\
\hline
&&&&&&&&&&&\\
2P/Encke                                                 &   \checkmark  &                       &      \checkmark  &                         &   \checkmark   &  \checkmark    &   \checkmark&                          &                           &                            &     \\
\hline
&&&&&&&&&&&\\
9P/Tempel 1                                            &   \checkmark  &   \checkmark                    &      \checkmark  &                        &    \checkmark     &   \checkmark &   \checkmark&                          &                             & \checkmark    &     \\
\hline
&&&&&&&&&&&\\
6P/d'Arrest                                            &  \checkmark   &                         &                           &                         &   \checkmark     &  \checkmark     &   \checkmark&                          &                           &                           &     \\
\hline
&&&&&&&&&&&\\
17P/Holmes                                            &   \checkmark  &   \checkmark &                           &                        &   \checkmark     &  \checkmark    &                          &                           &                           &     \\
\hline
&&&&&&&&&&&\\
21P/Giacobini-Zinner                              &  \checkmark   &  \checkmark  &                           &                         &      \checkmark                   &  \checkmark      &                          &                           &                           &     \\
\hline
&&&&&&&&&&&\\
67P summer hemisphere                       &  \checkmark   &  \checkmark  &      \checkmark &  \checkmark     &  \checkmark       &  \checkmark     &   \checkmark&   \checkmark   & \checkmark    &   \checkmark   &  \checkmark   \\
\hline
&&&&&&&&&&&\\
67P winter hemisphere                          &   \checkmark  &  \checkmark  &       \checkmark&    \checkmark    &   \checkmark     &  \checkmark      &   \checkmark&                          &   \checkmark  &   \checkmark  & \checkmark   \\
\hline
\end{tabular}
\label{overview}
\end{table*}

\section{Methods}
\label{methods}
This work makes use of both observed molecular abundances in comets and existing results from modelling of volatile chemistry in a protoplanetary disk midplane. The chemical modelling includes gas-phase chemistry, gas-grain-interactions, and grain-surface (ice) chemistry. The gas-phase chemistry is from the latest release
of the UMIST Database for Astrochemistry \citep[see][]{mcelroy13} termed \textsc{Rate12}, and the gas-grain interactions and grain-surface
chemistry are as described in \citet{walsh15}, and references therein. The statistical method used is described in Section \ref{stat}.

\subsection{Model description}
The modelled volatile ice abundances are taken from \citet{eistrup2018}. A physical disk model evolving in time was used featuring decreasing temperature and density structures from 0 to 30 AU, with the CO iceline residing inside 30 AU. Icelines
(or snowlines) mark the radius in the disk midplane beyond
which species exist solely in ice form and are thus depleted from
the gas. This occurs at the radius where the accretion rate onto
grain surfaces (or freezeout) exceeds the desorption rate from
grain surfaces due to the negative temperature gradient in the
midplane. The position of
the midplane iceline for a particular species will depend on its
volatility (that is, its binding energy).

\begin{figure}[!t]
\includegraphics[width=0.5\textwidth]{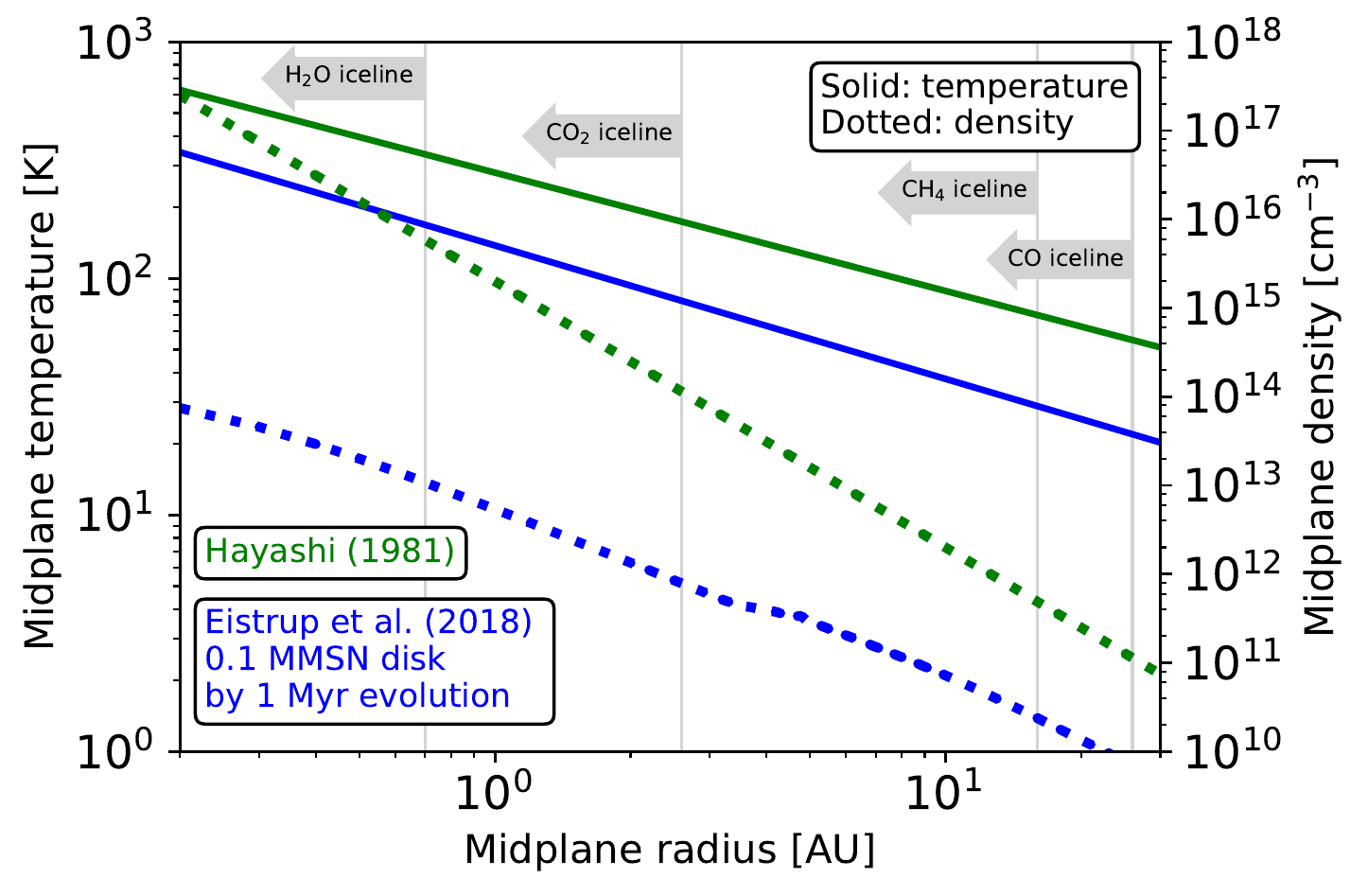}
\caption{Physical structures of disk midplanes for pre-solar nebula from \citet{hayashi1981} (in green), and for 0.1 MMSN disk by 1 Myr evolution from \citet{eistrup2018} (in blue). Solid profiles are for temperature. Dotted profiles are for number density. The vertical grey lines indicate the positions of the icelines of \ce{H2O}, \ce{CO2}, \ce{CH4} and CO by 1 Myr evolution for the 0.1 MMSN disk in \citet{eistrup2018}. The grey arrows on each of the vertical lines indicate which species each line is associated with, and how the iceline moves over time (all inwards).}
\label{phys_struct}
\end{figure}

The disk structure used is not the pre-solar nebula structure proposed by \citet{hayashi1981}, and parameterised in \citet{aikawa1997}. However, the utilised disk structure here is evolving in time, and chemical evolution model results are readily available from \citet{eistrup2018}. Besides, the locations in the disk important to comet formation are the volatile icelines, which are closer to the star for a colder disk, and further away from the star for a warmer disk. Tracking comet formation based on iceline positions therefore means that the exact choice of physical disk structure is less important, and in addition, the 30 AU range of this disk covers all important icelines (including the CO iceline which starts just inside 30 AU). 

Figure \ref{phys_struct} shows a comparison of the midplane temperature and density structures from \citet{hayashi1981}, and from \citet{eistrup2018} (for the 0.1 MMSN disk by 1 Myr evolution), extending to 30 AU. The structures are different, in that the structure from \citet{hayashi1981} is warmer (roughly twice as high temperature at any radius) and more massive (more than ten times higher mass) compared to the disk from \citet{eistrup2018} (hence the ``0.1 Minimum Mass solar Nebula''-designation for this disk). It is noted that the disk from \citet{eistrup2018} has already lost mass by 1 Myr evolution, so in Fig. \ref{phys_struct} the density difference between the disks is more than ten times. The disk mass of 0.1 MMSN from \citet{eistrup2018} is the initial mass of the disk, not the mass by 1 Myr (as plotted in Fig. \ref{phys_struct}).

Results for the ``high'' ionisation rate are taken, which includes contributions from galactic cosmic rays, as well as decay products of short-lived radionuclei in the disk midplane \citep[see e.g.][]{cleeves13crex,padovani2018}. A timescale up to 8 Myr is assumed for a long-lived gaseous protoplanetary disk. The high ionisation rate of typically $10^{-17}$s$^{-1}$ means that chemical changes occur after a few $10^{5}$ yrs. The disk structure cooling in time means that the volatile icelines move inwards in time.

Two different sets of initial abundances are assumed: inheritance and reset. Inheritance assumes all ices to have survived the trip from the parent molecular cloud to the disk midplane, thereby starting the chemical modelling with neutral molecules that are abundant in interstellar ices. The reset scenario assumes an energetic event to have dissociated all molecules into atoms (chemical reset) upon arrival in the midplane, thereby starting the chemical modelling with highly reactive atoms. Such a reset can occur for in-falling material that comes close to the protostar in the earliest stages of disk formation \citep[see e.g.][]{visser2009}. The inner solar system is assumed to have undergone some amount of chemical reset \citep[with evidence from studies of chondrules and CAIs, see e.g.][]{trinquier2009}, followed by a condensation sequence, depending on location in the disk. Comets, on the other hand, are often thought to be pristine, possibly because they formed, and mainly reside, in the outer solar system.

Colour maps of evolving abundances for different volatile ice species with respect to \ce{H2O} ice for both reset and inheritance scenarios are shown to the left in Figs. \ref{figure6}-\ref{figure10} in the Appendix. These plots allow an overview of when and where the different ice species are abundant. It can be seen that there are regions in the radius-time parameter space at which the modelled ice ratios well reproduce the observed ratios for most species. Based on these plots, the ice species considered in the statistical analysis are \ce{CO2}, CO, \ce{O2}, \ce{CH4}, \ce{C2H6}, \ce{H2S}, OCS, \ce{SO2}, \ce{H2CO}, and \ce{CH3OH}. It is noted in each panel which molecule is considered, and whether the inheritance or the reset scenario has been assumed. To the right of the colour maps are shown the observed abundances of the given ice species in different comets. The ice species \ce{NH3}, HCN, HNCO, \ce{CH3CN}, and \ce{C2H2} have all been detected and modelled, but they are excluded from the analysis. This exclusion is based on nitrogen ice chemistry in protoplanetary disks remaining poorly understood \citep[see e.g.][]{schwarz2014,walsh15}. Lastly, \ce{C2H2} is most likely to be a daughter species, and thus also not constraining the bulk cometary composition. 

Daughter species in general have been excluded because the parent species are expected to be dominant in cometary ices, even if the daughter species are abundant in the comae. Since the daughter species in the coma originate from dissociated parent species in the gas after sublimation, the exclusion of daughter species means that the detected abundances of parent species in the coma are likely lower than the actual abundance on the cometary surface.

The detected abundances of each molecule in each comet are taken from Tables 2, 3, 4, and 5 in \citet{leroy2015}. If, for a given molecule in a given comet in these tables, multiple abundance values and/or ranges are given, then the smallest and the largest of these values (and/or range limits) are taken as the error range for the abundance of that molecule in that comet, and the average of the smallest and largest values is taken as the measured abundance of that molecule in that comet. Upper limit values, or upper limit ranges given in the tables in \citet{leroy2015} are considered non-detections, and are thus not included in the analysis. If only one abundance value is available for a molecule in a comet, then a conservative error estimate of 50\% of the observed value is assumed. This estimate is reasonable when compared with observed errors (see right-hand panels of Figures A1 to A5). 

In addition to these cometary abundance measurements, the abundance measurement of CO in comet 17P by \citet{qi2015} is included, using a CO/HCN ratio of 40$\pm5$ as representative of the bulk cometary composition of comet 17P. The measured abundance of HCN relative to \ce{H2O} from \citet{dellorusso2008} (HCN/\ce{H2O}=0.538 (\%)) is then used to normalise the CO abundance in comet 17P to the \ce{H2O} abundance (assuming the aforementioned 50\% error on the HCN/\ce{H2O} measurement). Propagating both the error on the measurement by \citet{qi2015}, and the 50\% estimated error of the measurement by \citet{dellorusso2008} leads to CO/\ce{H2O}=21.52$\pm$11.09 (\%) for comet 17P, which is used in the analysis here. An overview of the comets included in the analysis is shown in Table \ref{overview}, which also indicates which molecules have been measured in each comet.

It is noted that this work will refer to the different comets listed in Table \ref{overview}, either by using their $P$-identifiers (e.g. comets 2P, 9P, and 67P-W, where the extension ``-W'' indicates the winter hemisphere of comet 67P), or by their commonly used names, as listed in parentheses in Table \ref{overview} (e.g. comets Halley or Hyakutake). Therefore, please refer to this table for the cometary body identifier codes.

\subsection{Statistical comparison between observations and models}
\label{stat}
With observed abundances of several molecules available for all comets, along with the evolving spatial midplane abundances of those molecules, it can now be quantified how likely it is for a comet to have formed at a given time and place, assuming that the comets acquired all their ices at one time and place, and that the ices remained unaltered thereafter. For each comet, a log-space $\chi^{2}$-surface in time $t$ and radius $r$ is computed.  We do the analysis in log-space because of the large dynamical range (orders of magnitude) in the modelled abundances; hence, we consider a good agreement to lie within an order of magnitude of the observed ratio.  For each set of radius $r$ and time $t$, the $\chi^{2}$ value for a given comet is given by

\begin{equation}
\chi^{2}(r,t)=\sum_{i=1}^{n} \frac{\left(\log(n_{i, \rm{obs}}(r,t))-\log(n_{i, \rm{mod}}(r,t))\right)^{2}}{\left(\sigma'_{i, \rm{obs}}(r,t)\right)^{2}},
\label{chi2}
\end{equation}

where $n_{i, \rm{obs}}(r,t)$ and $n_{i, \rm{mod}}(r,t)$ are the observed and modelled abundances of species $i$ with respect to \ce{H2O} ice at $(r, t)$. $\sigma'_{i, \rm{obs}}(r,t)$ is defined as

\begin{equation}
\begin{split}
\sigma'_{i, \rm{obs}}(r,t)=0.434\frac{\sigma_{i, \rm{obs}}(r,t)}{n_{i, \rm{obs}}(r,t)}
\label{sigmaprime}
\end{split}
\end{equation}

where $\sigma_{i,\rm{obs}}(r,t)$ is the observed or estimated error on the abundance, and $\sigma'_{i,\rm{obs}}(r,t)$ is the propagated error appropriate when using the log-function.

The $\chi^{2}$-surface for each comet is then transformed into a maximum likelihood function $P(r,t)$ using

\begin{equation}
P(r,t)= e^{-\chi^{2}(r,t)/2}
\label{likely}
\end{equation}

and subsequently all sets of $P(r,t)$ for each comet are normalised by the maximum $P$-value for that comet. The resulting surfaces of maximum likelihood show contours of different colours in different regions of parameter space, with colours depending on how close a region is to the maximum likelihood value for the comet. Regions of parameter space close to the maximum likelihood value for a comet are in turn the regions showing best agreement between the chemical models and the observed cometary abundances.


These maximum likelihood ($P$-value) surfaces are shown for all comets in Fig. \ref{P_res_evol_percent}, which compares observations with models of the reset scenario, and in Fig. \ref{P_inh_evol_percent} for models of the inheritance scenario. The $x$-axes in each panel in each figure are radial distance from the star in AU, and the $y$-axes are chemical evolution time in Myr. For each contour level the value of the contour indicates where the fraction of the local $P$-value to the maximum $P$-value is above a certain level, for each comet. Regions of yellow contour indicate good agreement between models and observations, whereas darker colours indicate poorer agreements.

The comet names, and dynamical types from \citet{cochran2012}, are listed in all panels. In each panel is also given the number of molecular detections for each comet \citep[from][]{leroy2015}, with the panels from left to right, and top to bottom featuring decreasing numbers of molecular detections per comet. This way, the first panel with ten molecular detections (Comet 67P-S) can be distinguished from the last panel with only two molecular detections (Comet 21P), in that more molecular detections in a comet should make the comparison between the models and the observations more robust.

\section{Results}
\label{results}

\subsection{Full sample of species: reset scenario}
Figure \ref{P_res_evol_percent} features the maximum likelihood surfaces for the reset scenario with the full sample of species. For all comets, there are regions of the parameter space that show good agreement between models and observations. All comets are in good agreement with formation between 11-13 AU by $\sim$8 Myr evolution, and most comets, excluding comets 67P-W, LINEAR, and 17P, also show good agreement with formation between 27-30 AU by $\sim$0.5-1 Myr evolution. For comets Lemmon, 103P, 9P, 6P, 2P, and 21P, the two aforementioned regions of parameter space are connected with the contours at the chosen levels (down to 10$^{-4}$ relative to the maximum likelihood value for each comet). For comets 103P and 2P there is a high degree of degeneracy in radius and time, as the maximum contours for these comets follow a trail spanning from the aforementioned 30 AU by $\sim$1 Myr inwards to $\sim$12 AU by 8 Myr evolution. This trail is marked by the red shaded region overplotted for comet 2P, and this trail overlaps with the regions of highest likelihood for all comets. 

This trail, in turn, roughly traces the vicinity of the CO iceline (at $T\sim$ 21 K), as is seen in the left panel of Fig. 4 in \citet{eistrup2018}. That figure, amongst others, shows the changing location of the CO iceline in the physically evolving disk midplane utilised in that work. This is interesting, because it points to all the comets here agreeing well with formation in the vicinity of the CO iceline in the pre-solar nebula. Additionally, all comets show high likelihood of formation at 6-13 AU by 7-8 Myr, a range encompassing also the \ce{CH4} iceline \citep[see again Fig. 4, left panel, in][]{eistrup2018}.

Comet 103P is the only comet showing some likelihood of formation at $\sim$30 AU by 2-3 Myr of evolution. However, the maximum likelihood of formation for this comet is at 10-15 AU by $\sim$8 Myr, which is consistent with the rest of the comets. 


\subsection{Full sample of species: inheritance scenario}
Turning to the inheritance scenario the maximum likelihood surfaces for all comets are presented in Fig. \ref{P_inh_evol_percent}. All comets in this scenario share good agreement with formation at $\sim$12 AU by 8 Myr, similar to the case for the reset scenario in Fig. \ref{P_res_evol_percent}. All comets also show agreement with late formation (7-8 Myr) from small radii out to 10 AU, and some (103P, 6P, Lovejoy and 21P) also agree well with having formed inside 5 AU at various times during the evolution. The trail along the CO iceline that was seen for the reset scenario is not reproduced for all comets in the inheritance scenario. However, Hale-Bopp, LINEAR, Lemmon, 103P, 9P, 6P, and 21P do somewhat agree with parts of the trail (103P and 6P are both in good agreement with the trail). Most of the comets for the inheritance scenario also agree with formation at 25-30 AU by <1 Myr evolution. However, the best agreements between models and observations are for late evolutionary times.

Lastly, addressing both the reset scenario in Fig. \ref{P_res_evol_percent} and the inheritance scenario in Fig. \ref{P_inh_evol_percent} it is generally seen that the more species that are observed in a comet (the top panels of each figure), the smaller the regions in parameter space over which the models well reproduce the observed ratios, thus better constraining the potential formation location and time of formation.

\subsection{Correlation for C- and O-carrying species only}

Given the interlinked chemical nature of carbon and oxygen, and the fact that the modelled nitrogen and sulphur chemistry is less well understood, it is interesting to have a look at maximum likelihood surfaces for the comets excluding sulphur species, and considering only ice species with carbon and oxygen. In the sample of molecules from \citet{leroy2015} these species are: \ce{C2H6}, \ce{CO2}, CO, \ce{H2CO}, \ce{CH3OH}, \ce{CH4}, and \ce{O2}, thus seven in total.

For these seven ice species maximum likelihood surfaces are shown in Fig. \ref{P_res_evol_no_sn_percent} for the reset scenario and in Fig. \ref{P_inh_evol_no_sn_percent} for the inheritance scenario. Since sulphur species have only been observed for comets 1P, Hale-Bopp, Hyakutake, LINEAR, 73P, 9P, and 67P (both seasons), only these comets are relevant to analyse for differences compared to the analysis with the full sample of species. 

\begin{figure*}[h!]
\subfigure{\includegraphics[width=0.33\textwidth]{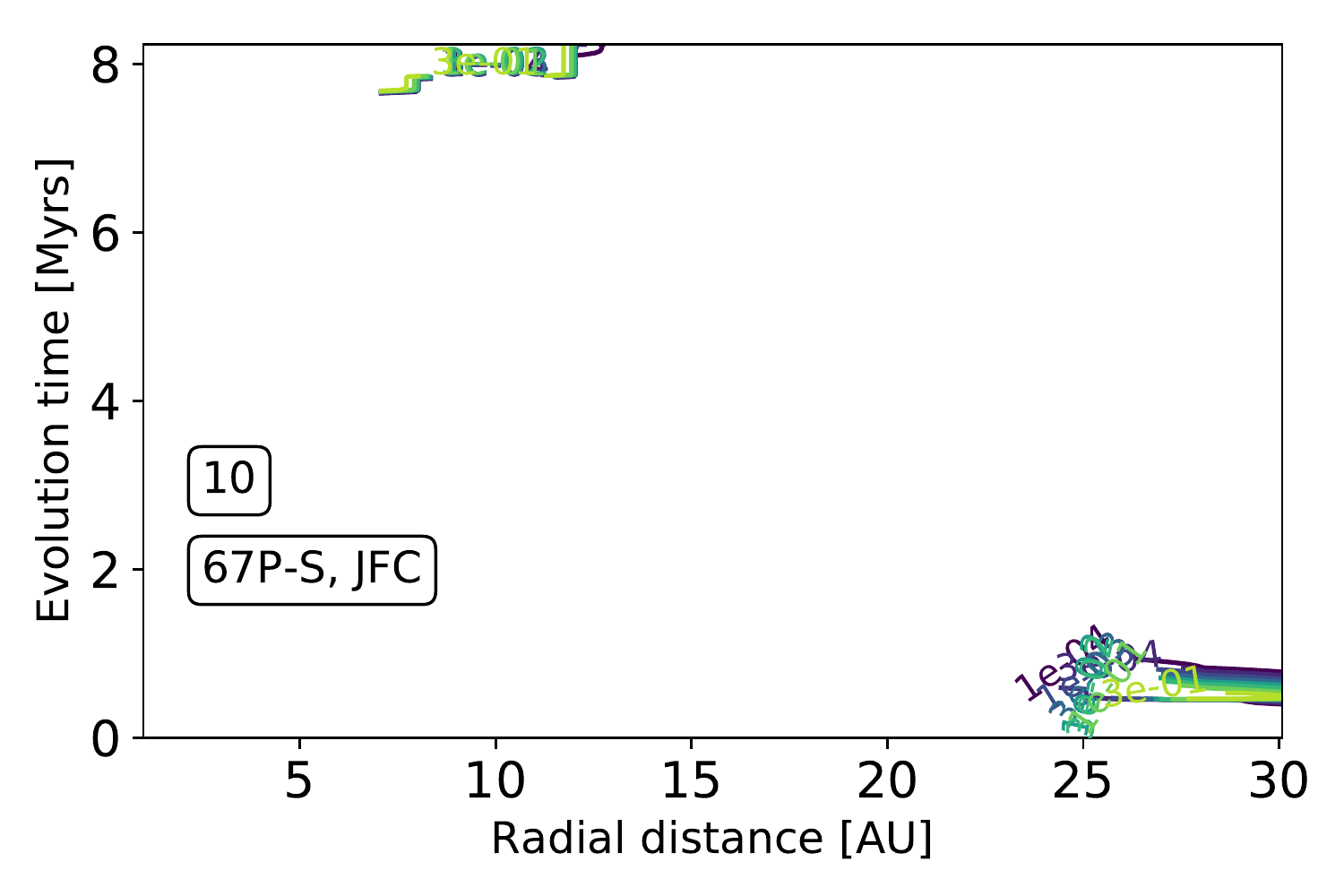}}
\subfigure{\includegraphics[width=0.33\textwidth]{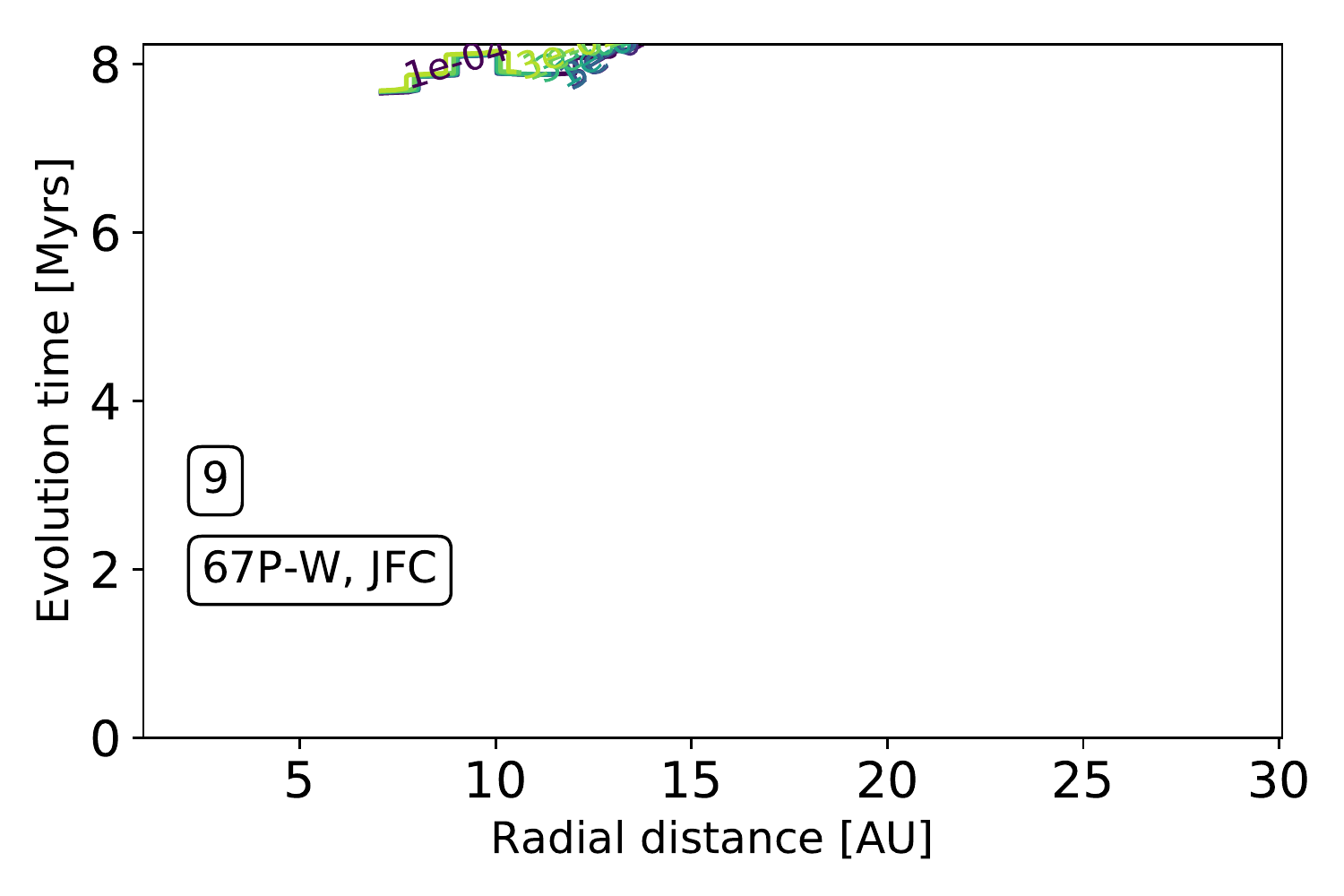}}
\subfigure{\includegraphics[width=0.33\textwidth]{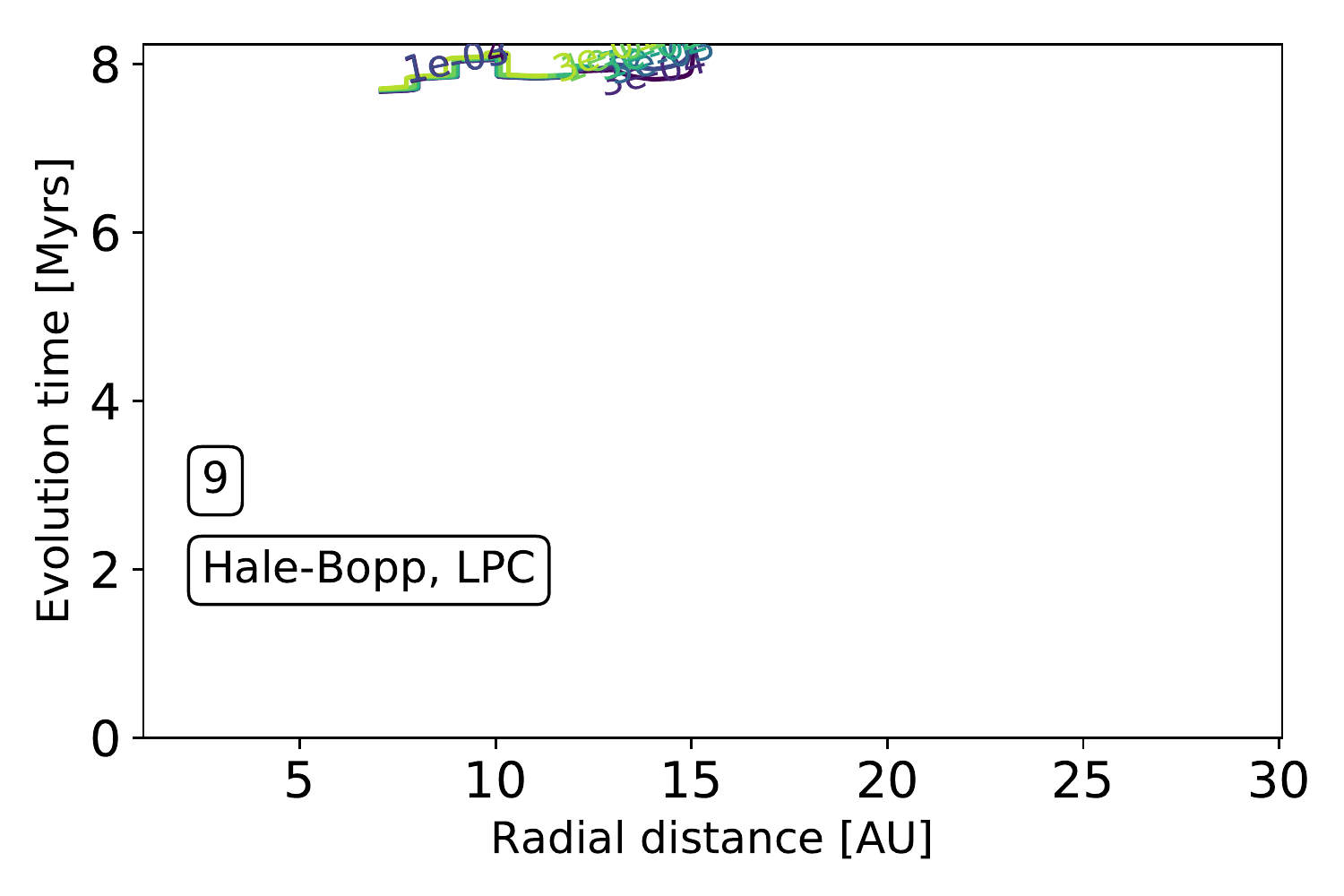}}\\
\subfigure{\includegraphics[width=0.33\textwidth]{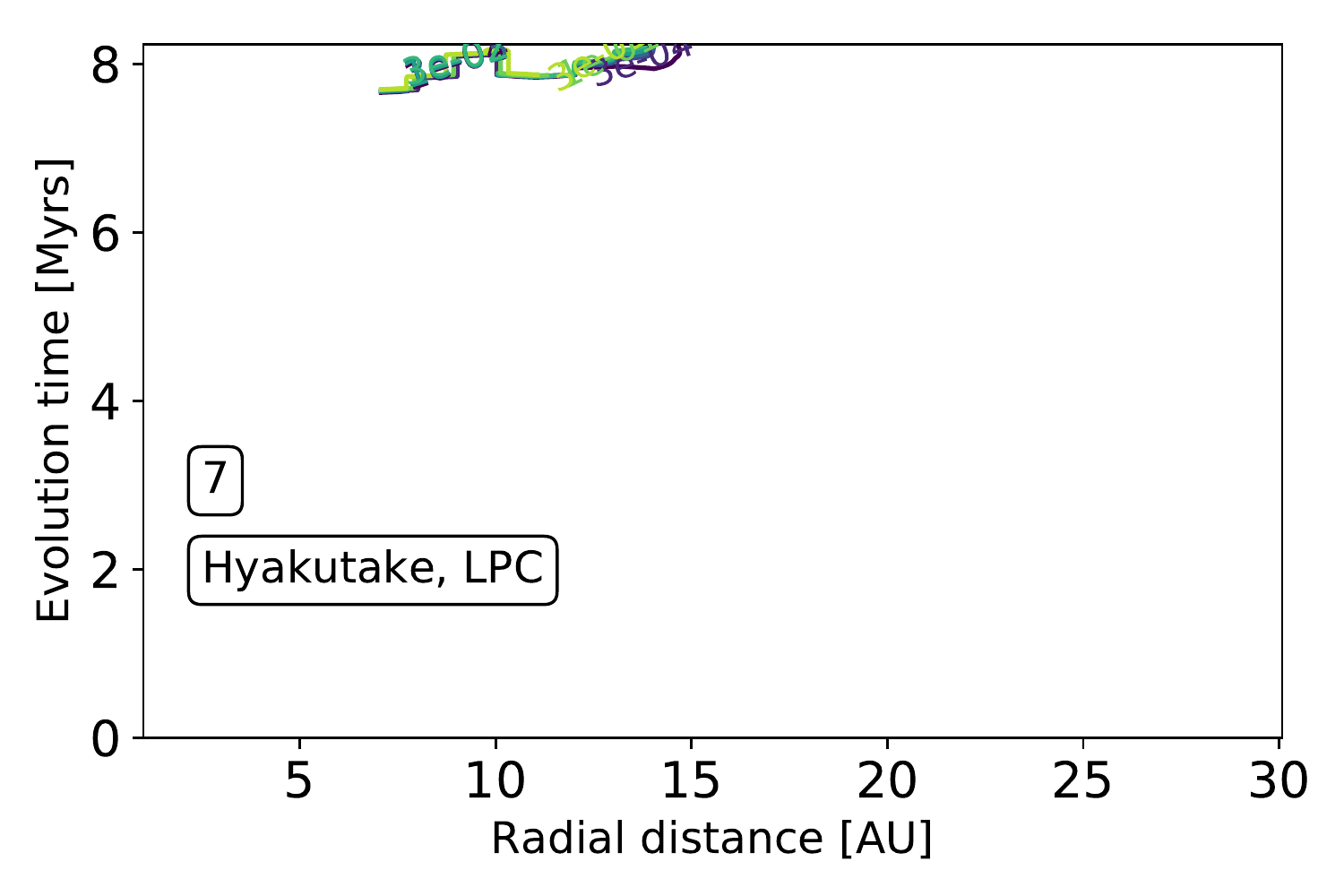}}
\subfigure{\includegraphics[width=0.33\textwidth]{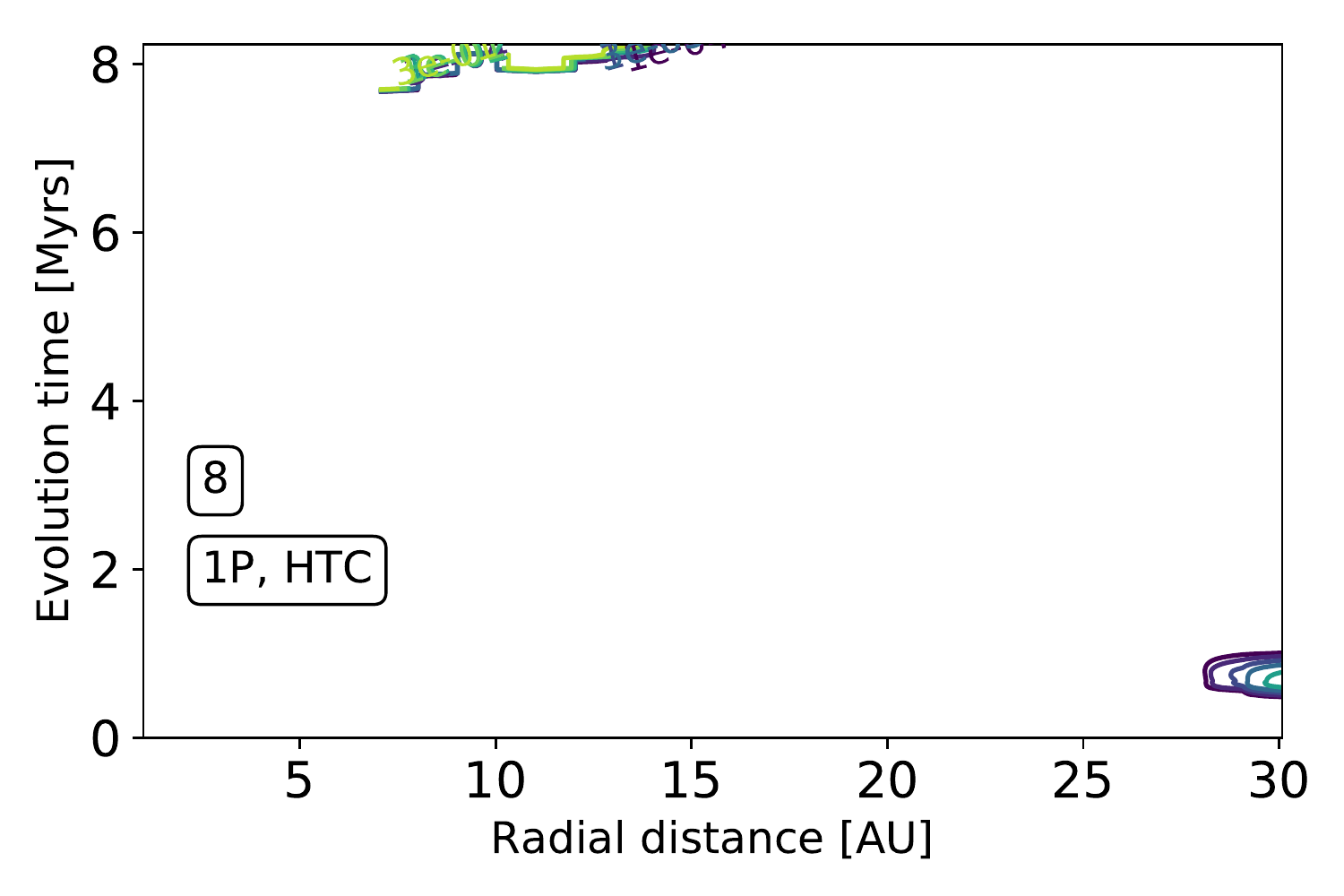}}
\subfigure{\includegraphics[width=0.33\textwidth]{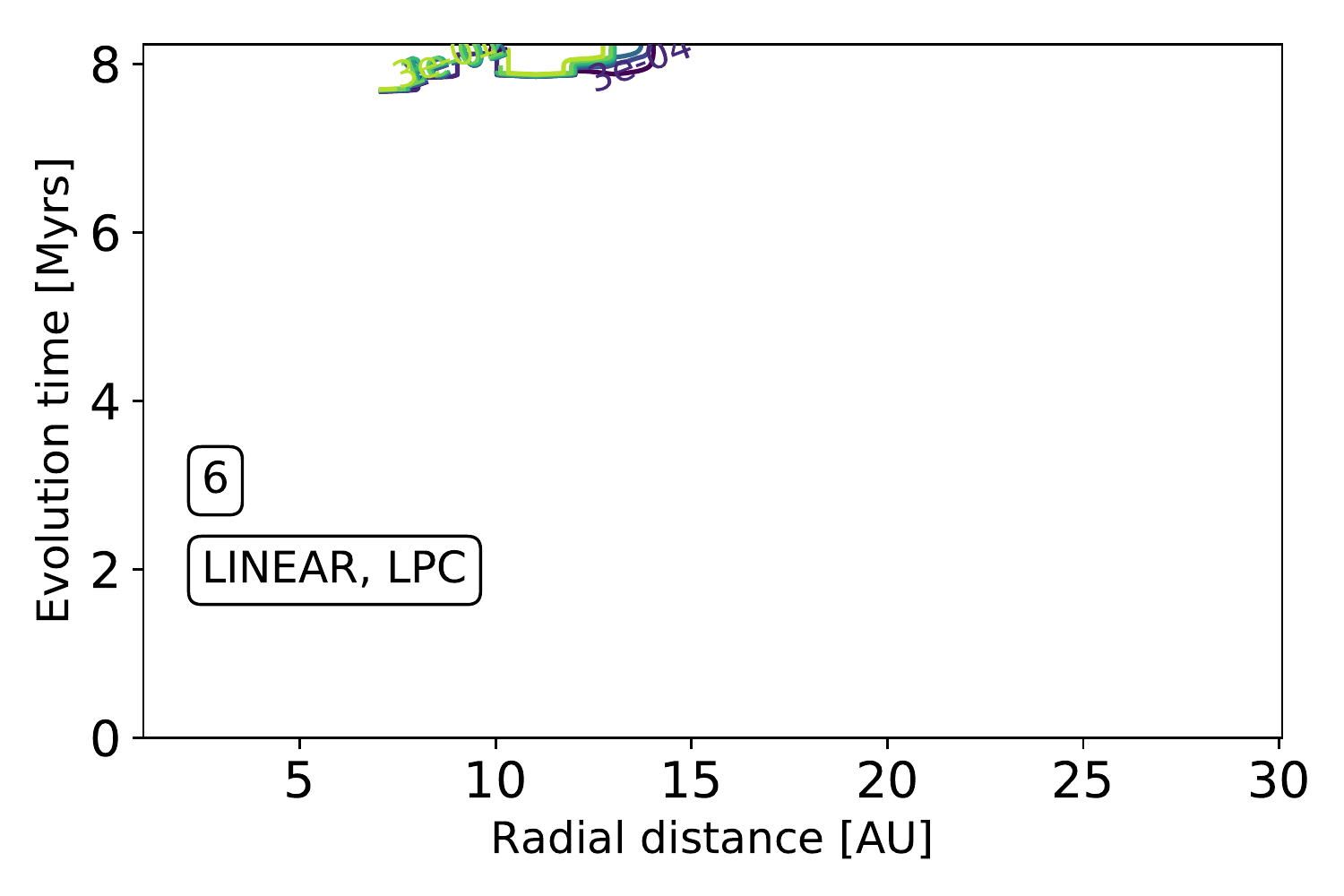}}\\
\subfigure{\includegraphics[width=0.33\textwidth]{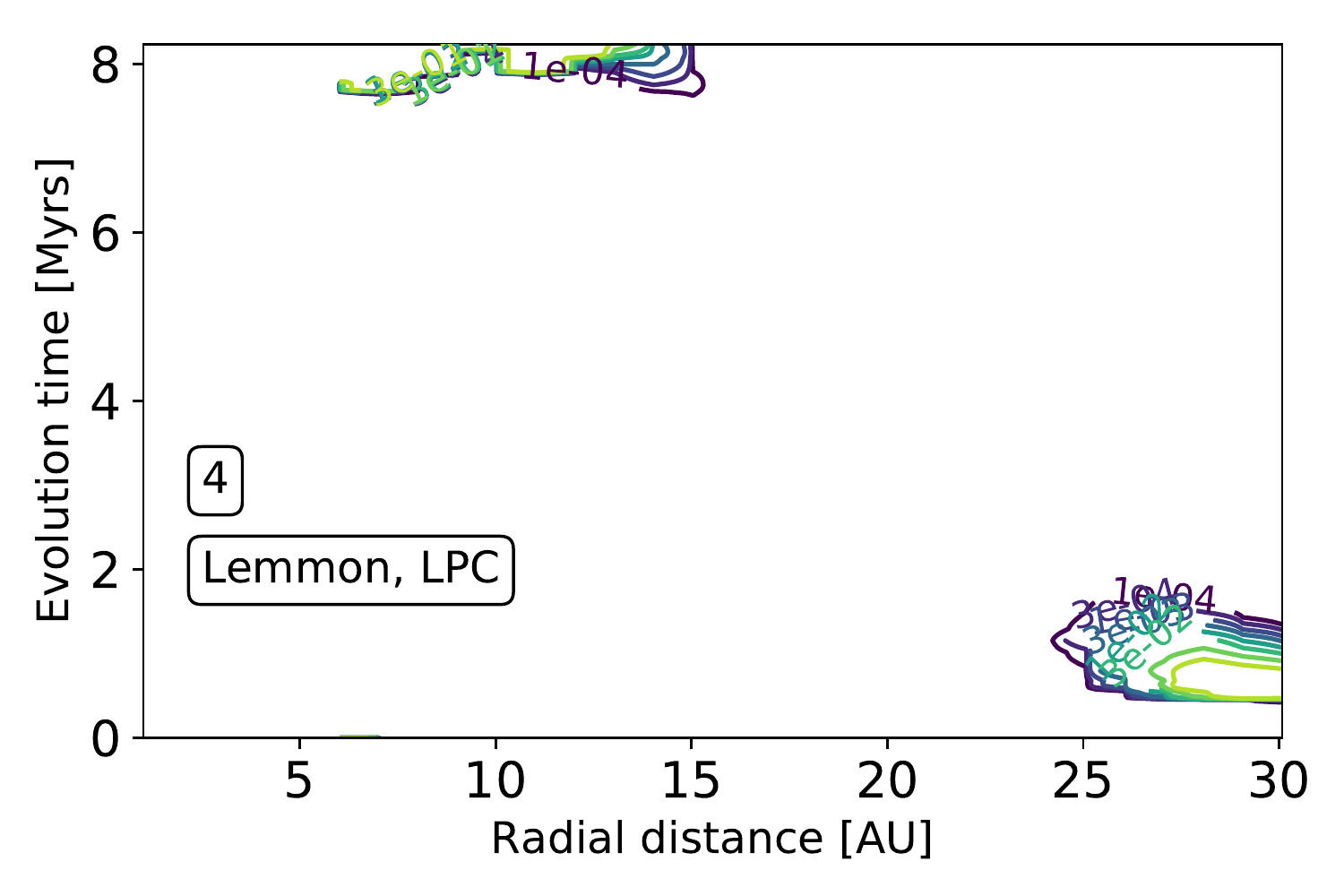}}
\subfigure{\includegraphics[width=0.33\textwidth]{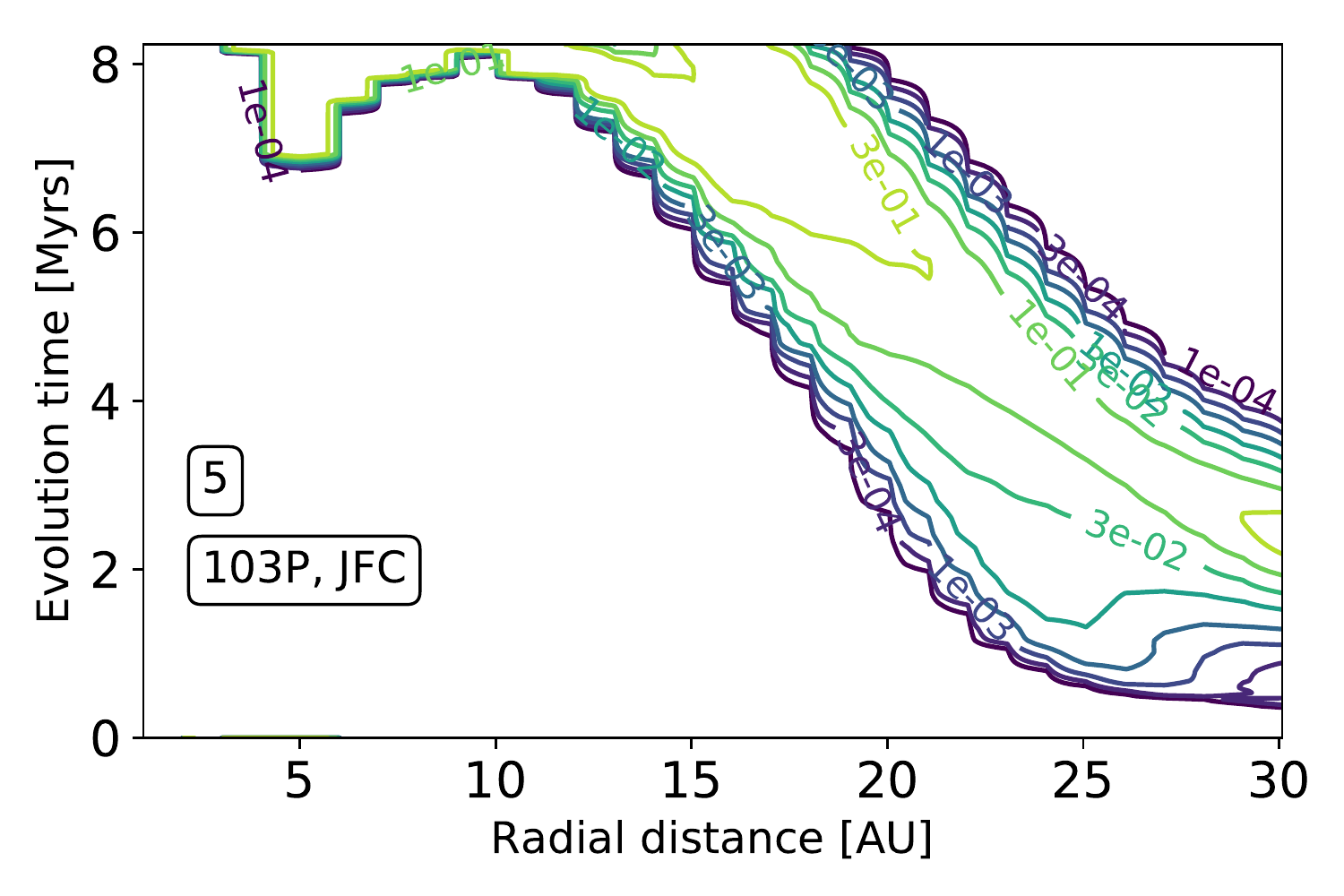}}
\subfigure{\includegraphics[width=0.33\textwidth]{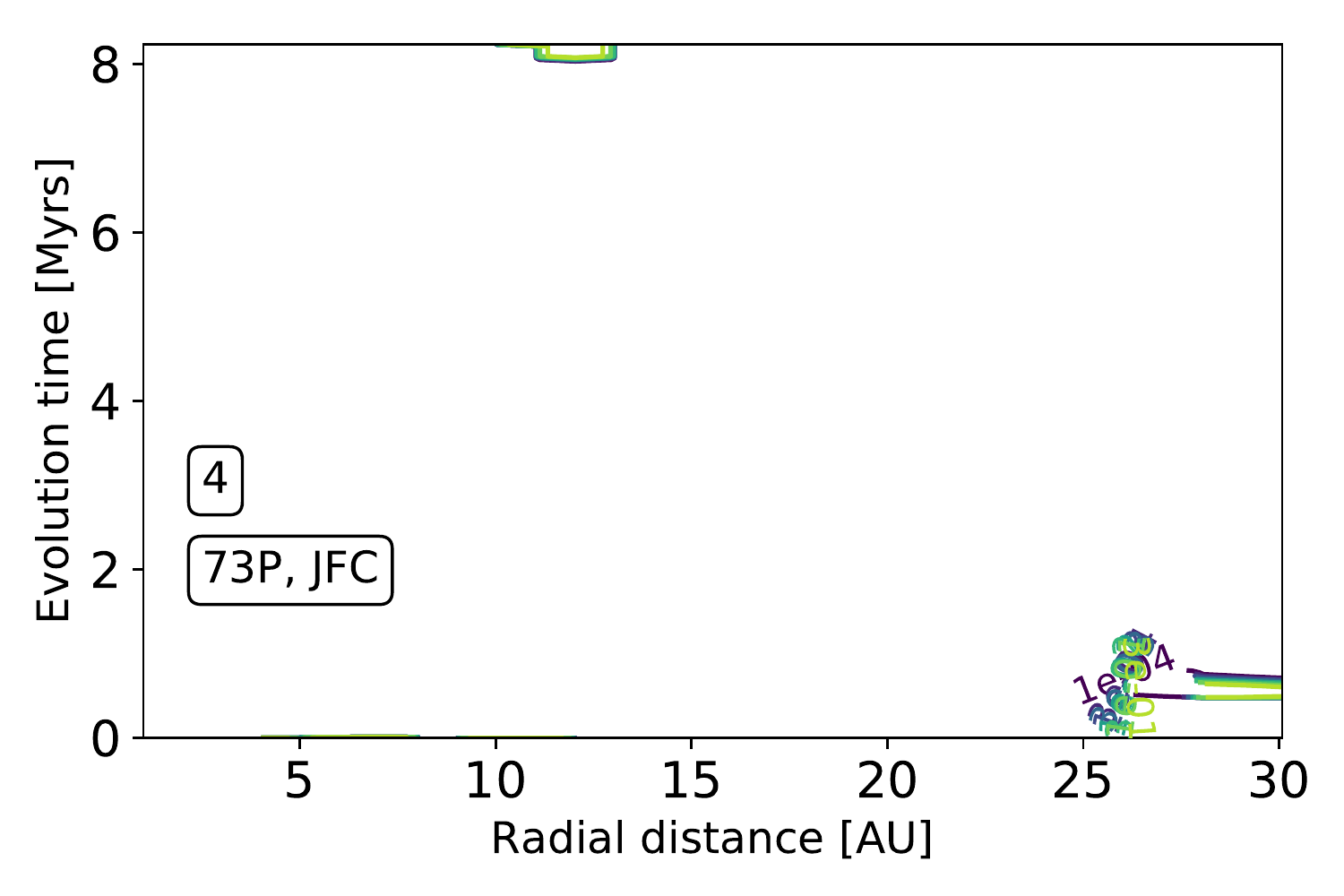}}\\
\subfigure{\includegraphics[width=0.33\textwidth]{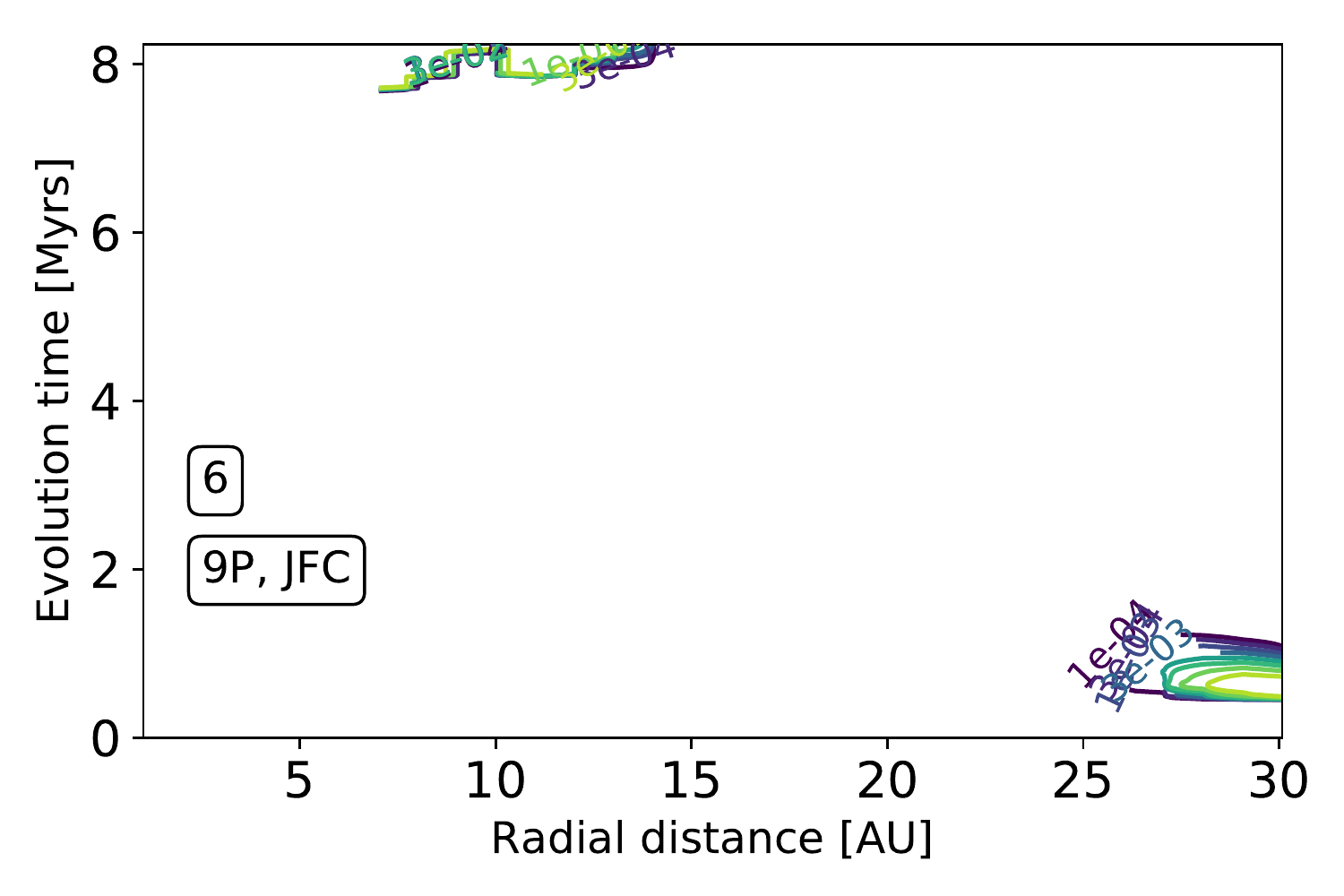}}
\subfigure{\includegraphics[width=0.33\textwidth]{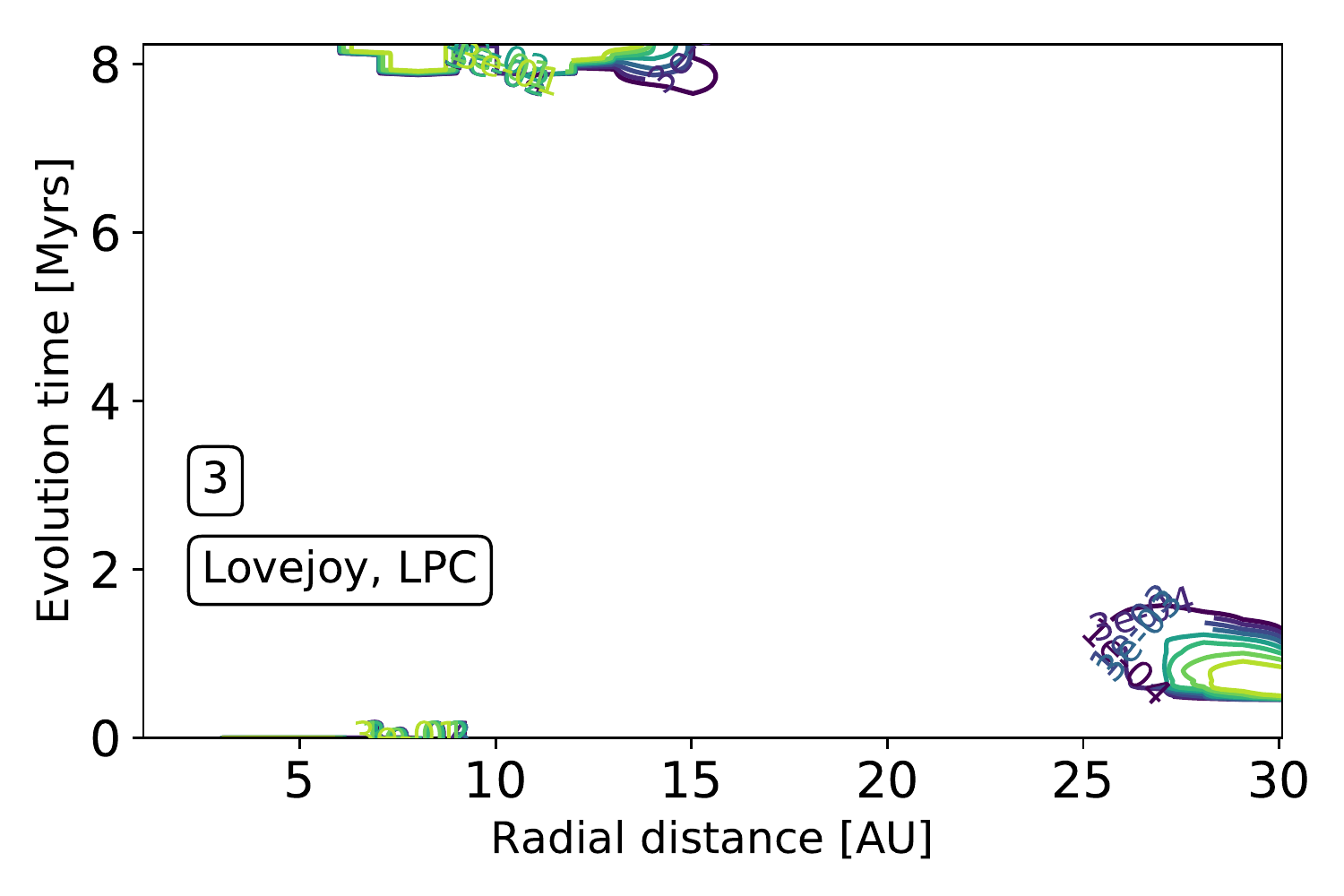}}
\subfigure{\includegraphics[width=0.33\textwidth]{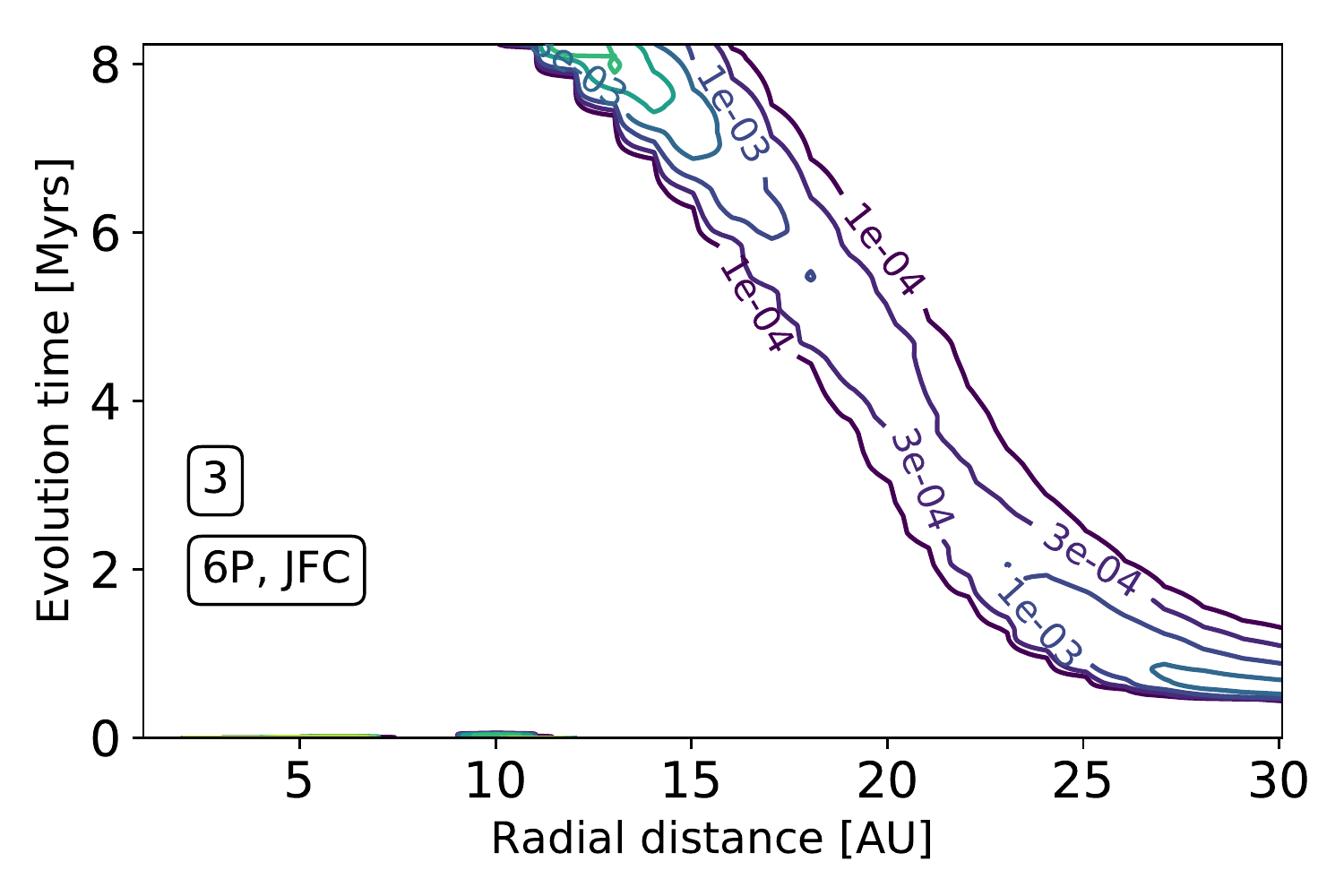}}\\
\subfigure{\includegraphics[width=0.33\textwidth]{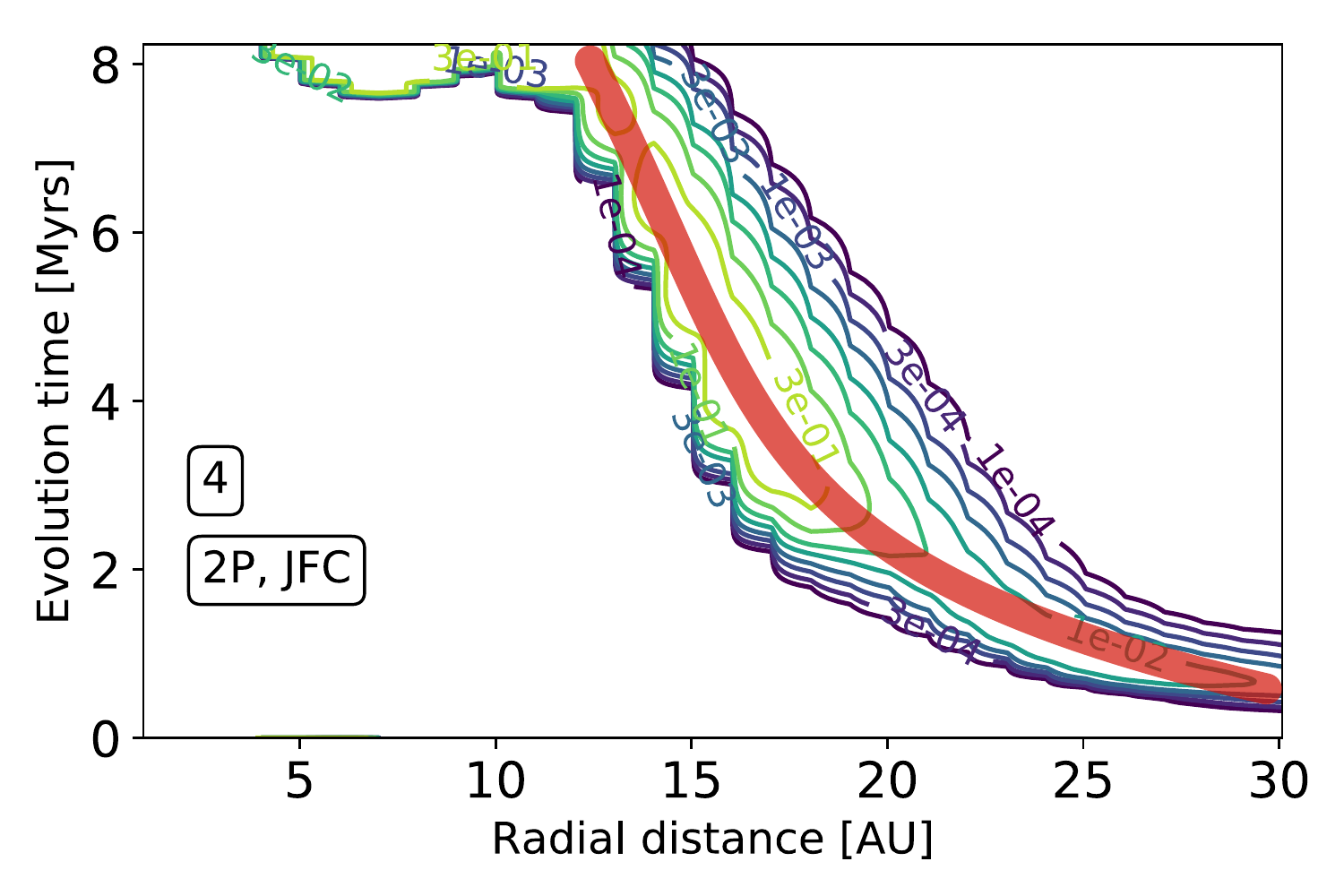}}
\subfigure{\includegraphics[width=0.33\textwidth]{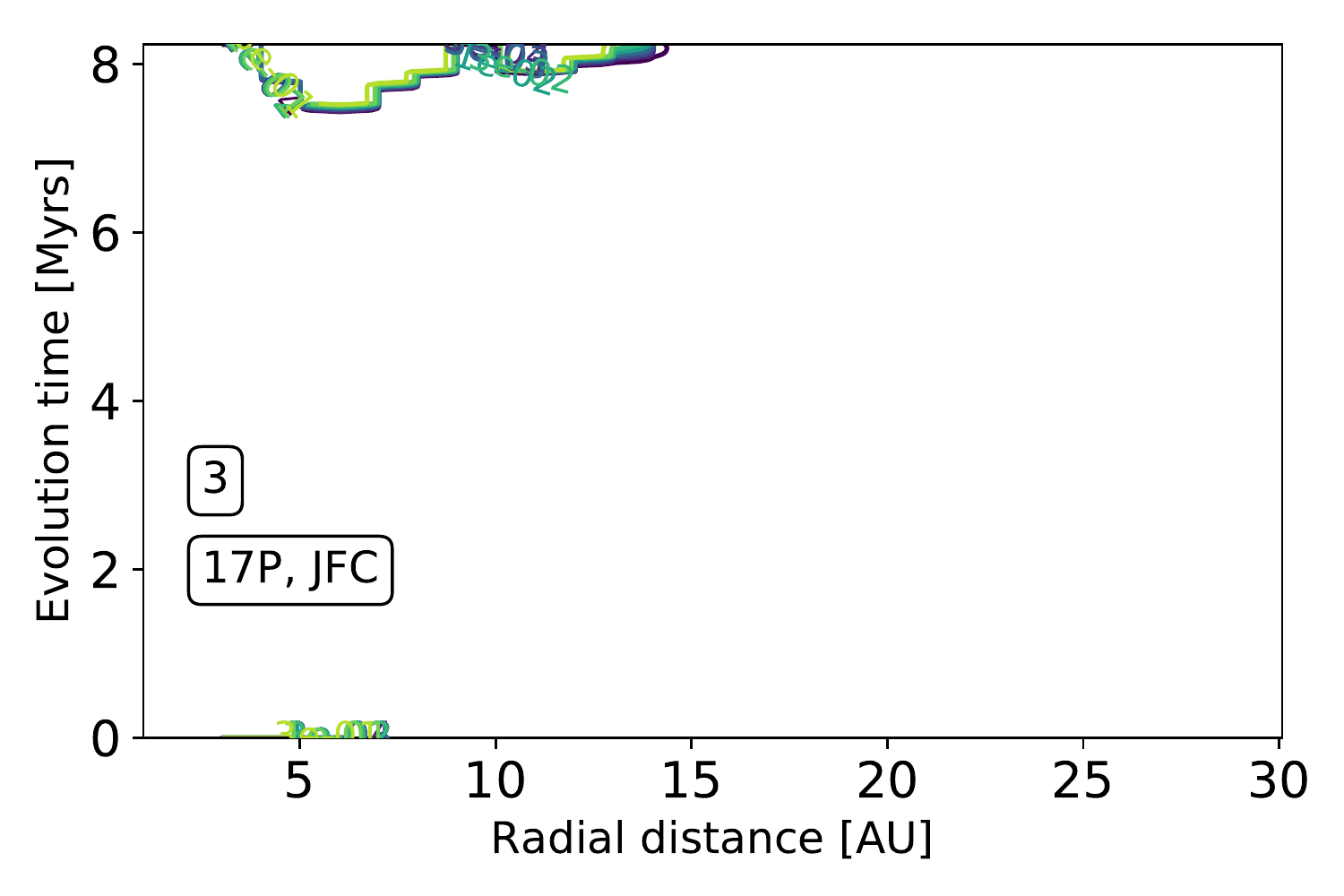}}
\subfigure{\includegraphics[width=0.33\textwidth]{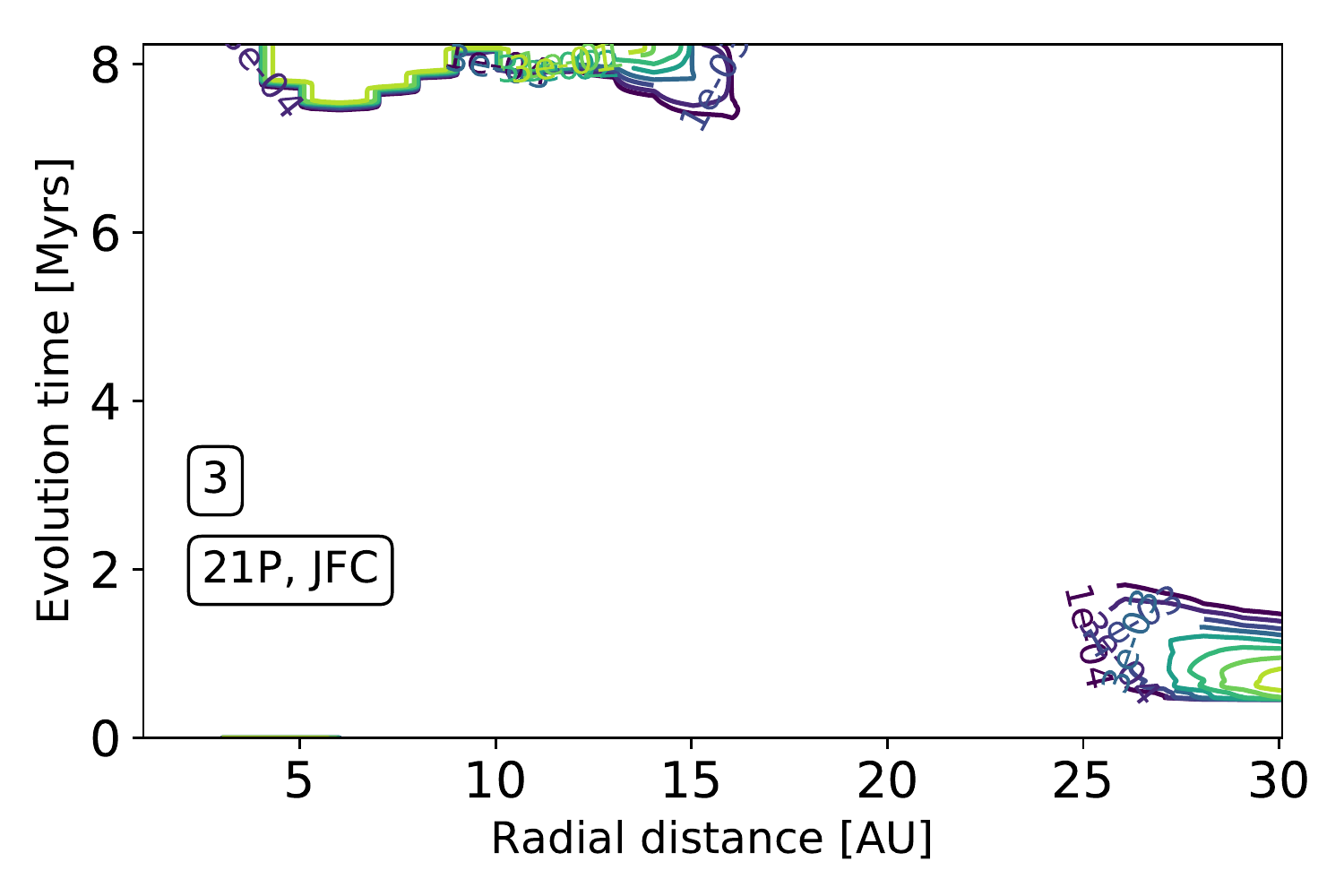}}\\
\caption{Maximum likelihood surfaces for reset scenario, for full sample of molecular species. The maximum likelihood increases from dark purple contours to light green contours. Darkest purple contour represents maximum likelihood of $10^{-4}$, and increases with each lighter contour level to 3x$10^{-4}$, $10^{-3}$, 3x$10^{-3}$, $10^{-2}$, 3x$10^{-2}$, $10^{-1}$, and finally 3x$10^{-1}$ for the lightest green contour. Radius in AU in the physically evolving protoplanetary disk midplane is on the $x$-axis, and evolution time in Myrs is on the $y$-axis. The red shaded region for comet 2P indicates the trail through parameter space (largely tracing the vicinity of the CO iceline) on which most of the comets show good agreement with formation.}
\label{P_res_evol_percent}
\end{figure*}

\begin{figure*}[h!]
\subfigure{\includegraphics[width=0.33\textwidth]{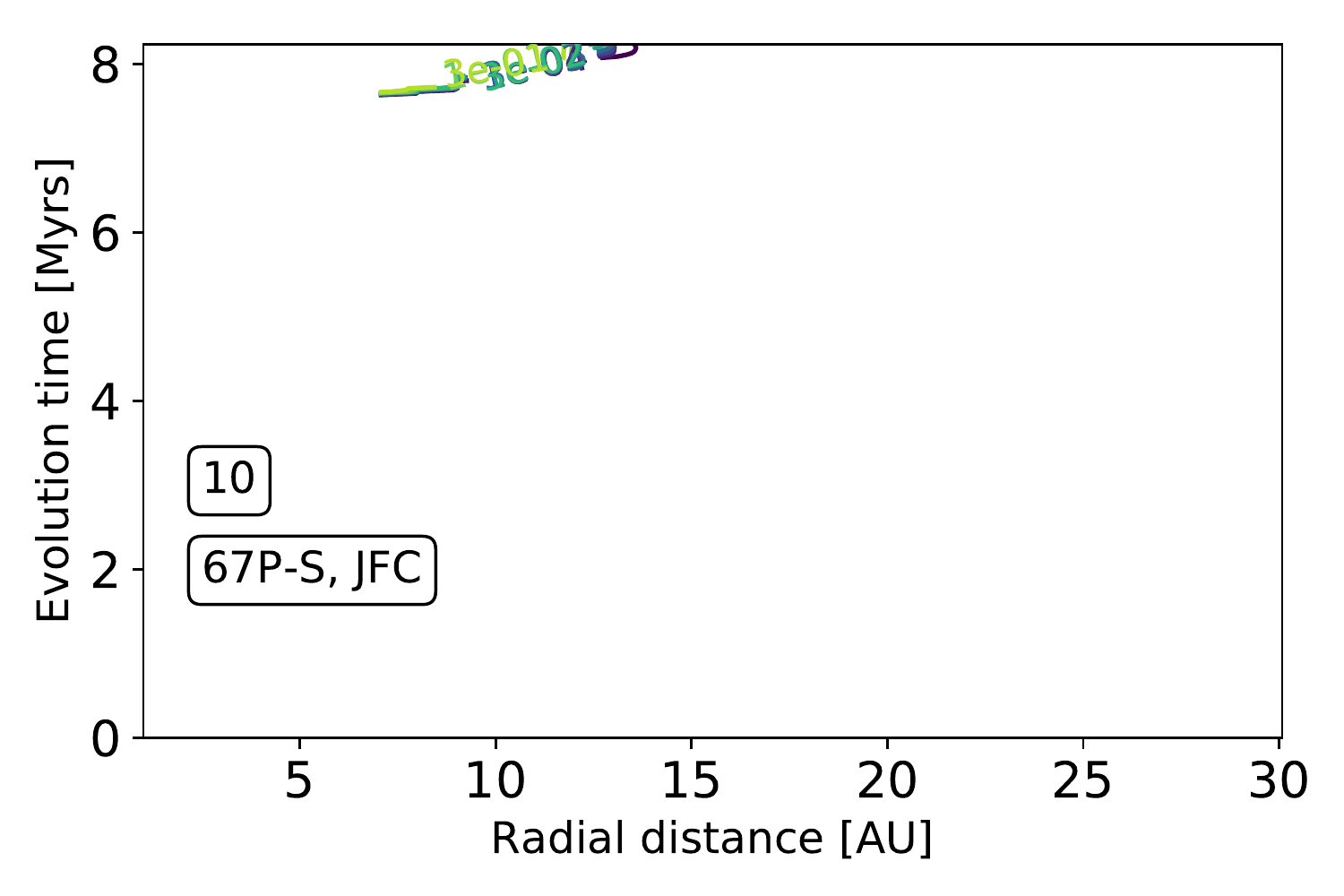}}
\subfigure{\includegraphics[width=0.33\textwidth]{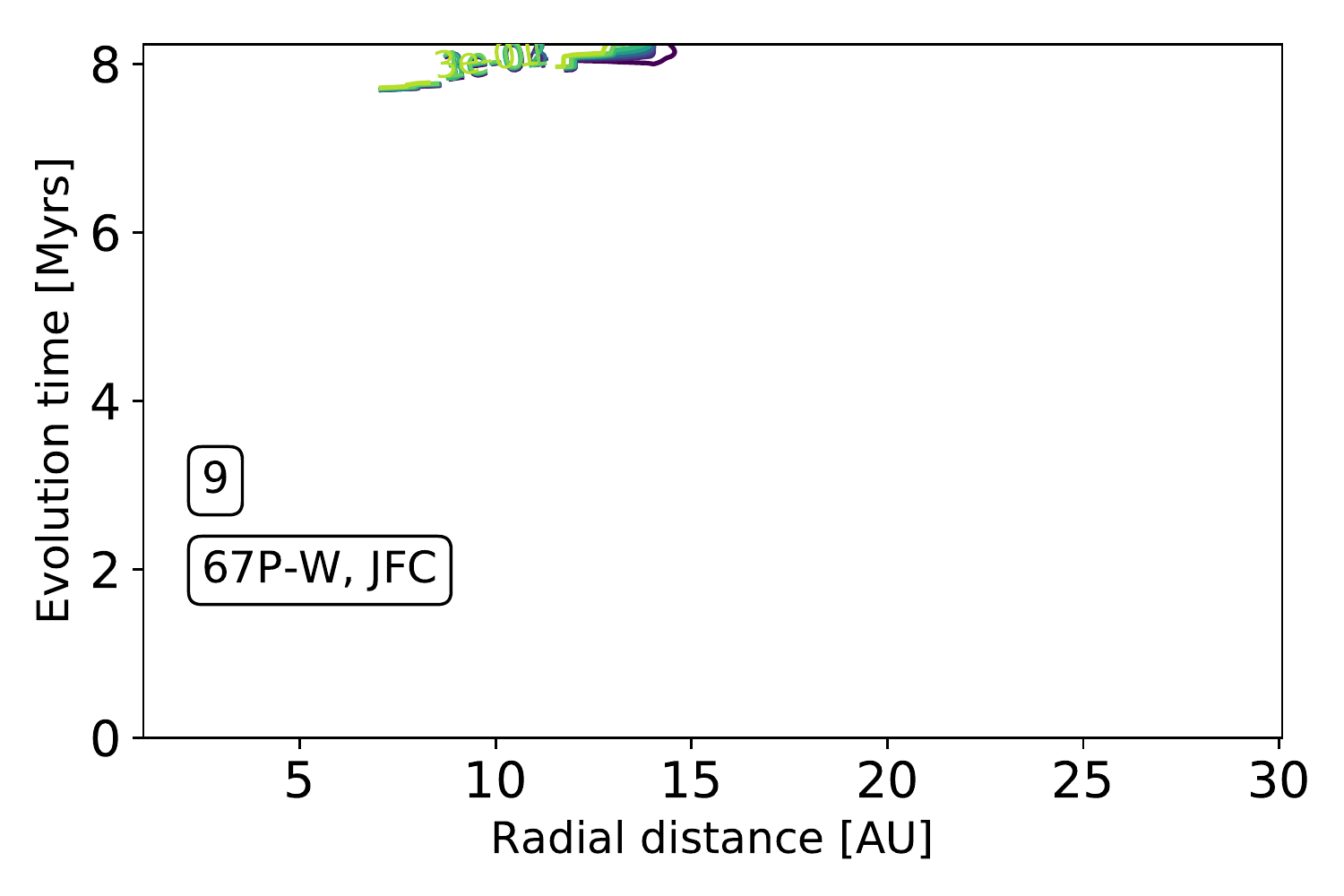}}
\subfigure{\includegraphics[width=0.33\textwidth]{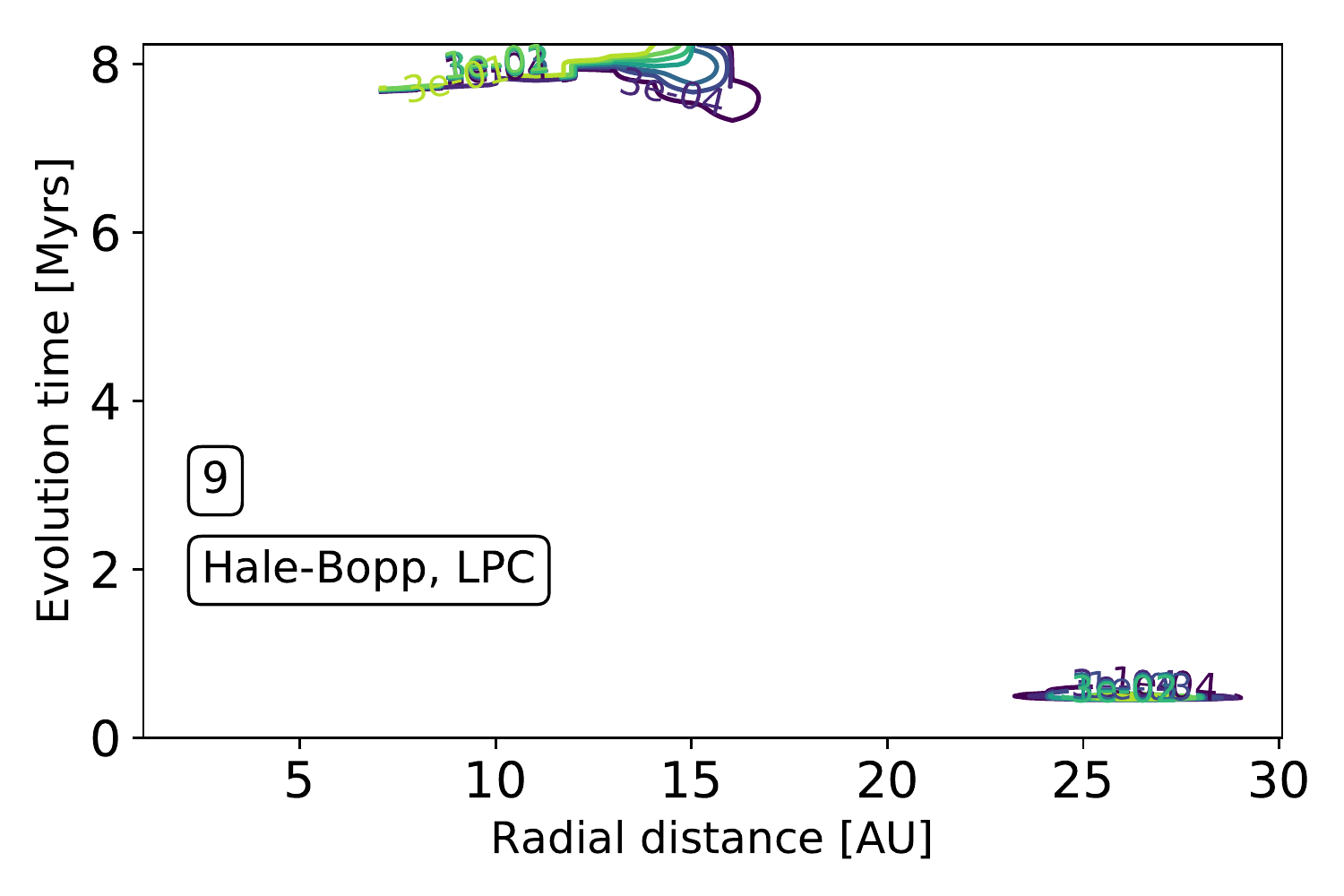}}\\
\subfigure{\includegraphics[width=0.33\textwidth]{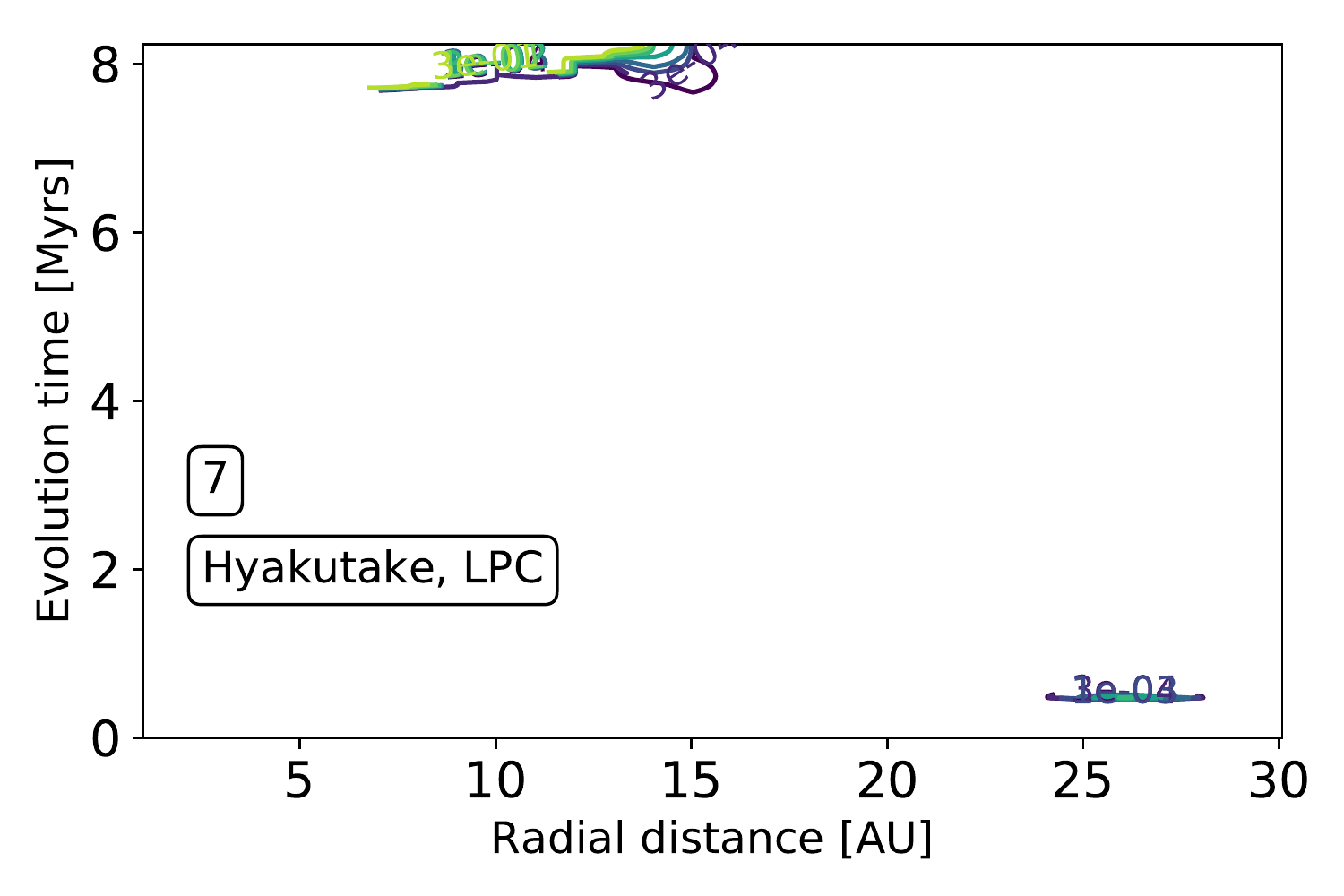}}
\subfigure{\includegraphics[width=0.33\textwidth]{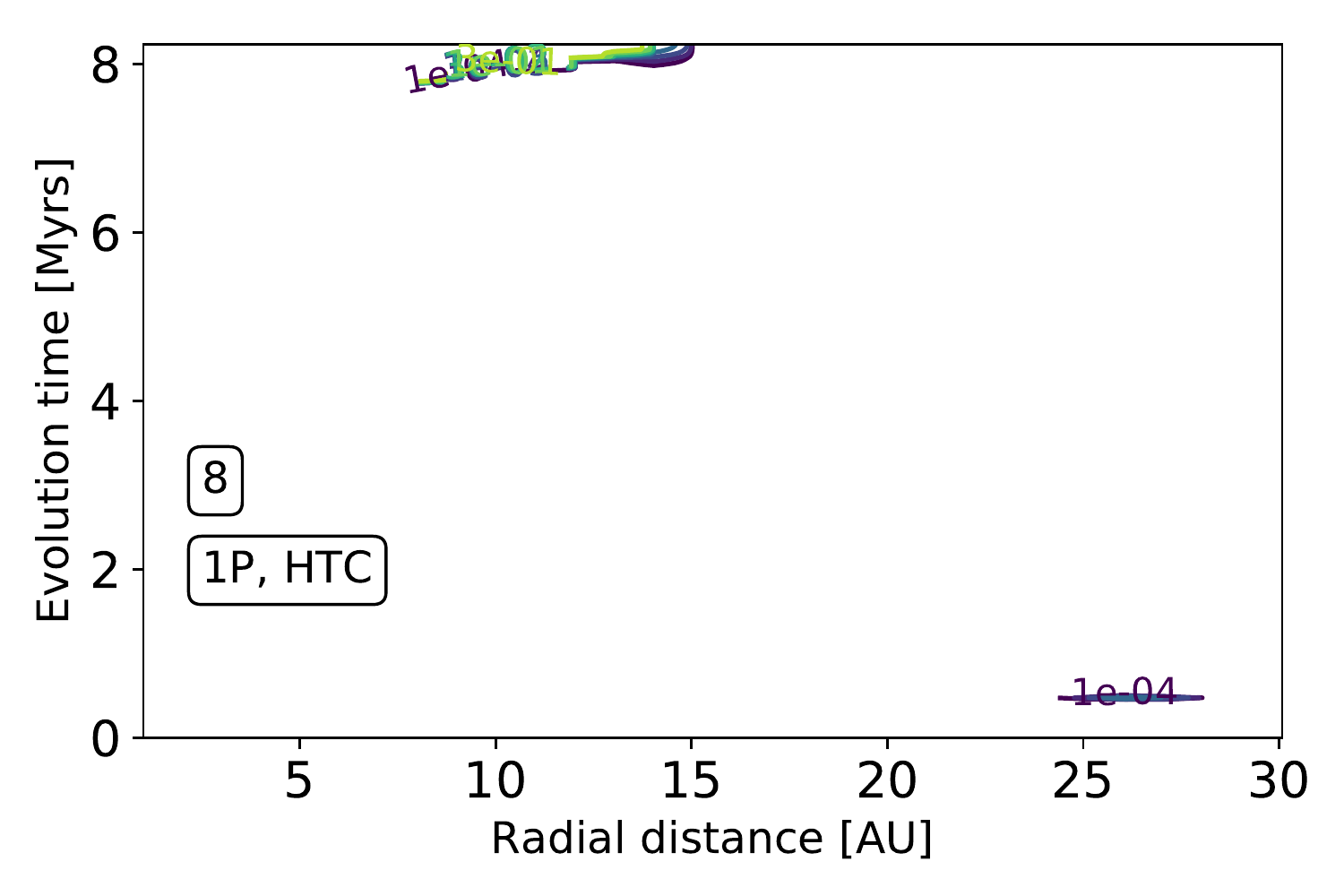}}
\subfigure{\includegraphics[width=0.33\textwidth]{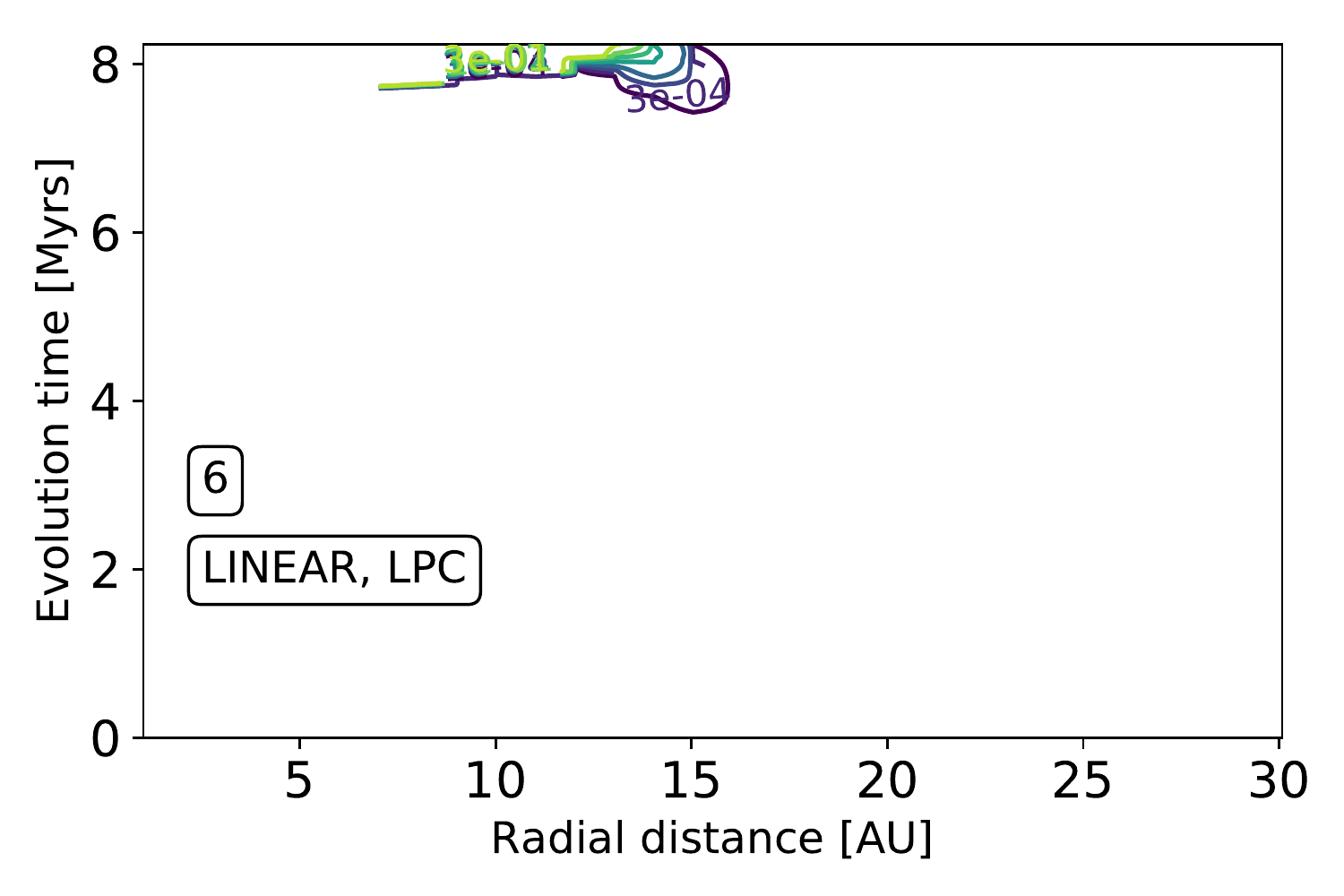}}\\
\subfigure{\includegraphics[width=0.33\textwidth]{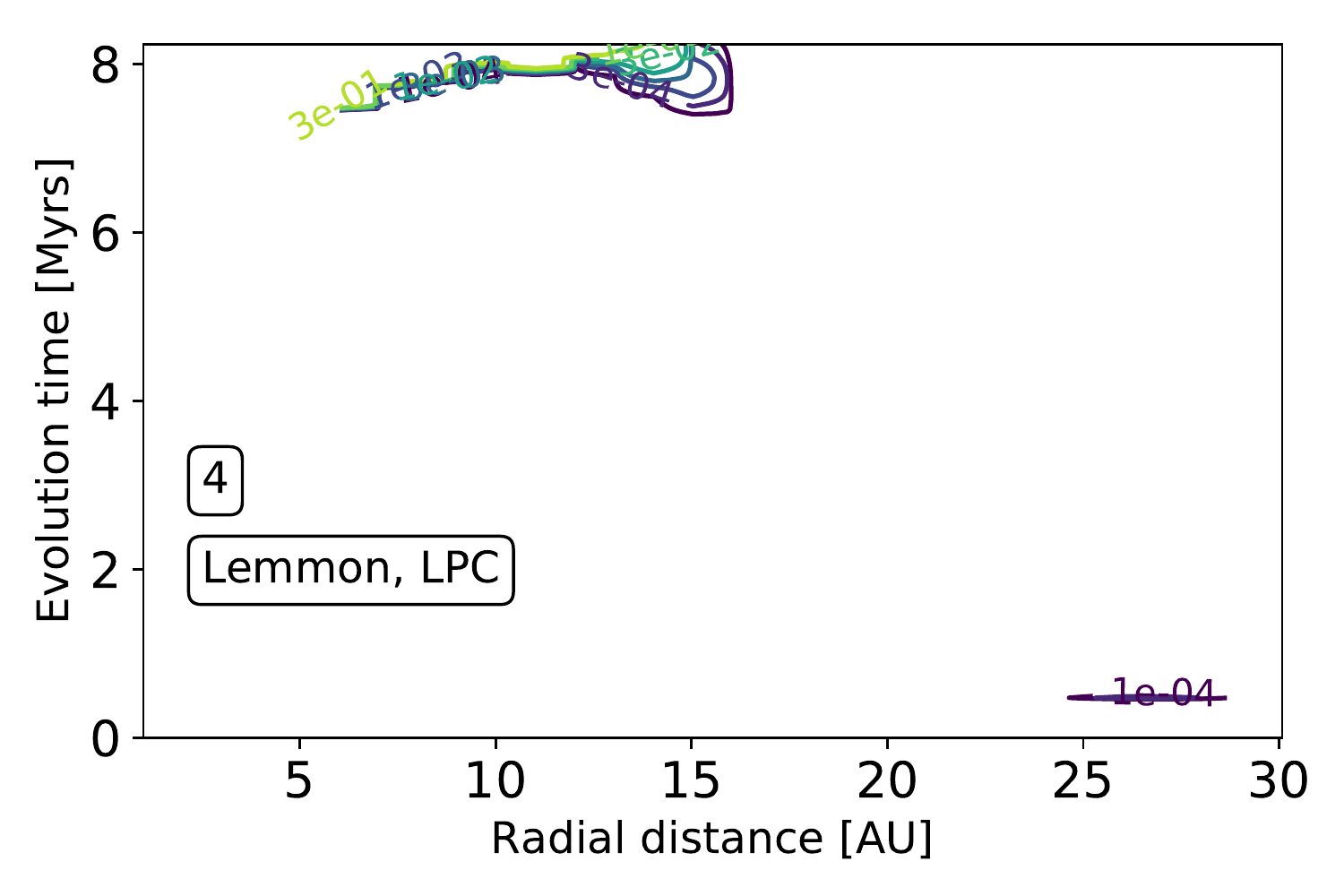}}
\subfigure{\includegraphics[width=0.33\textwidth]{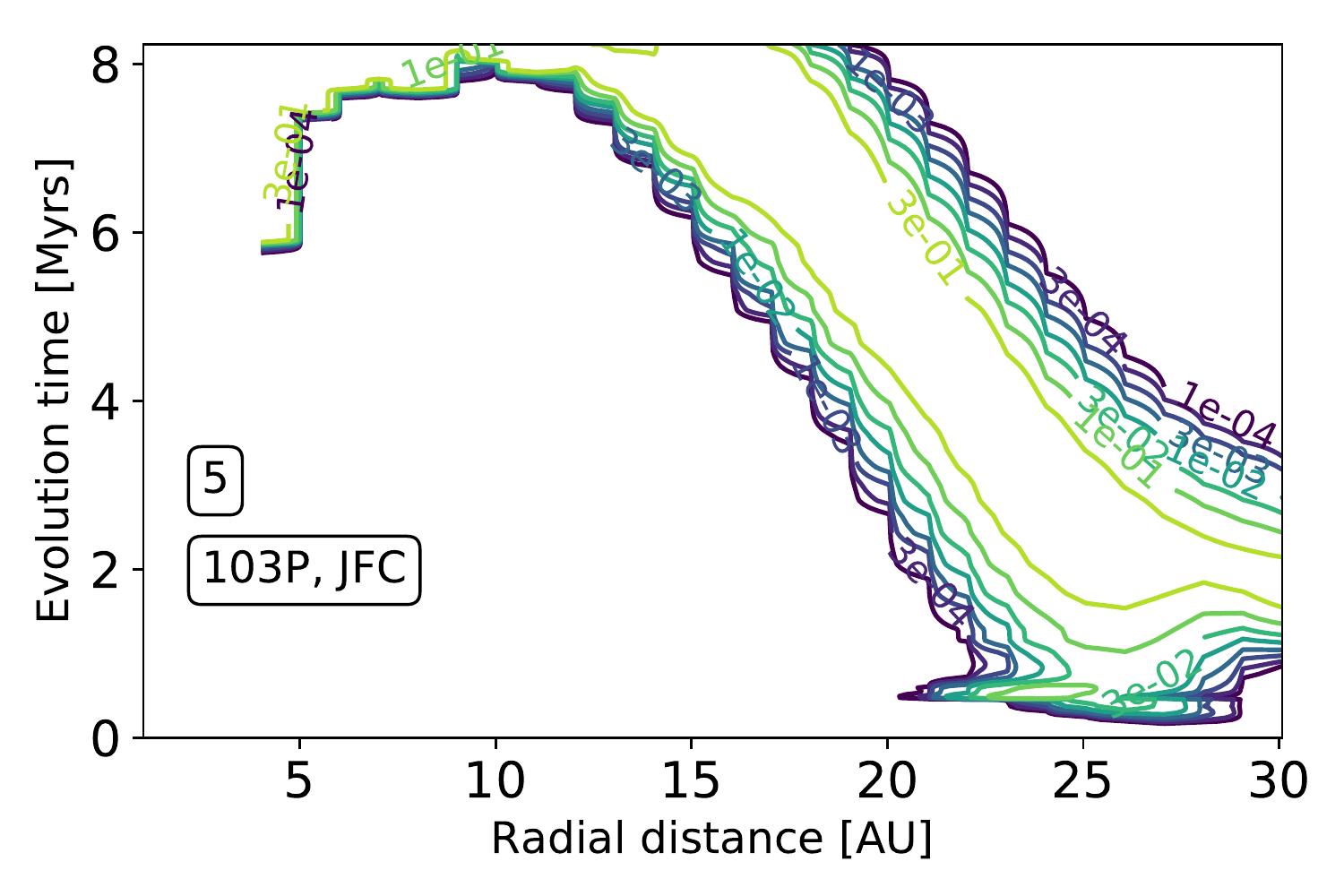}}
\subfigure{\includegraphics[width=0.33\textwidth]{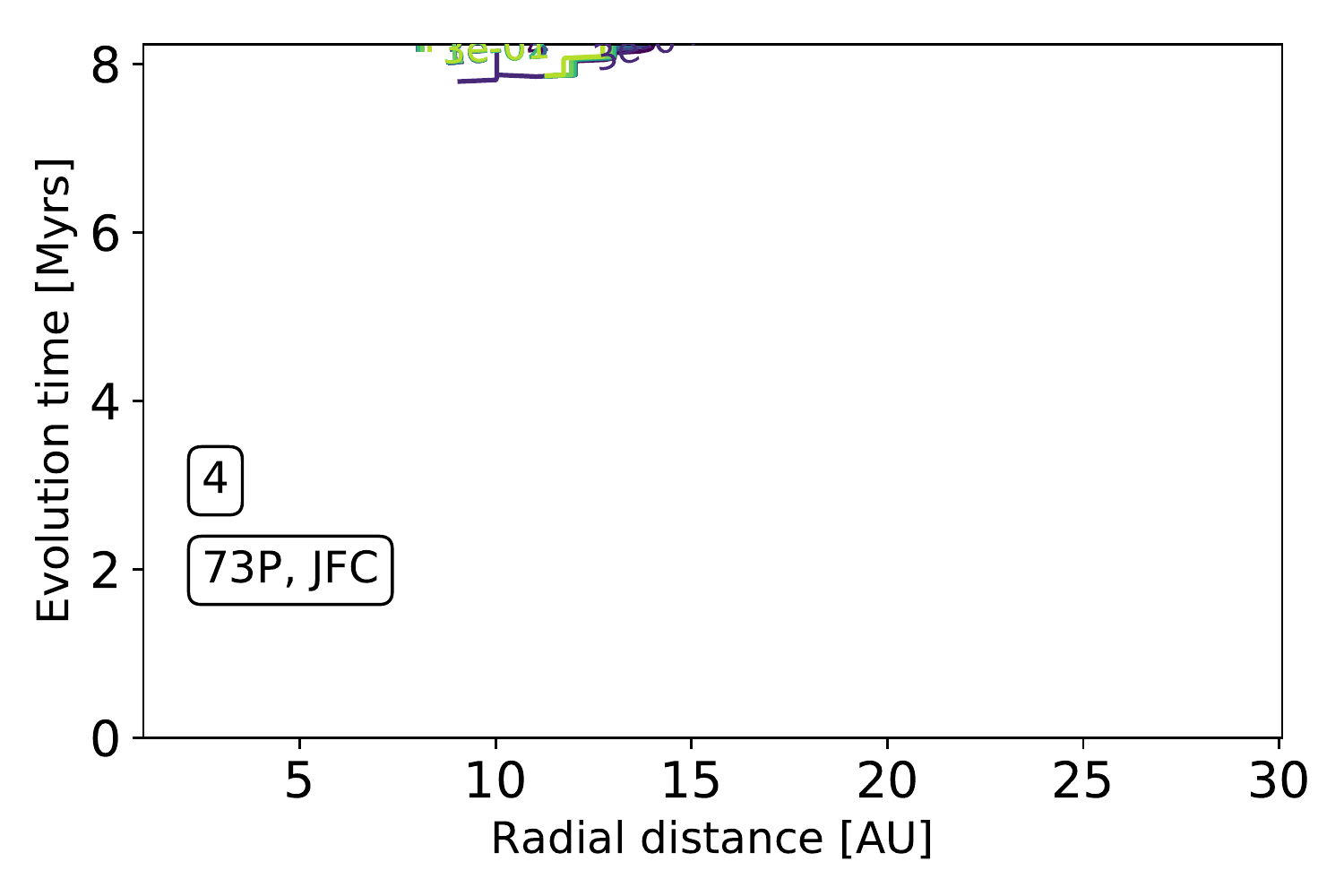}}\\
\subfigure{\includegraphics[width=0.33\textwidth]{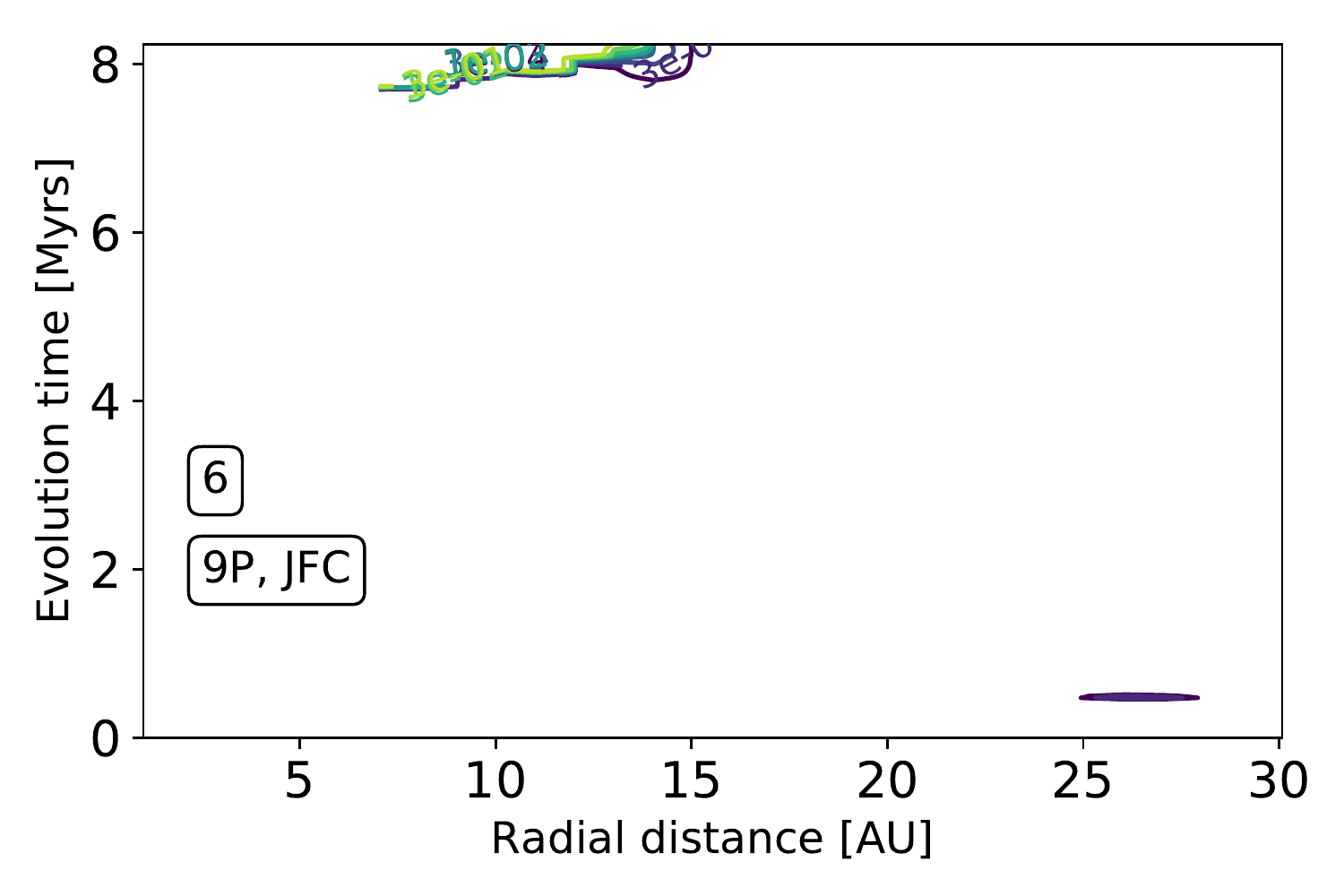}}
\subfigure{\includegraphics[width=0.33\textwidth]{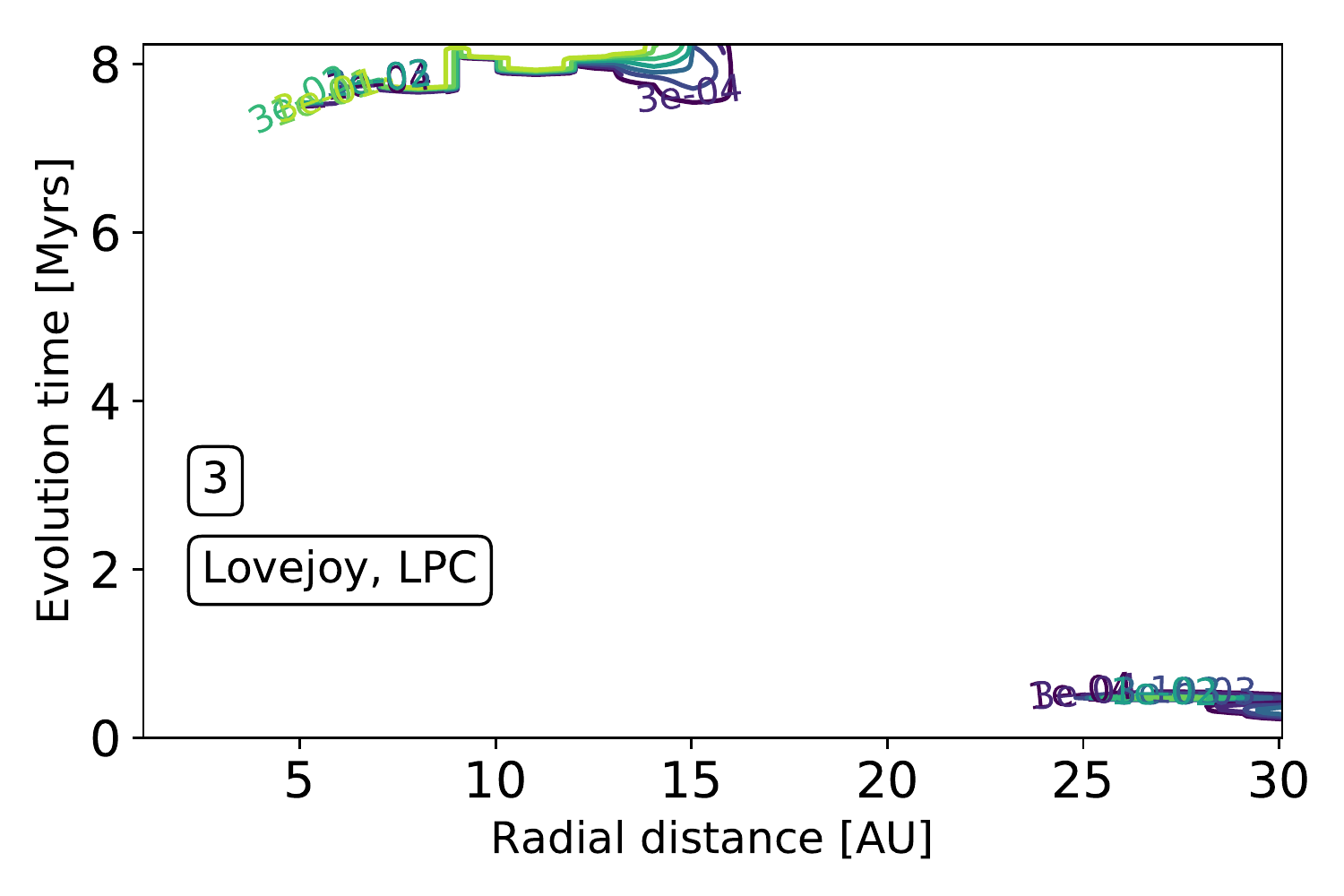}}
\subfigure{\includegraphics[width=0.33\textwidth]{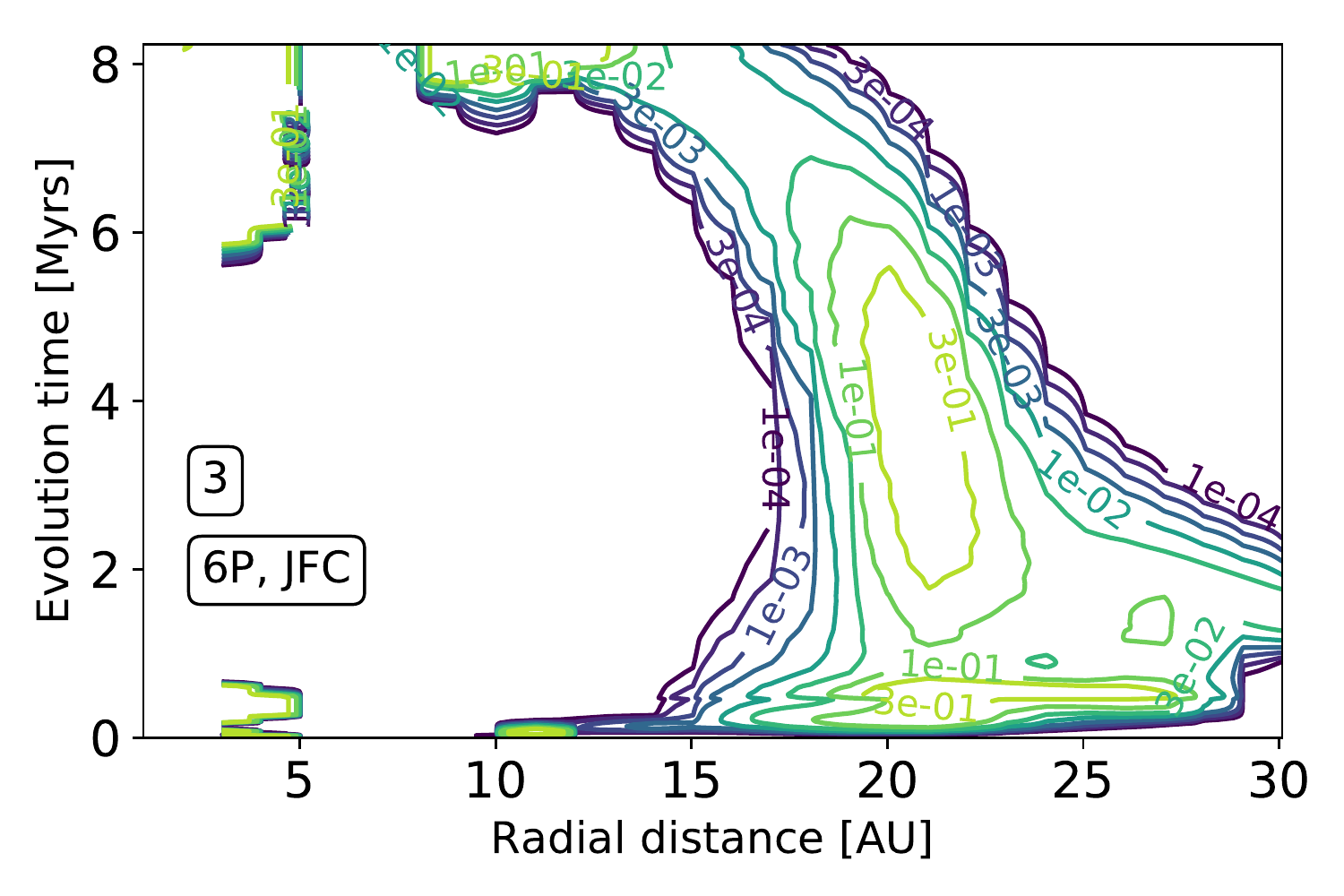}}\\
\subfigure{\includegraphics[width=0.33\textwidth]{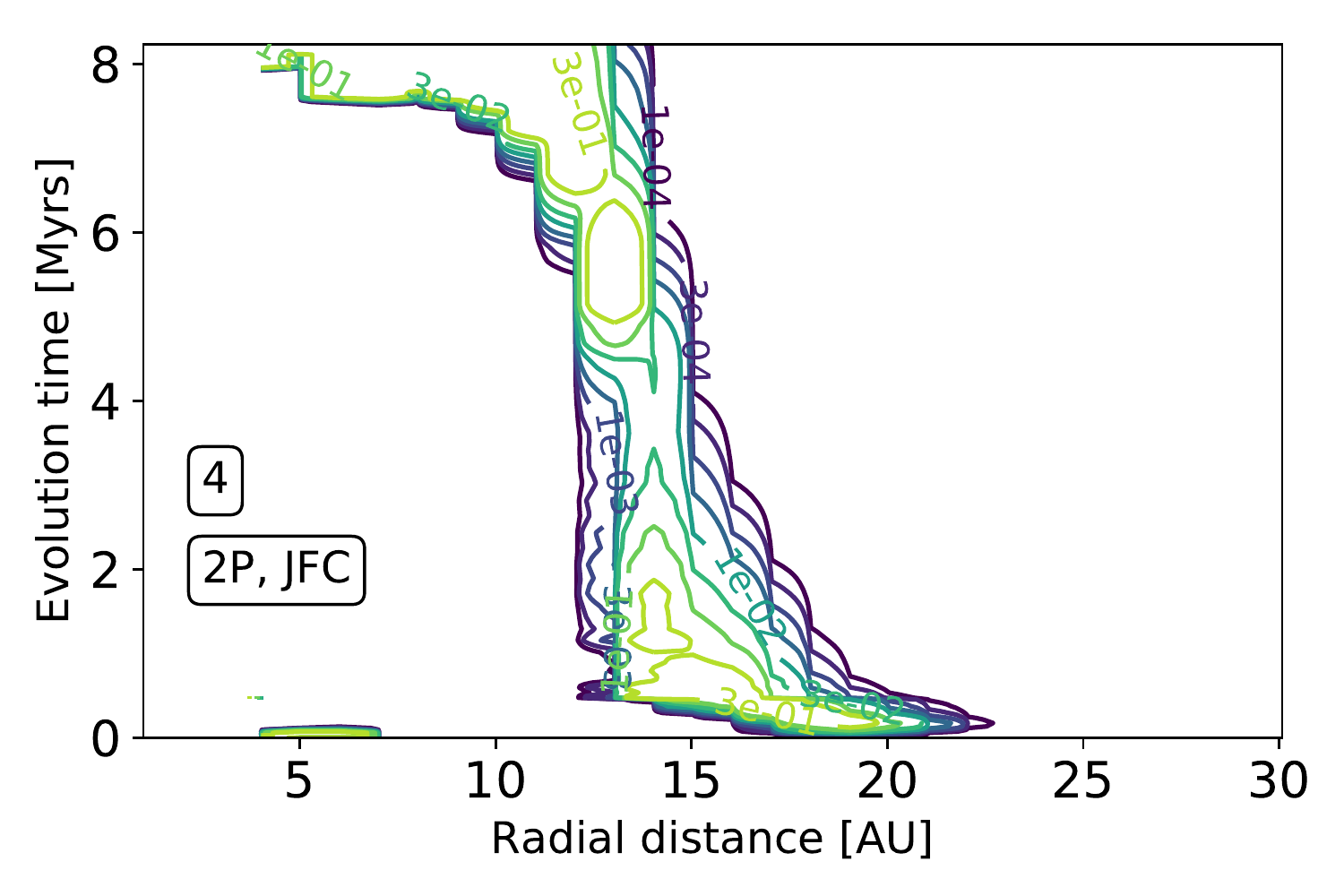}}
\subfigure{\includegraphics[width=0.33\textwidth]{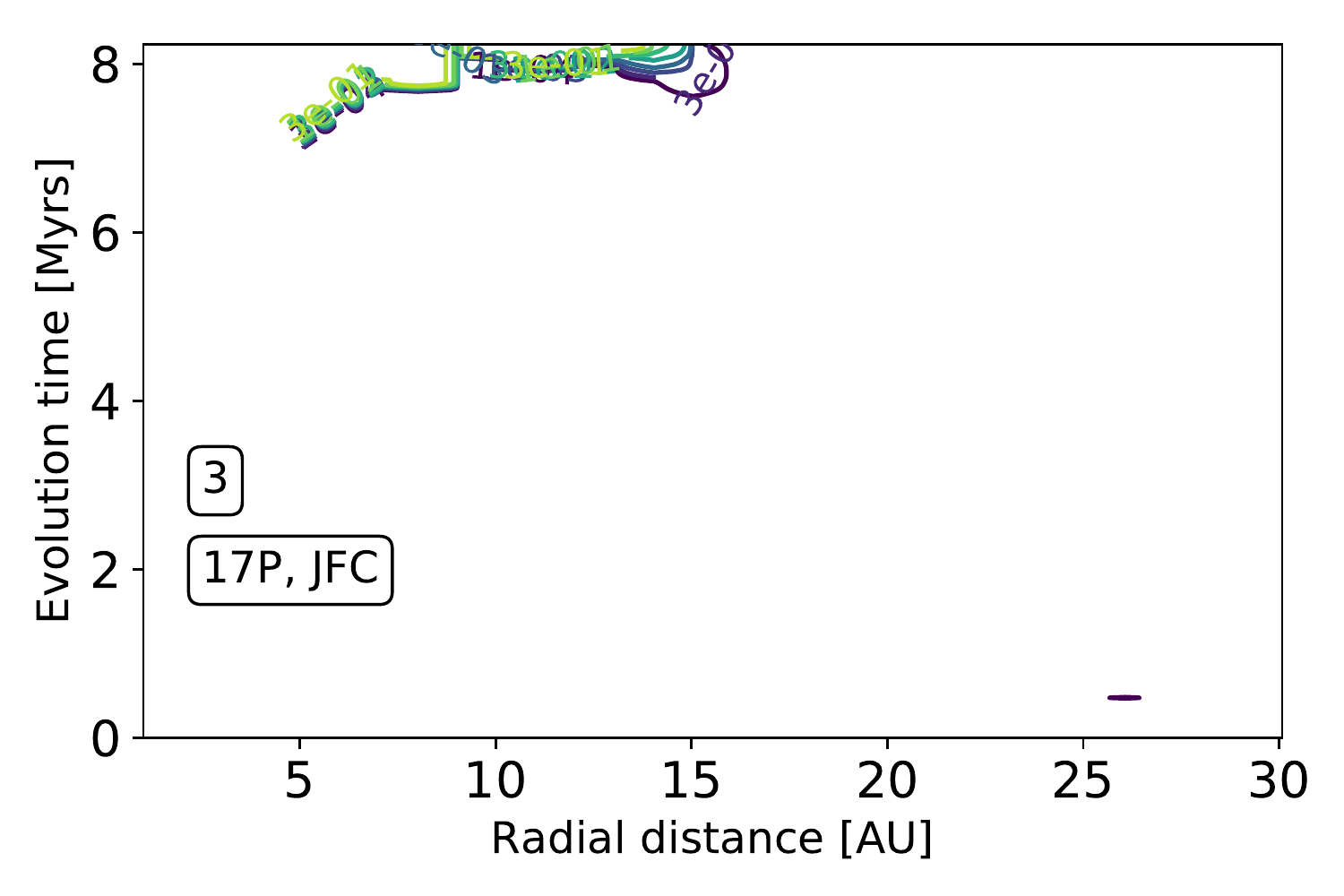}}
\subfigure{\includegraphics[width=0.33\textwidth]{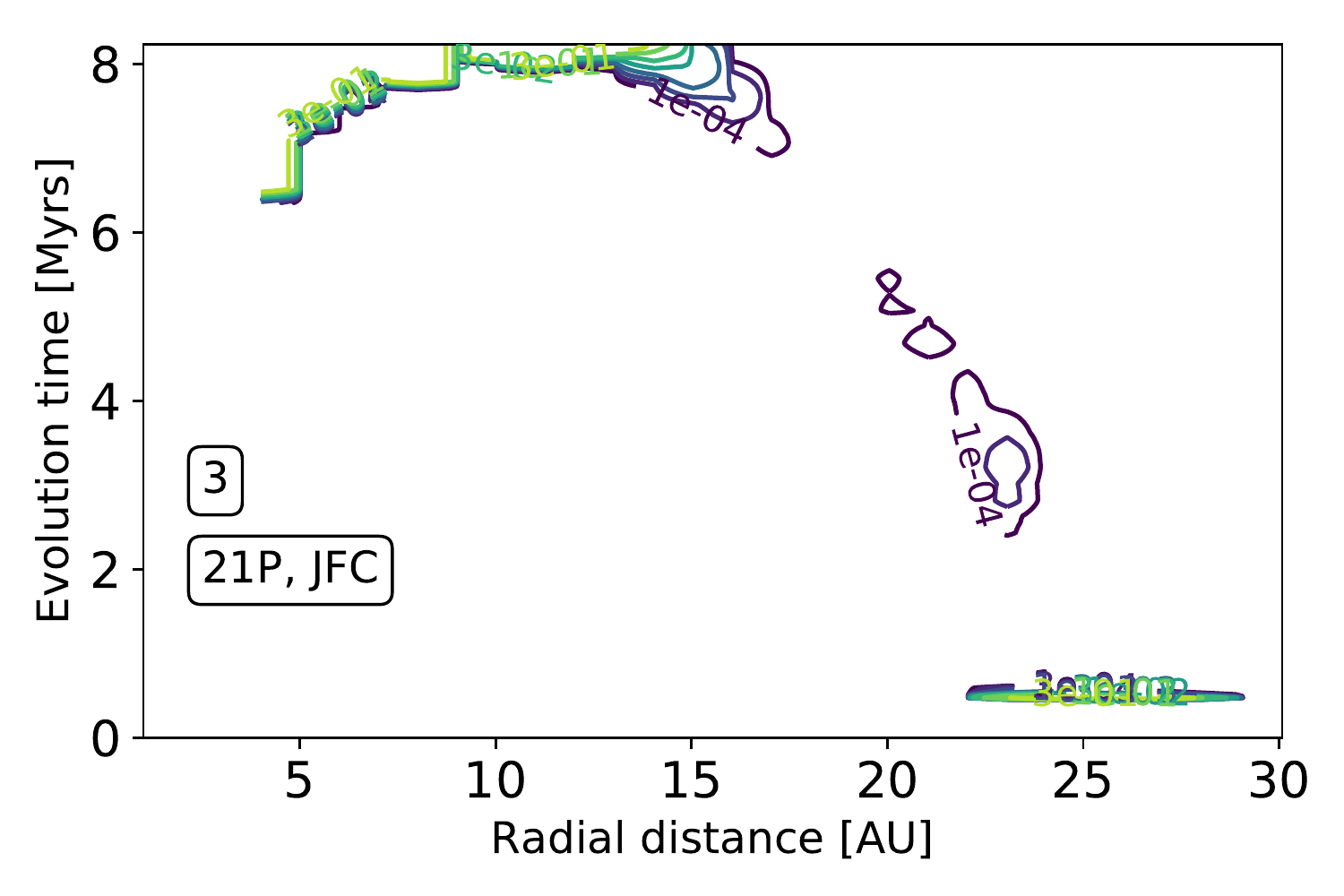}}\\
\caption{Maximum likelihood surfaces for inheritance scenario, for full sample of molecular species. The maximum likelihood increases from dark purple contours to light green contours. Darkest purple contour represents maximum likelihood of $10^{-4}$, and increases with each lighter contour level to 3x$10^{-4}$, $10^{-3}$, 3x$10^{-3}$, $10^{-2}$, 3x$10^{-2}$, $10^{-1}$, and finally 3x$10^{-1}$ for the lightest green contour. Radius in AU in the physically evolving protoplanetary disk midplane is on the $x$-axis, and evolution time in Myrs is on the $y$-axis.}
\label{P_inh_evol_percent}
\end{figure*}

\begin{figure*}
\subfigure{\includegraphics[width=0.24\textwidth]{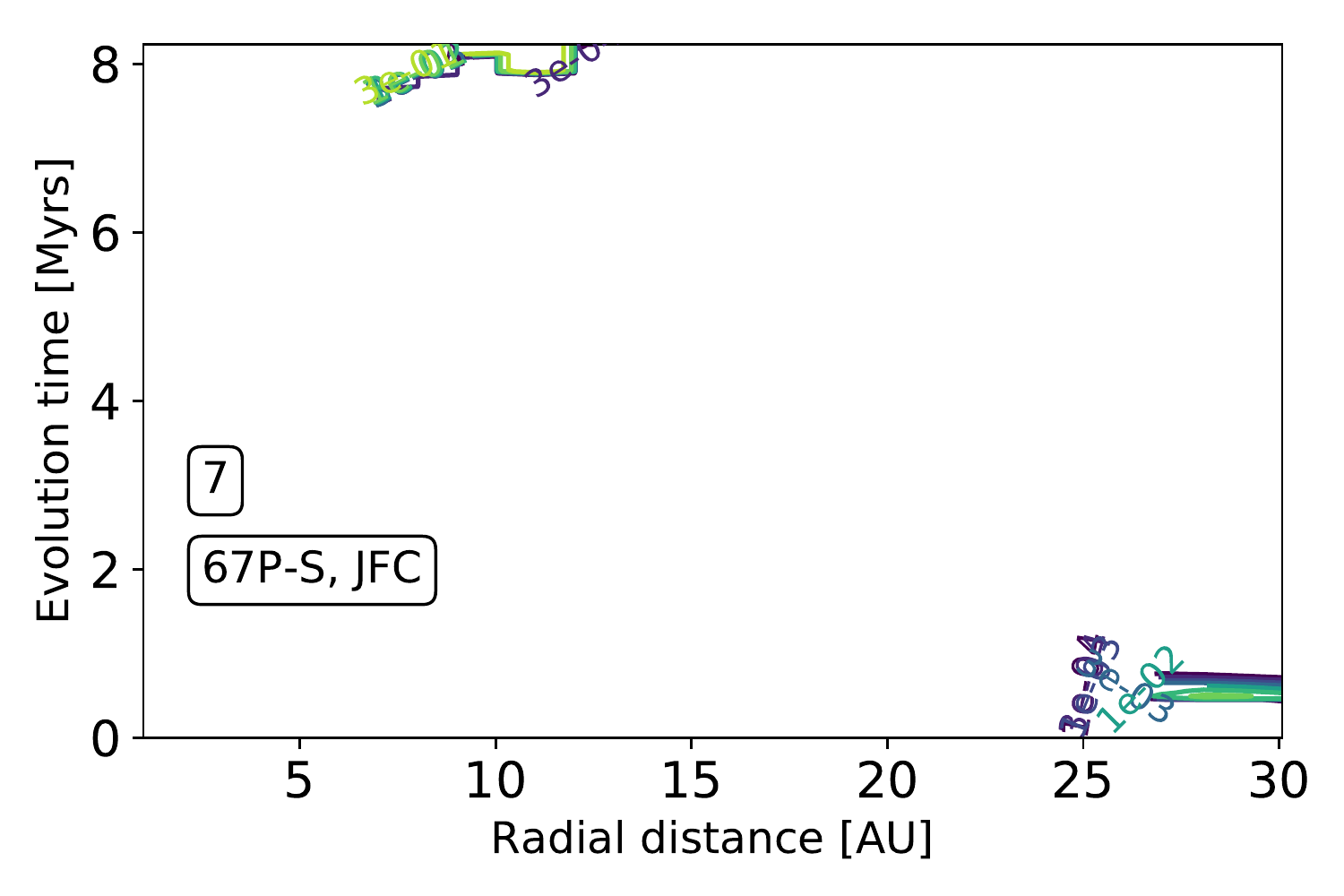}}
\subfigure{\includegraphics[width=0.24\textwidth]{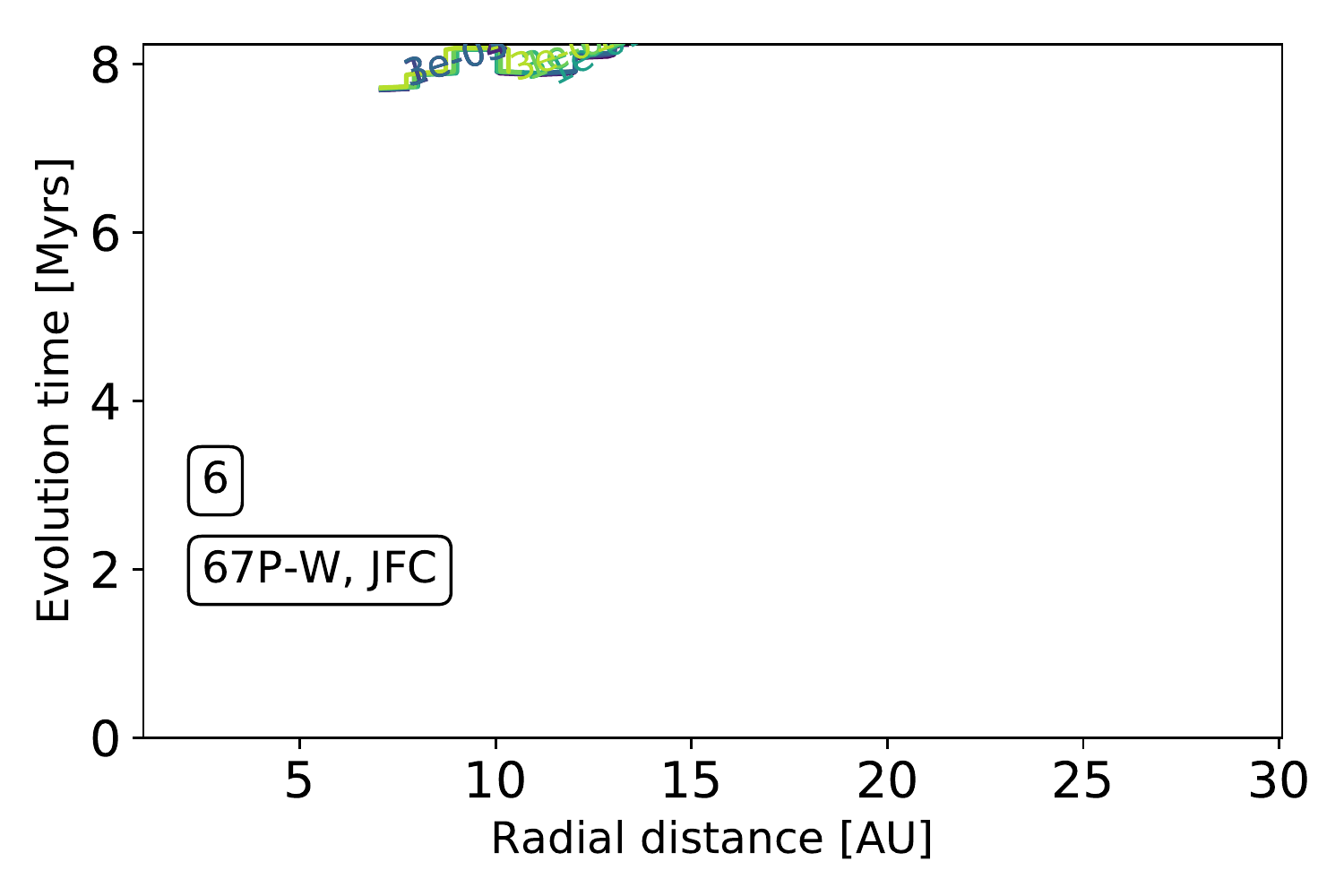}}
\subfigure{\includegraphics[width=0.24\textwidth]{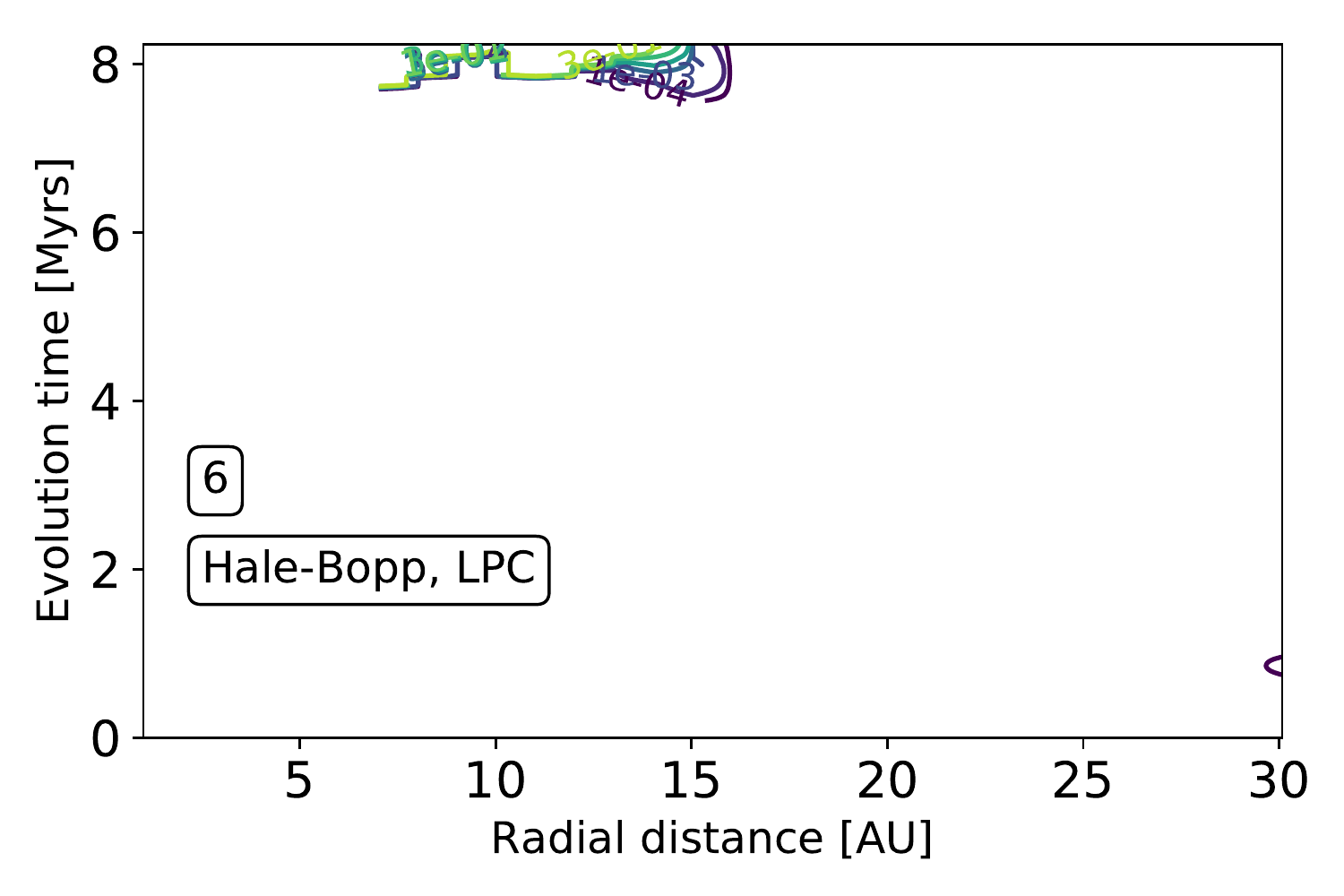}}
\subfigure{\includegraphics[width=0.24\textwidth]{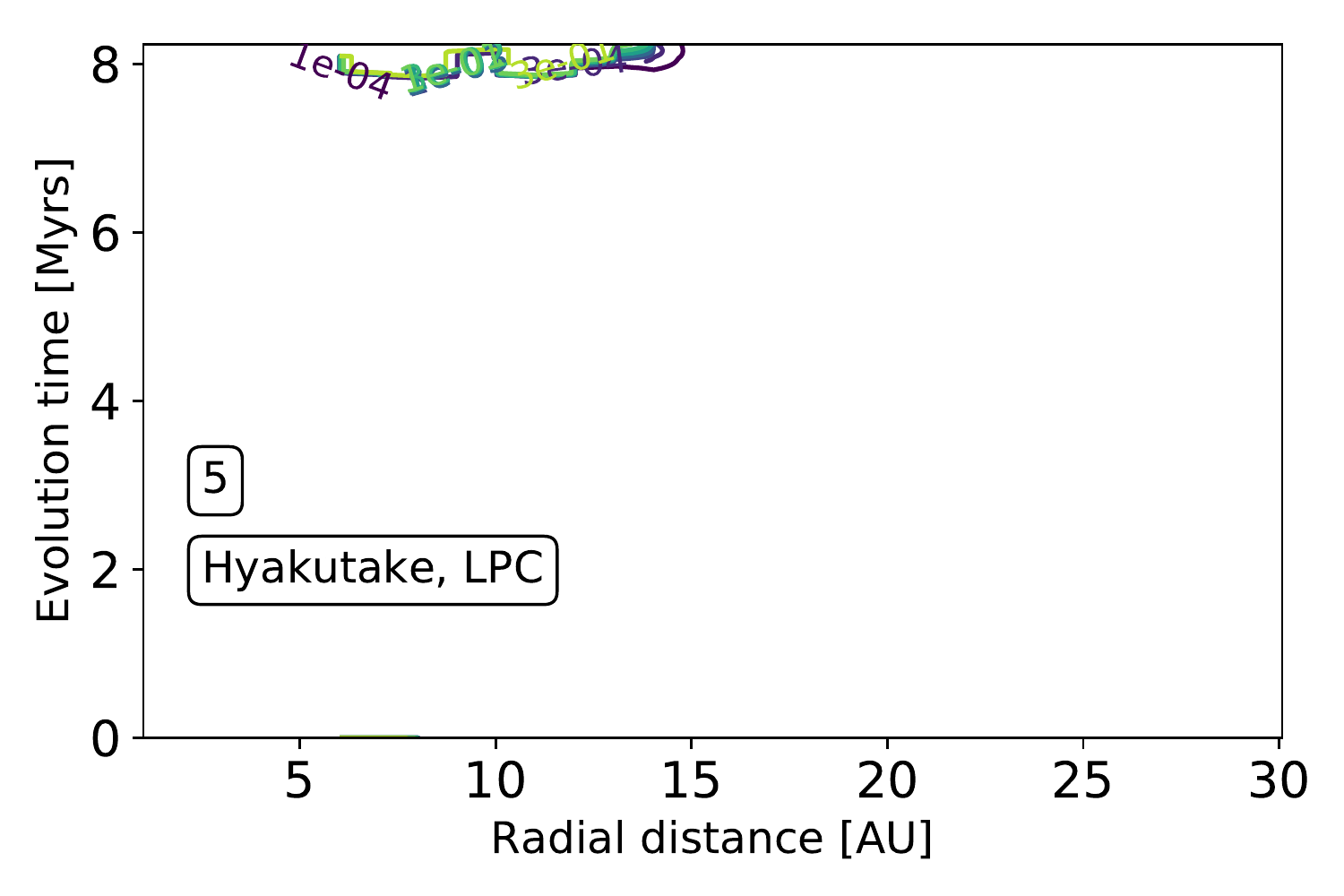}}\\
\subfigure{\includegraphics[width=0.24\textwidth]{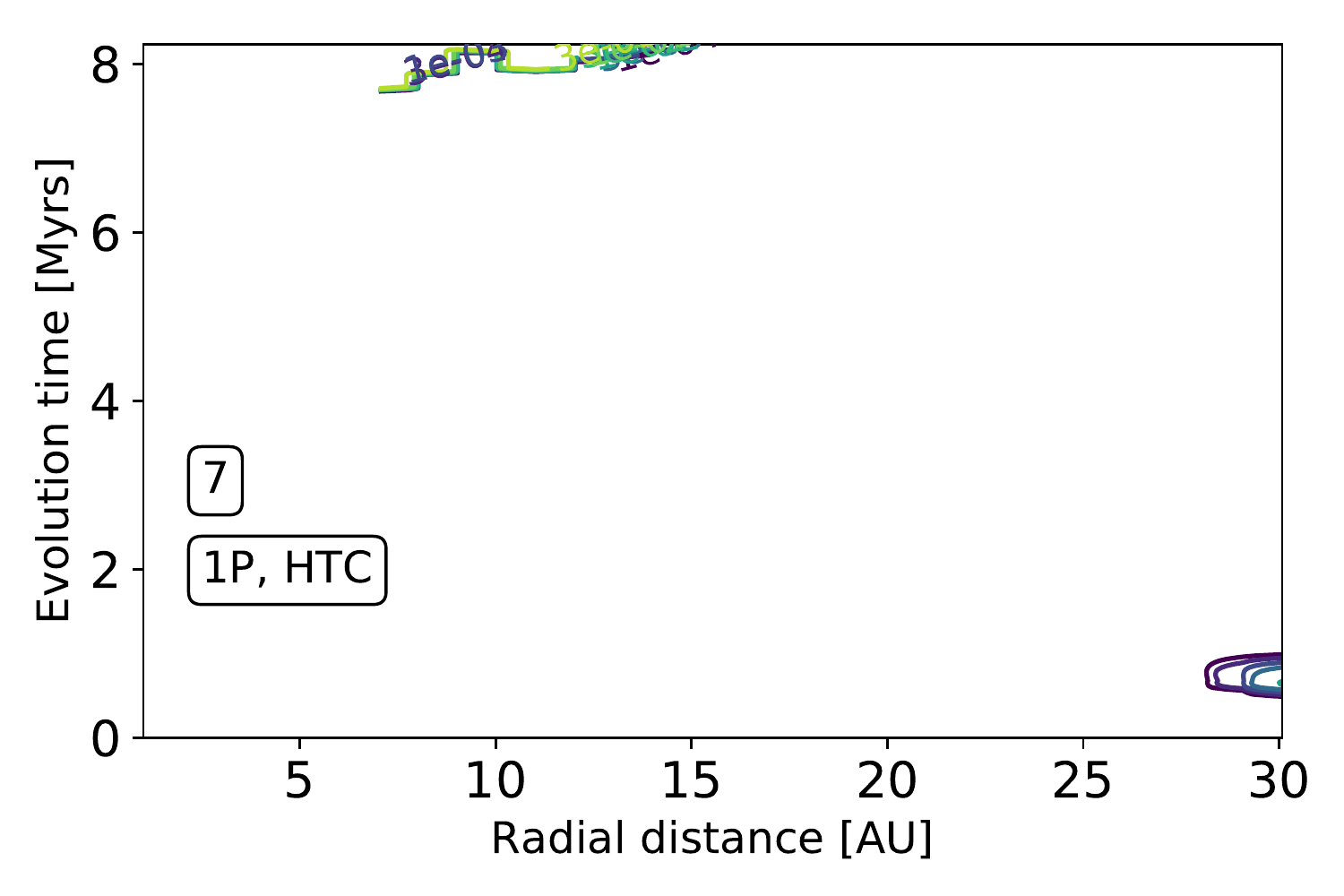}}
\subfigure{\includegraphics[width=0.24\textwidth]{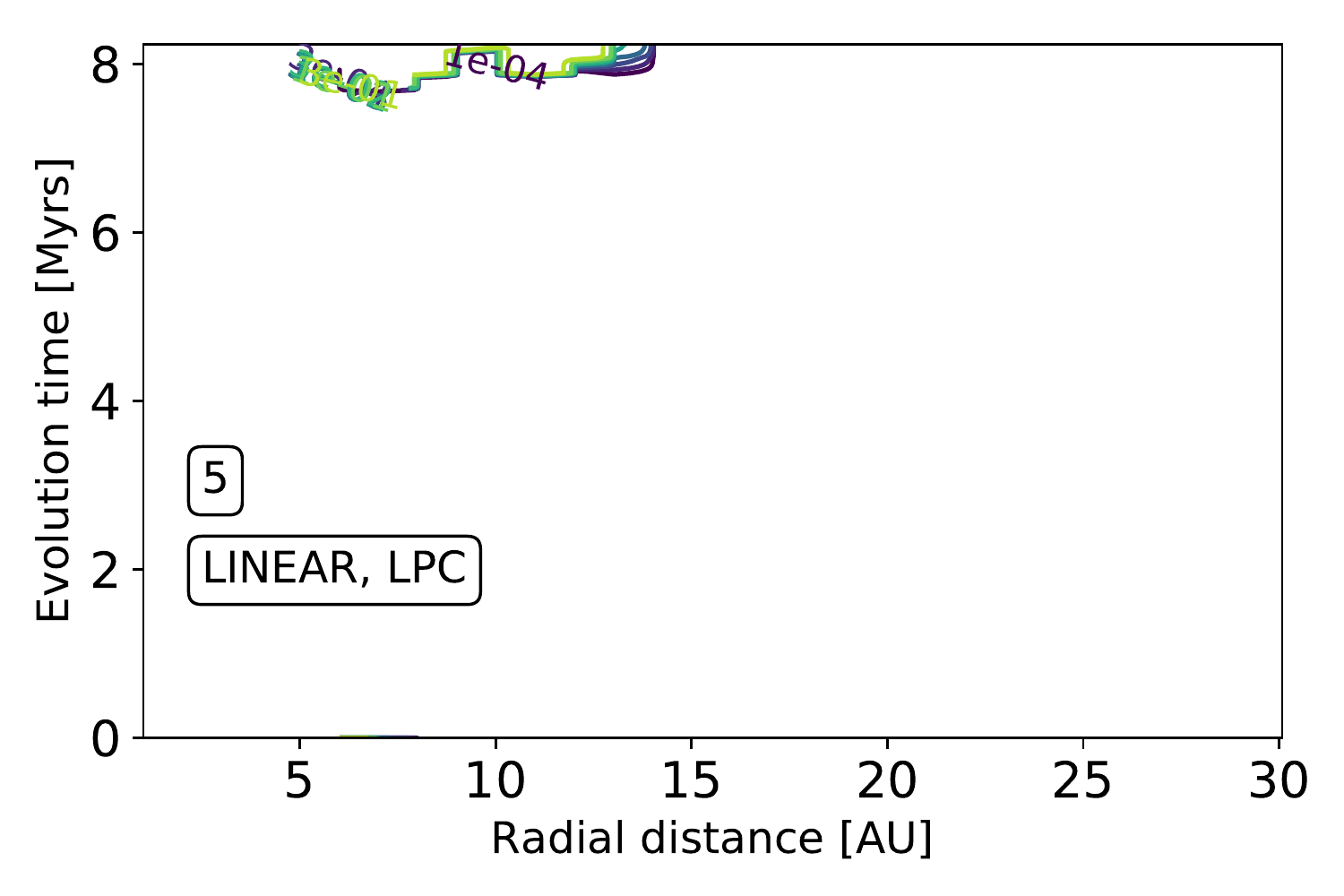}}
\subfigure{\includegraphics[width=0.24\textwidth]{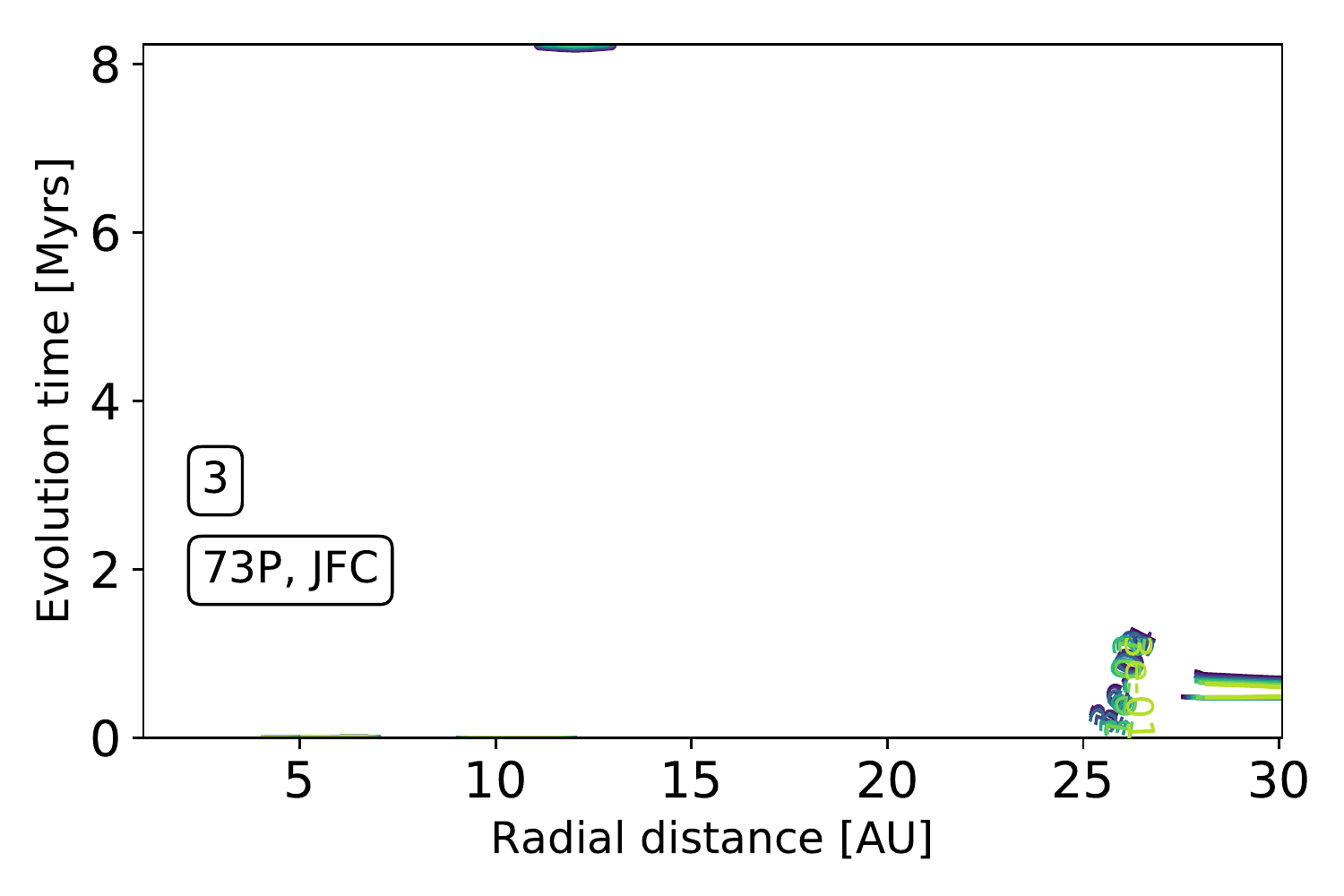}}
\subfigure{\includegraphics[width=0.24\textwidth]{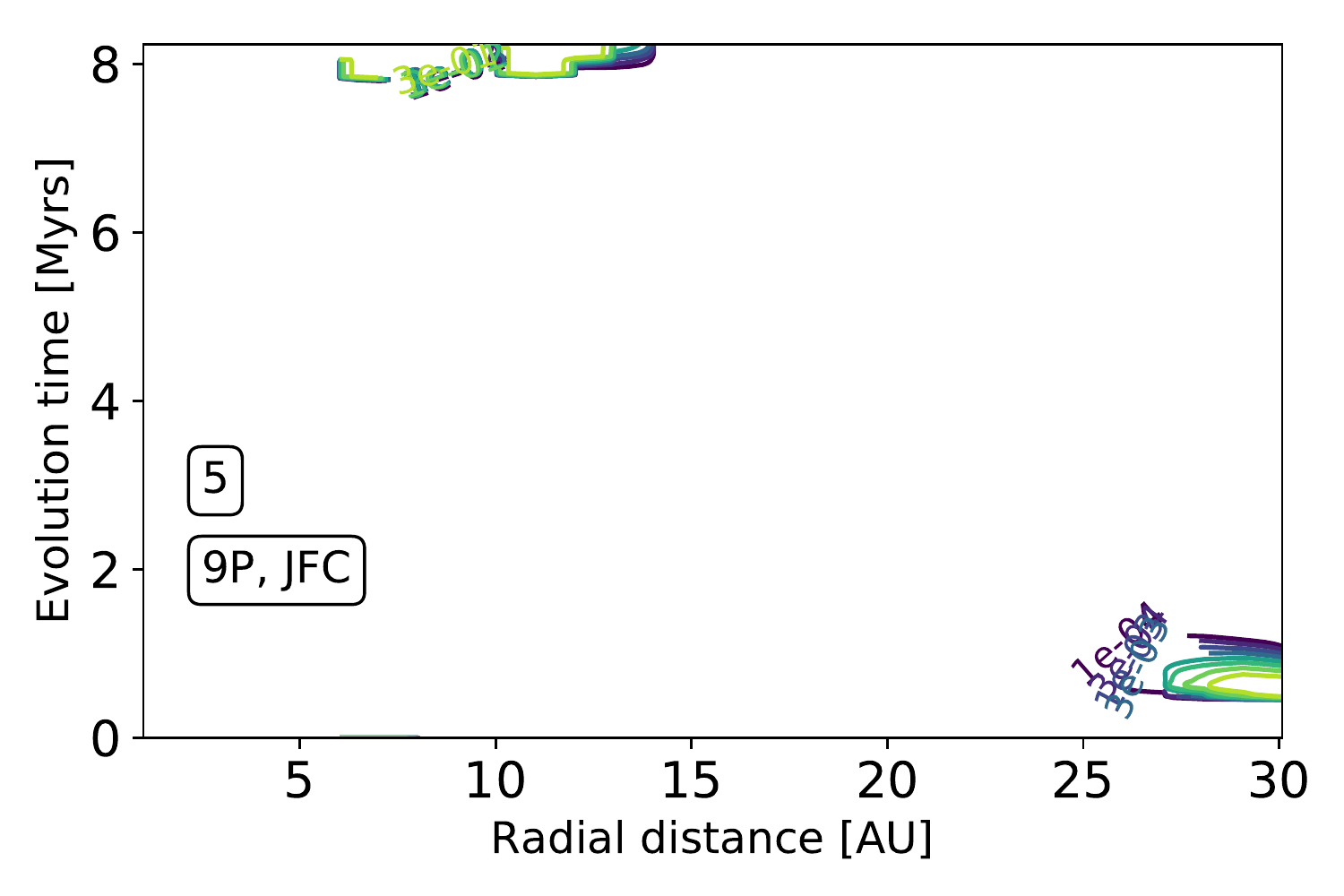}}\\
\caption{Maximum likelihood surfaces for reset scenario, considering C- and O-bearing species only. The maximum likelihood increases from dark purple contours to light green contours. Darkest purple contour represents maximum likelihood of $10^{-4}$, and increases with each lighter contour level to 3x$10^{-4}$, $10^{-3}$, 3x$10^{-3}$, $10^{-2}$, 3x$10^{-2}$, $10^{-1}$, and finally 3x$10^{-1}$ for the lightest green contour. Radius in AU in the physically evolving protoplanetary disk midplane is on the $x$-axis, and evolution time in Myrs is on the $y$-axis.}
\label{P_res_evol_no_sn_percent}
\end{figure*}

\begin{figure*}
\subfigure{\includegraphics[width=0.24\textwidth]{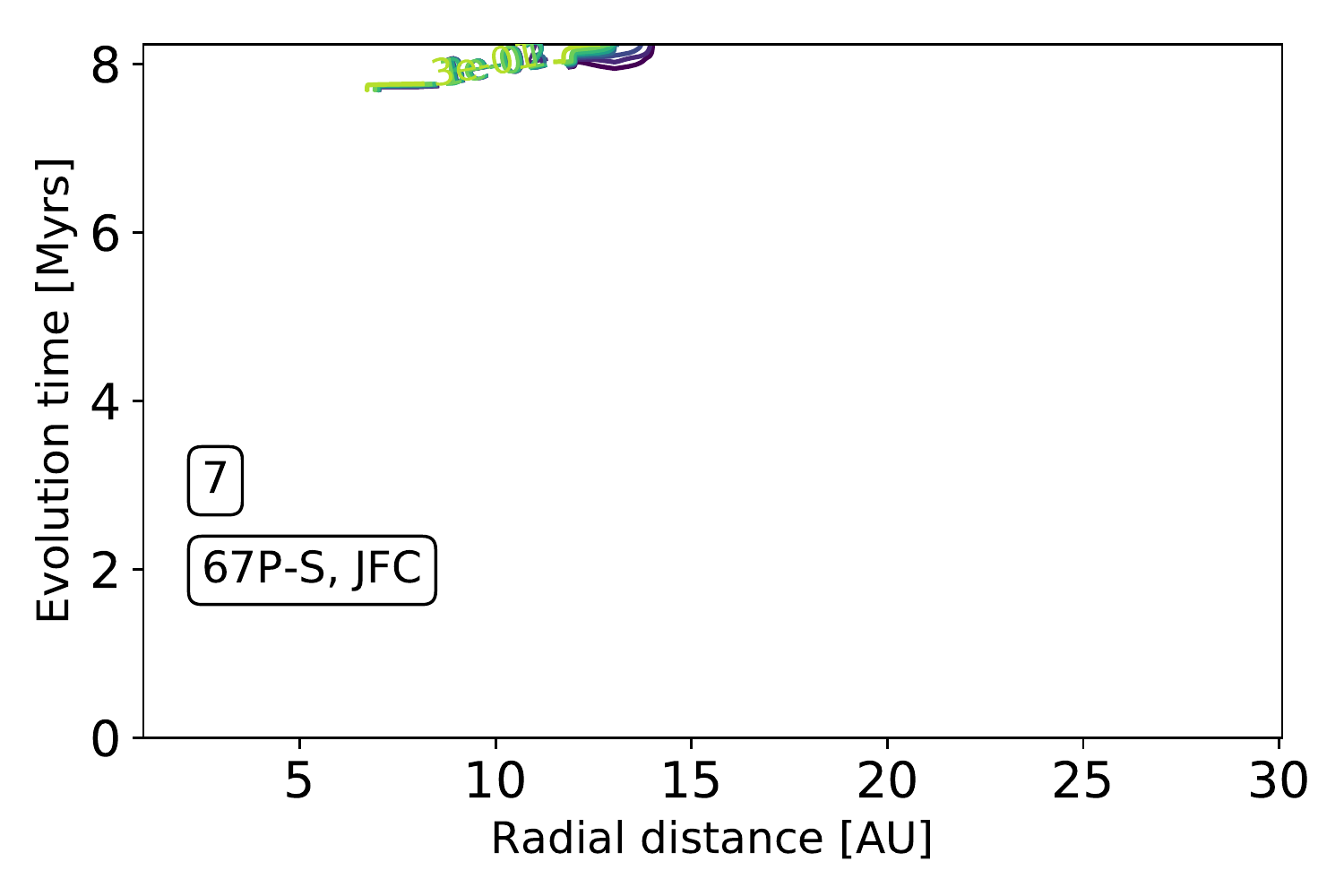}}
\subfigure{\includegraphics[width=0.24\textwidth]{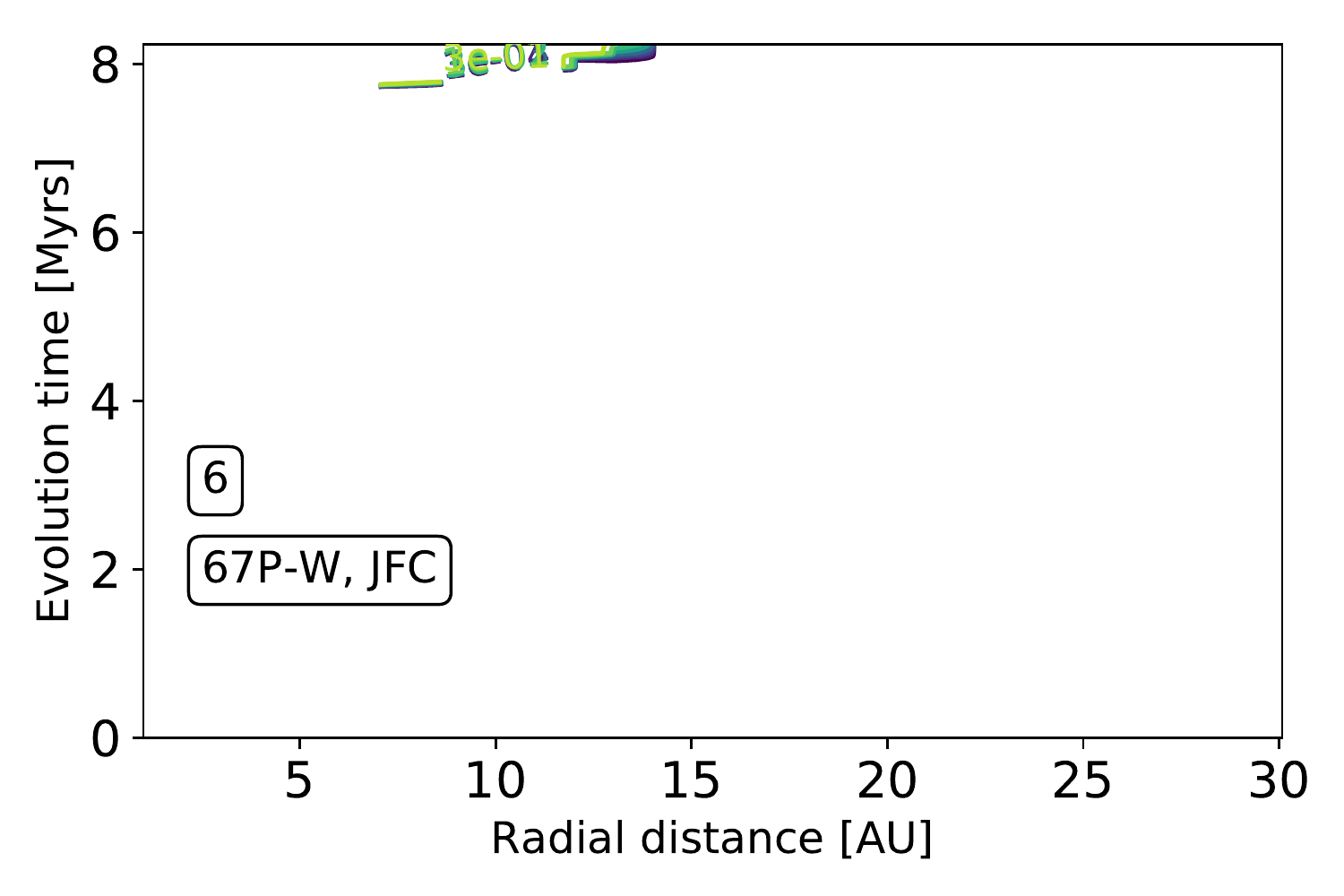}}
\subfigure{\includegraphics[width=0.24\textwidth]{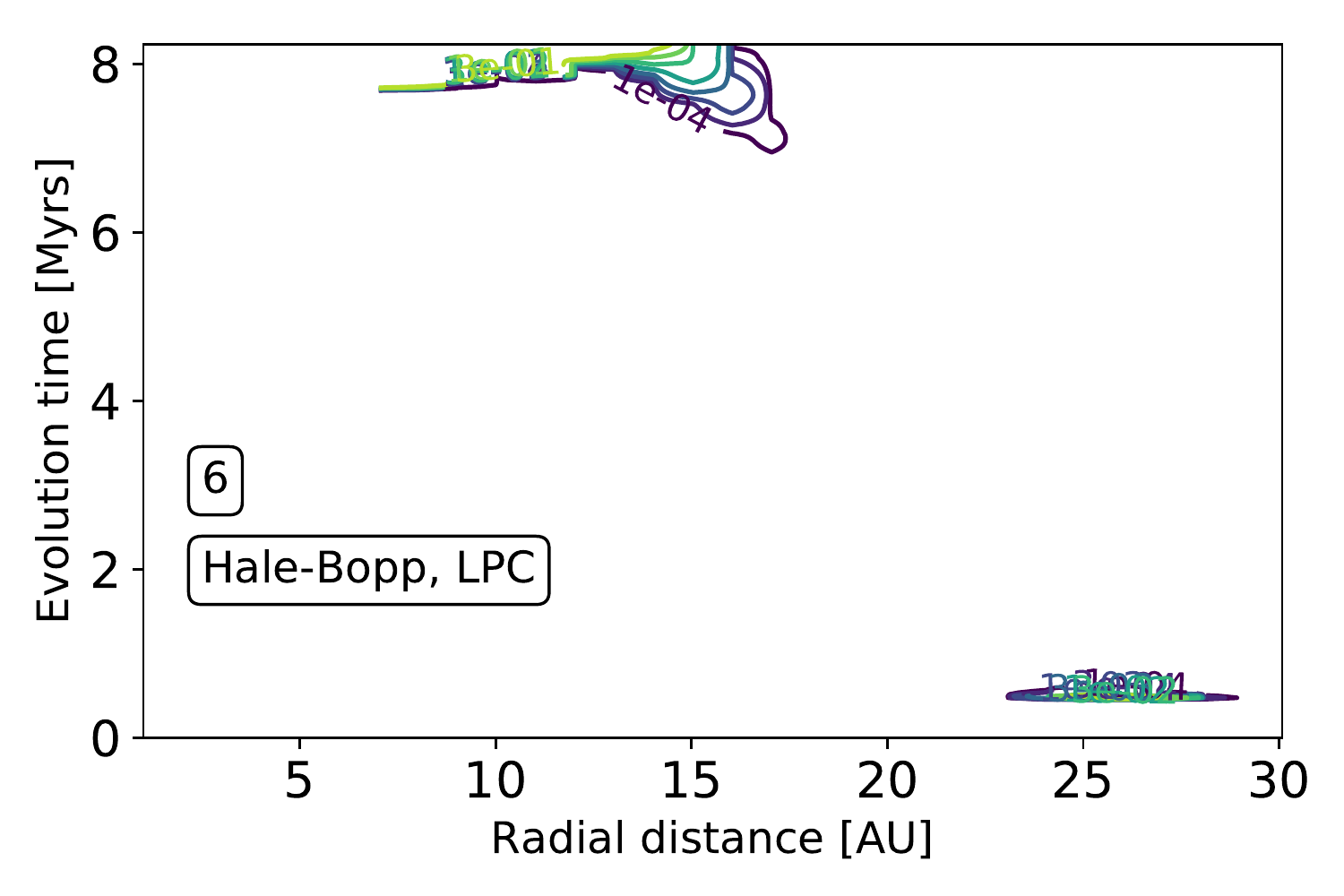}}
\subfigure{\includegraphics[width=0.24\textwidth]{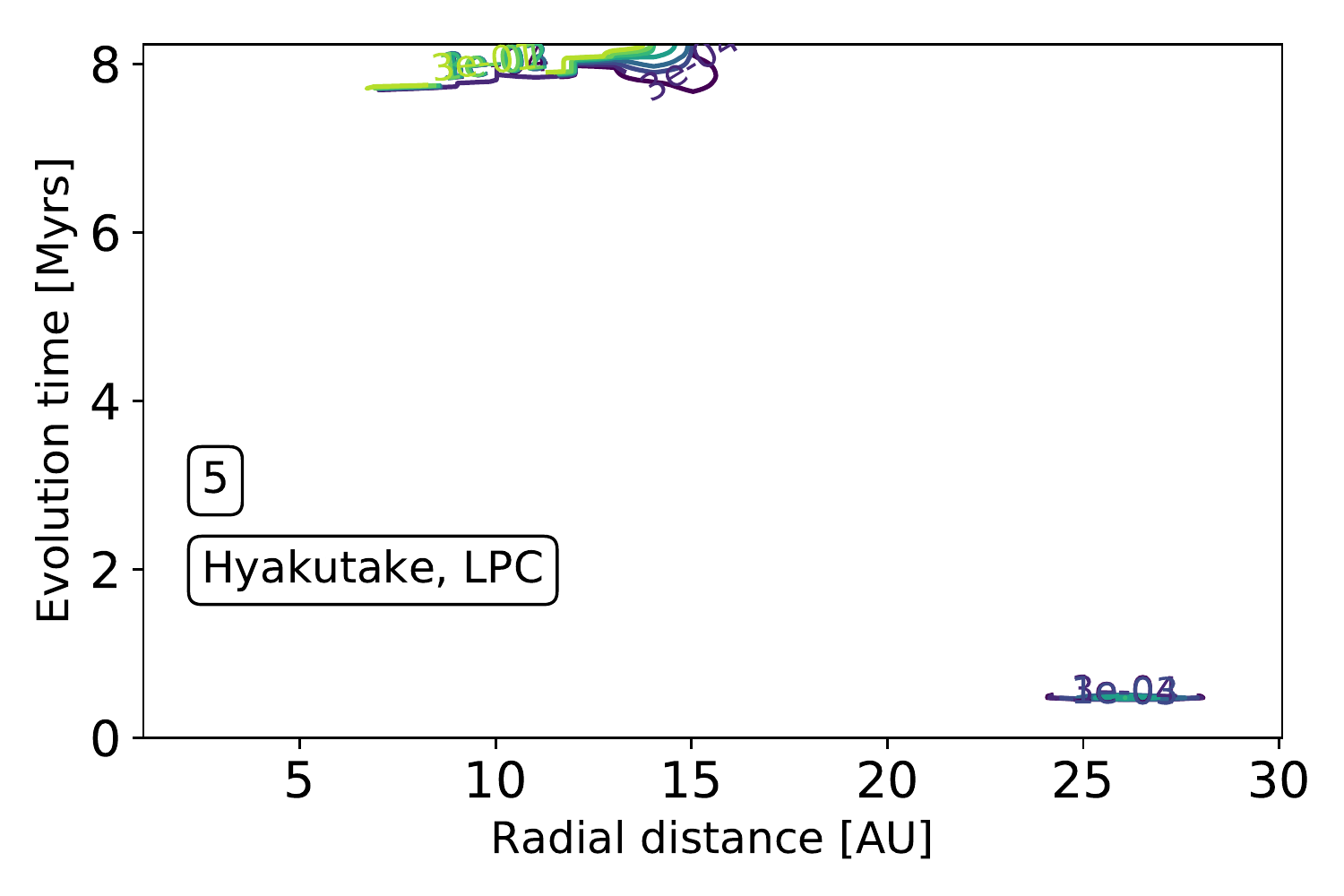}}\\
\subfigure{\includegraphics[width=0.24\textwidth]{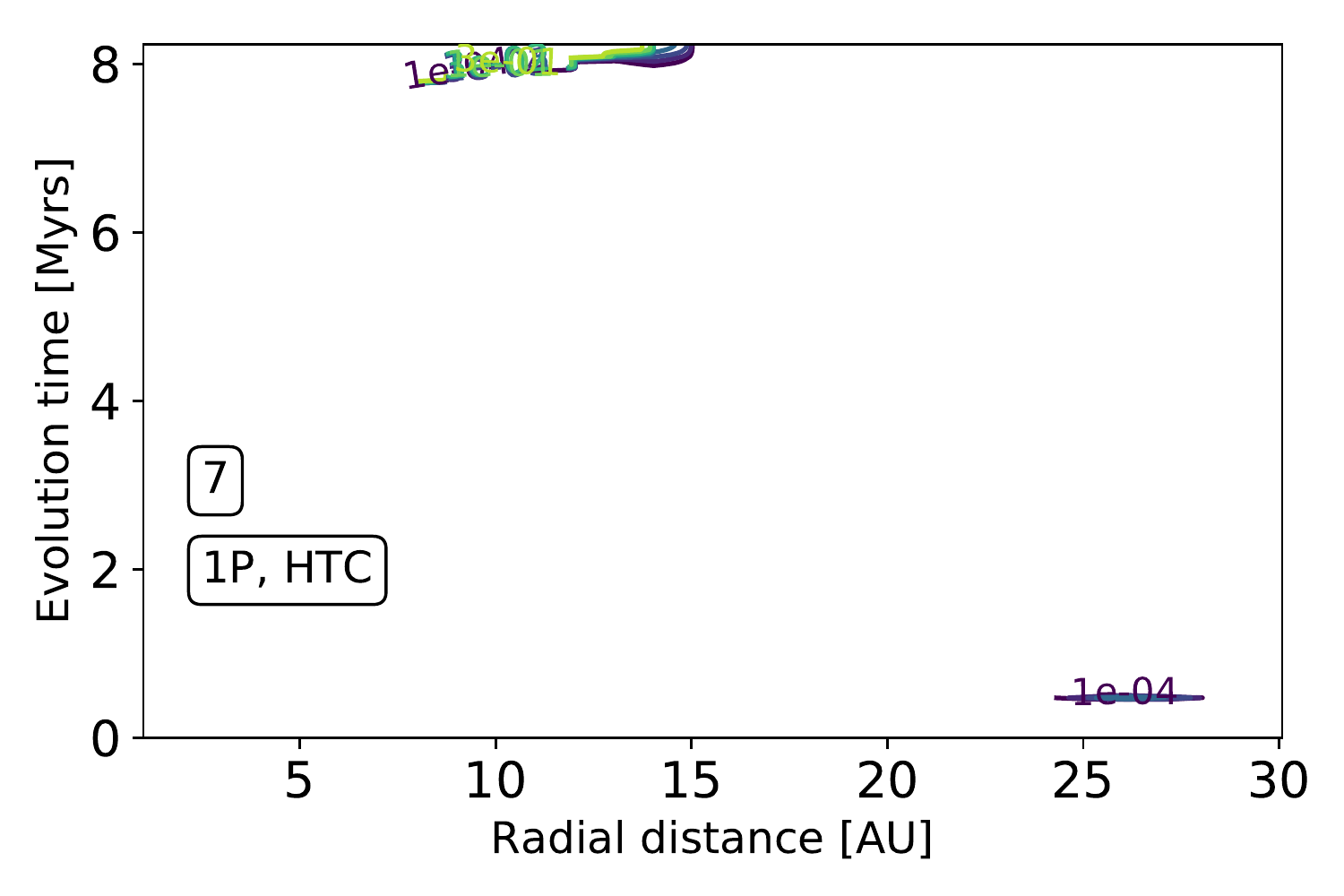}}
\subfigure{\includegraphics[width=0.24\textwidth]{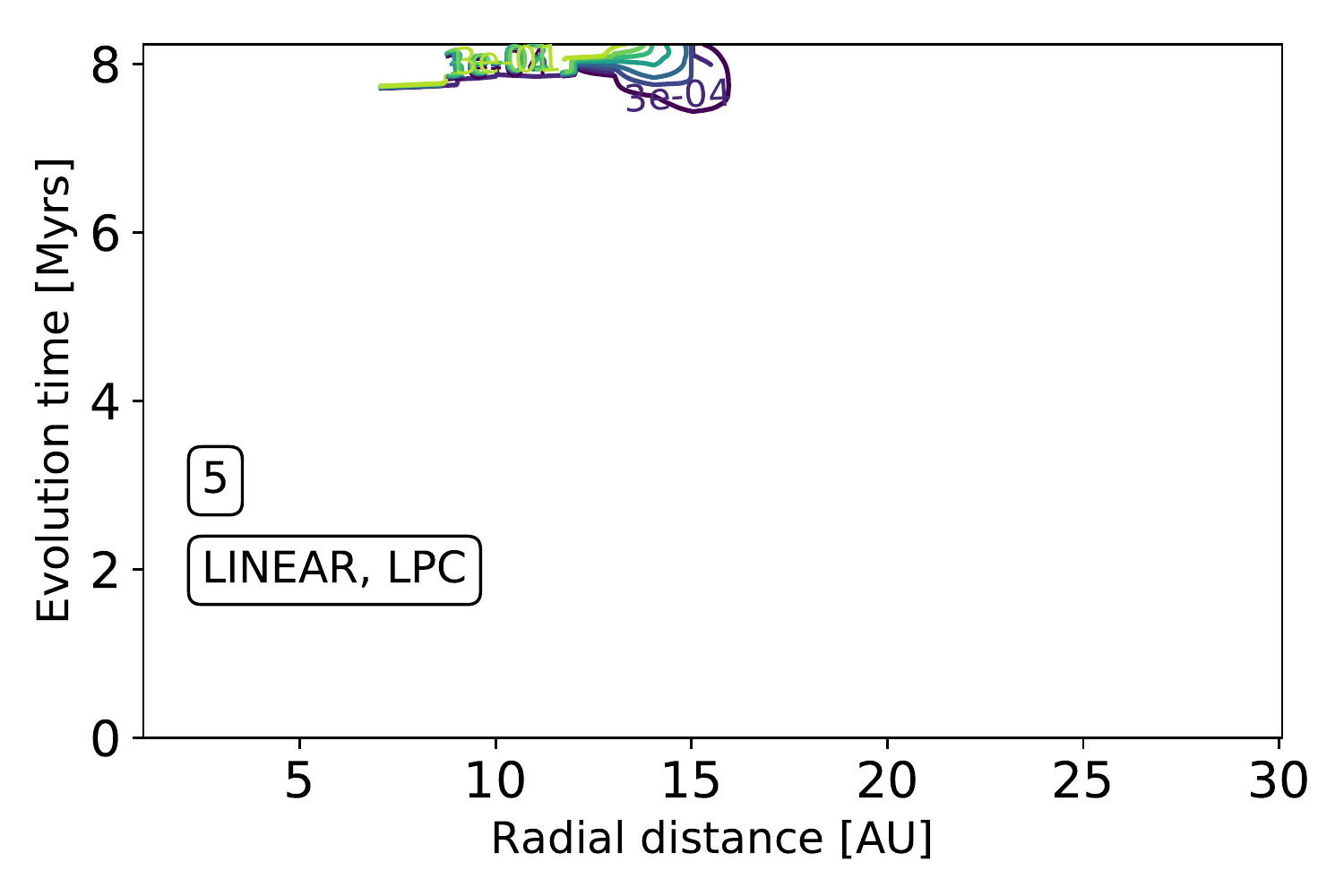}}
\subfigure{\includegraphics[width=0.24\textwidth]{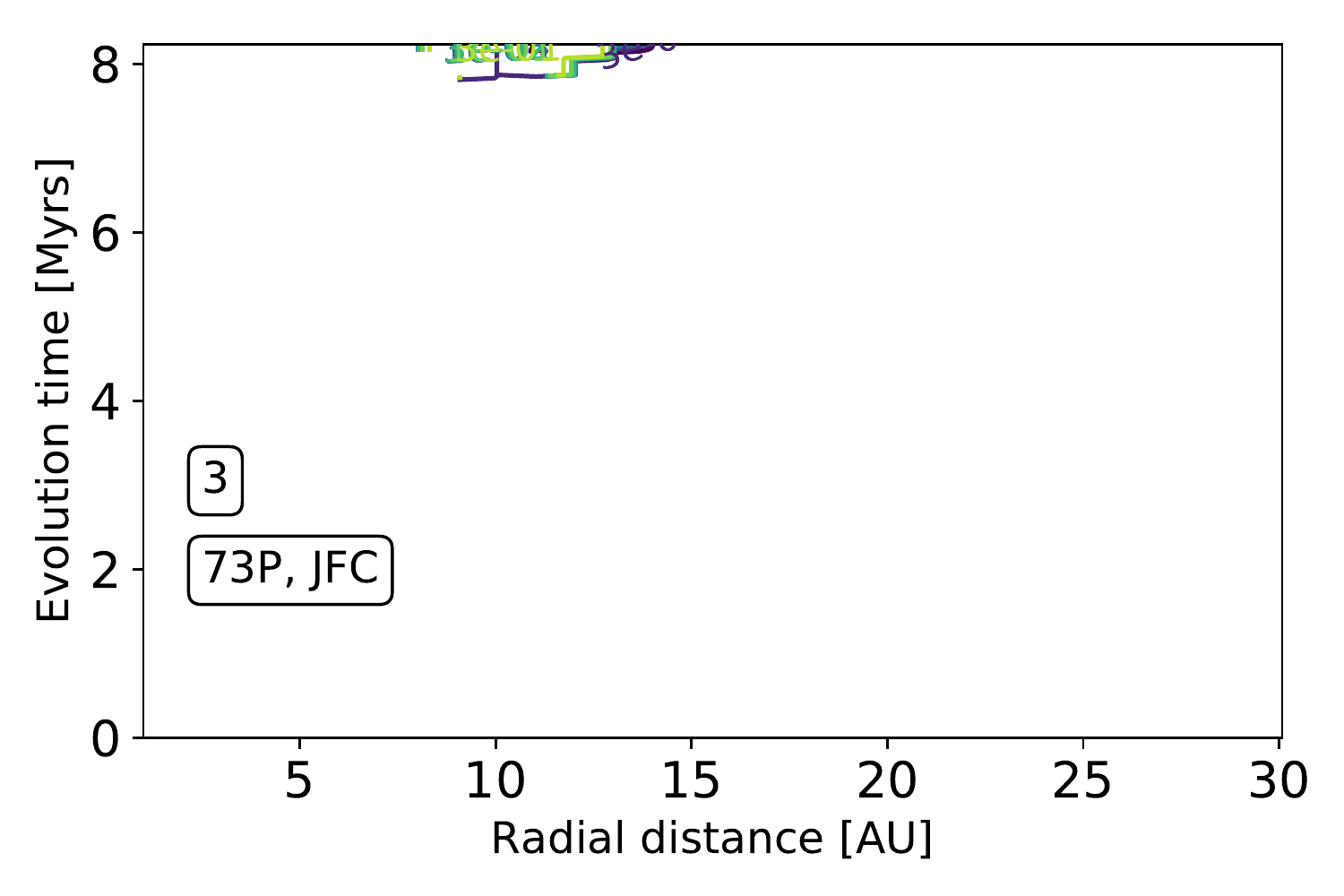}}
\subfigure{\includegraphics[width=0.24\textwidth]{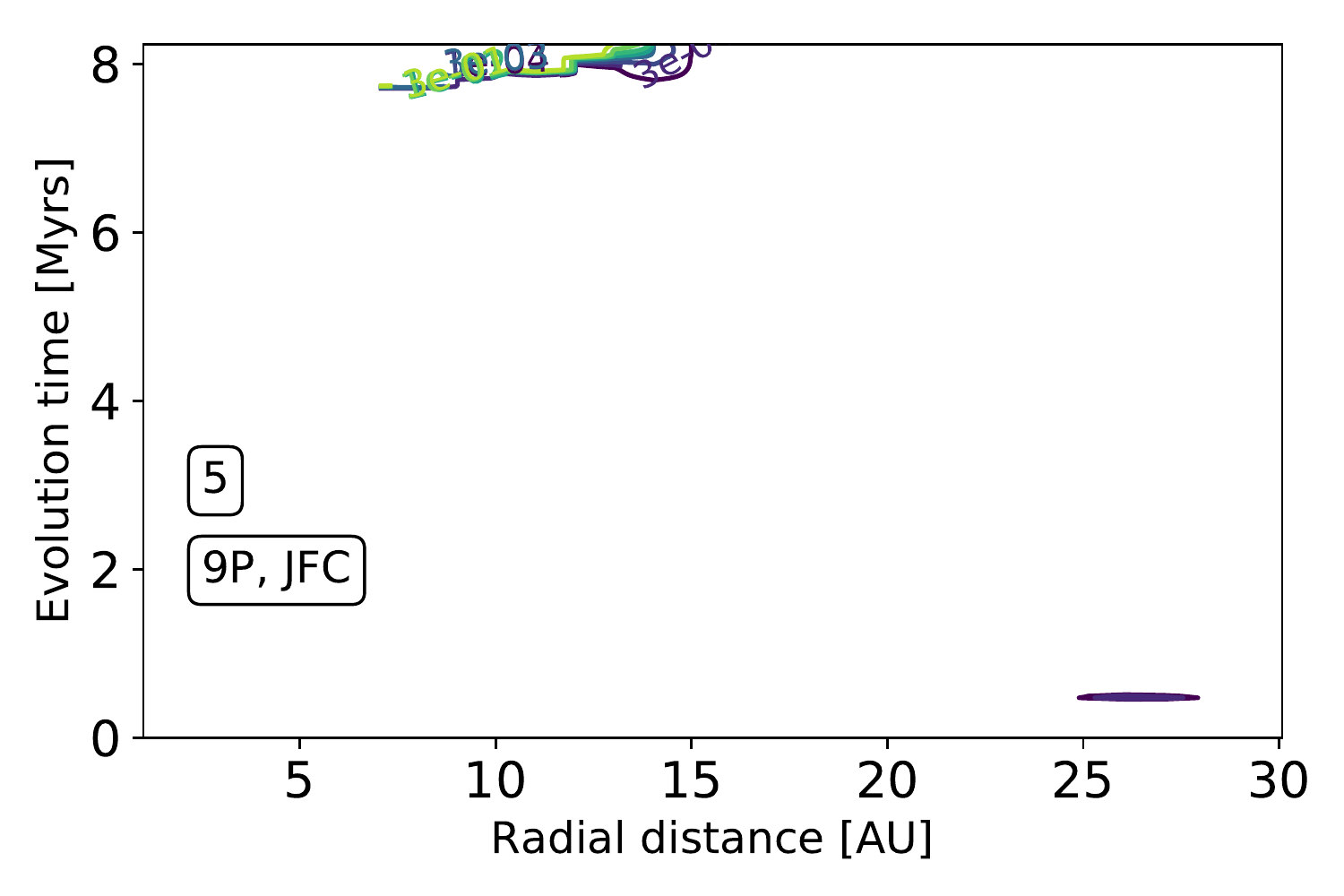}}\\
\caption{Maximum likelihood surfaces for inheritance scenario, considering C- and O-bearing species only. The maximum likelihood increases from dark purple contours to light green contours. Darkest purple contour represents maximum likelihood of $10^{-4}$, and increases with each lighter contour level to 3x$10^{-4}$, $10^{-3}$, 3x$10^{-3}$, $10^{-2}$, 3x$10^{-2}$, $10^{-1}$, and finally 3x$10^{-1}$ for the lightest green contour. Radius in AU in the physically evolving protoplanetary disk midplane is on the $x$-axis, and evolution time in Myrs is on the $y$-axis.}
\label{P_inh_evol_no_sn_percent}
\end{figure*}

Comparing these maximum likelihood surfaces to their counterparts in Fig. \ref{P_res_evol_percent} for the reset scenario and Fig. \ref{P_inh_evol_percent} for the inheritance scenario reveals that excluding sulphur species does not significantly change the behaviour in the results. The only apparent difference is found for comet Hale-Bopp, which in the reset scenario in Fig. \ref{P_res_evol_no_sn_percent} without sulphur agrees more broadly with formation at various evolutionary times and locations (though still along the CO iceline) than it did when including sulphur. However, this makes sense because excluding species from the analysis should cause less constraints on the most likely time and location of formation.

\section{Discussion}
\label{discussion}

Studies of comets and cometary compositions so far have been grouping them by either dynamical characteristics (length of orbit or inclination), or by their molecular contents \citep[see][]{cochran2012}. \citet{ahearn1995} and \citet{cochran2012} defined standards for molecular abundances (typical or depleted), such that cometary measurements could then indicate either enhanced or depleted abundances for a given molecule. While this approach does provide a grouping for comets that fit with the standard, it fails to include the possible chemical evolution of the comet-forming material in setting that standard.

This work attempts to trace the formation histories of 15 comets, by comparing cometary abundances with evolving ice abundances in a protoplanetary disk midplane through statistical tests. This way, rather than simply comparing similarities in the cometary contents, the potential formation times of the comets can also be addressed, since the chemical composition of the comet-forming disk midplane evolves over time. Four different comparison setups were investigated, with two different sets of observed abundances (either including or excluding sulphur-bearing species), and two different models for evolving abundances for the midplane (the inheritance or the reset scenario). For the reset scenario, good agreement is seen for formation in the vicinity of the CO iceline for all 15 comets.


The comets with more species observed in the coma are constrained to either formation at 27-30 AU by $\sim$1 Myr evolution, or to formation at $\sim$12 AU by 8 Myr evolution. Comets with fewer observed species agree well with formation along the CO iceline, as indicated by the red shaded region overplotted on the panel for comet 2P in Fig. \ref{P_res_evol_percent}. From the left panel in Fig. 4 from \citet{eistrup2018}, 30 AU is by 0.5 Myr just outside the CO iceline (at 27 AU), which moves to 12 AU by 8 Myr. The \ce{CH4} iceline is found at 10 AU by 8 Myr. This suggests that for all comets, their formations can be constrained to lie roughly between the icelines of \ce{CH4} and CO, with some degeneracy remaining in the formation time. It is noted that comet 17P does not follow the trend for formation between the CO and \ce{CH4} icelines, unless the CO abundance measured by \citet{qi2015} is included in the analysis.

For the inheritance scenario, both with and without sulphur-bearing species, the results do not point as strongly to formation along the CO iceline, as found for the reset scenario. There is a similarity between the reset and inheritance scenarios with good agreement for formation at 11-13 AU by 8 Myr, but this should be seen in the light of the results from \citet{eistrup2018}. It is shown there that the ice abundances for both scenarios tend towards icy disk midplane steady state at late evolutionary times (>5 Myr). That means that the ice abundances after 5 Myr evolution are largely similar across the two scenarios, resulting in similar maximum likelihood surfaces for both (see also the abundance evolution surfaces in Figs \ref{figure6} -\ref{figure10}).

Generally, the inheritance scenario does not point to a single formation history for all comets, but rather good agreement is seen for multiple formation regions in parameter space, and overall this scenario seems less constraining. This may indicate that our starting molecular abundances for the inheritance scenario do not resemble well those for the particular molecular cloud from which the Sun formed. Although this set of abundances are motivated by observations of ices in protostellar (or interstellar) environments \citep[see][]{eistrup2016}, it is possible that the Sun formed in a different (warmer) environment to that for nearby well-studied protostellar sources \citep[see e.g.][]{adams2010,taquet2018}. It would be worthwhile to explore the impact of a warmer interstellar environment on the initial inherited molecular abundances for a comet-forming disk.

Using the analysis here as a test of which chemical starting conditions and evolution best agree with the observations, the reset scenario is seen to generally agree best with the cometary observation, by constraining all the comets to most likely having formed between the icelines of \ce{CH4} and CO. As these icelines reside at temperatures ($T_{\rm{CO,}ice}\sim 21$K and $T_{\rm{\ce{CH4},}ice}\sim 28$K) at which grain-surface (ice) chemistry is particularly active, large changes in the relative abundances of the volatile ice species are seen over time \citep[see][]{eistrup2018}. This active chemistry could in turn explain the diversity of observed cometary abundances. 

That the reset scenario provides better constraints on location and time for comet formation, points to the comet-forming region in the pre-solar nebula having been seeded with chemically processed material. This chemical processing can have several origins including an accretion shock \emph{en route} into the forming disk, turbulent mixing within the disk once formed, or an accretion outburst caused by material in-falling from the disk onto the star. If any (all) of these processes have occurred, then this supports an early formation of comets (<1 Myr). However, if the pre-solar nebula formed and evolved in a quiescent manner, a late formation of comets ($\sim$8 Myr) is also supported. In this latter case, it cannot be distinguished whether or not the cometary material has an interstellar origin.

Addressing the formation times of the comets also leads to the question of whether or not each individual comet formed quickly (e.g. assembly took place within 0.5 Myr after the comet started forming), or if each comet formed slowly, such that its assembly could have lasted throughout the lifetime of the pre-solar nebula disk midplane. In the former case, a quick assembly of the icy material at a given radius into a larger cometary body could have acted to impede chemical processing of the ices inside the comet, such that the composition inside the comet would remain unaltered, representing the composition of the icy material at the specific time and radius at which the assembly took place in the disk midplane. In the latter case, on the other hand, if the comet assembled slowly during the lifetime of the pre-solar nebula disk midplane, then the solid material incorporated into the comet would have experienced different degrees of chemical evolution in the disk midplane, and hence have different ice compositions, depending on what time that material was incorporated into the comet. Assuming again that, subsequent to the comet's final assembly, the ice chemistry inside it is impeded, then the observed cometary abundances in this slowly-formed comet would likely be a mix of different, say, \ce{H2O} ice abundances from different times during its formation. When observed, such a possible mix of different abundances of one ice species inside the comet would then likely manifest as the average over these different abundances, thereby not carrying a fingerprint of any specific time and radius of formation. The results presented here highlight specific formation times and radii for individual comets which suggests that quick assembly of comets is possible. However, further calculations are needed to ascertain to what degree a slower rate of assembly can influence the bulk comet composition over time.

Comparing the grouping of 15 comets here with the previous cometary classifications from \citet{cochran2012} as typical, depleted or ``mixed classification'' \citep[as of Table 2 in][]{leroy2015} they fall under all these three classes. These 14 comets have thus never been grouped chemically together before. With these 14 comets now grouped together based on likely formation in the vicinity of the CO iceline, it is also possible to propose a formation sequence for this group, based on the peak of the maximum likelihood surface for each comet in Fig. \ref{P_res_evol_percent}.

Based on this, the following formation time classes from early to late for the 15 comets can be proposed: 9P, 73P, Lemmon, and Lovejoy (all comets feature maximum likelihood peaks from 0.4-0.64 Myr at 28-30 AU), 2P (by 4.3 Myr at 16 AU), and 1P, 6P, 17P, 21P, 67P-S, 67P-W, 103P, Hyakutake, LINEAR, and Hale-Bopp (by 8.23 Myr at 11-13 AU). It is noted that some of these comets have their maximum likelihood peaks located outside of the formation region around the CO iceline. For the sake of focusing on the possible formation time classification around this iceline, only likelihood peaks happening between 0.4 and 8.23 Myr were used for the analysis, thereby ignoring peaks outside of the CO iceline region of the parameter space. This formation time classification creates one group of four comets that possibly formed early, between 0.4-0.64 Myr, one comet that possibly formed at an intermediate time by 4.3 Myr, and another group of ten comets that possibly formed later after 8 Myr. This hints that comet formation may occur in tandem with disk evolution over $\sim$8 Myr timescales.

This formation classes should be considered in light of which comets have more detections. As is evident in Fig. \ref{P_res_evol_percent}, the maximum likelihood peaks are very localised (either $\sim$13 AU by 8 Myr or at $\sim$30 AU by 1 Myr) for the comets that have more detections. The comets with fewer detections are less constrained and thus have larger regions of parameter space with good agreement with the models. The comets with more detections are thus constrained to having formation times similar to each other.

An interesting additional consideration regards which specific species have been detected in which comets, and how that may relate to which regions of parameter space the comets are in best agreement with. Comets 73P, 2P, and 6P have no CO detections in them, yet in the analysis here these are all constrained to have formed in the vicinity of the CO iceline. Detection of CO in a comet is thus not a pre-requisite for constraining the comet formation (roughly) to this region. It is noted that a comet that has formed inside of the CO ice line should naturally be CO poor.

Lastly, it is seen from all plots of comet 67P-S and 67P-W that the summer and winter sides of the comet feature different compositions. Which side, or which relative proportions of the sides are representative of the bulk composition of the comet is still unclear, as was noted in \citet{leroy2015}. Classifying the formation time of the comet as a whole should be done with caution, although the summer side observations of the comet are likely more comparable to the observations of the other comets. One reason for the differences between the two seasons around 67P, could be that the temperature on its winter side is too low for \ce{H2O} ice to sublimate. This, in turn, can lead to increased abundances of the rest of the molecular species, because they all have a lower sublimation temperature than \ce{H2O} ice. This is important evidence that cometary activity is a crucial factor to take into consideration when extrapolating abundances measured in the cometary coma to the bulk composition. In future work, it would be worthwhile to explore how the ice ratios vary relative to a different species that is less susceptible to summer and winter effects. 

\section{Conclusion}
\label{conclusion}

In this work a statistical $\chi^{2}$ method has been used to perform a quantitative comparison between observed cometary abundances and modelled chemical evolution in a protoplanetary disk midplane, by computing maximum likelihood surfaces, as a function of disk midplace location (radius) and time. The best agreements between observed and modelled abundances were found when considering the chemical evolution models to be chemically reset at the start, thus assuming that the volatile content of the pre-solar nebula was (perhaps partially) atomised before dust grains starting building larger bodies. This is consistent with the traditional idea about the chemical start of the inner pre-solar nebula \citep{grossman1972}, but not the outer nebula.

All 15 comets (14, when counting 67P-W and 67P-S as one) were found to have high likelihoods of formation along a trail in time and radial parameter space which is in the vicinity of the CO iceline. Since CO is the molecule (next to \ce{N2} and \ce{H2}) with the lowest binding energy ($E_{b}$=855 K) this means that most molecules are in the ice at the point where the comets are most likely to have formed. We do not consider formation radii here that are outside of the CO iceline, as the abundances in this region are found to be very different from those around the iceline \citep[see][]{eistrup2018}. There, it is found that the grain-surface chemistry is mainly driven by hydrogenation reactions leading to high abundances of, for example, \ce{H2O}, \ce{CH4}, \ce{C2H6}, and \ce{CH3OH} ices.

Based on the maxima of the likelihood functions for each comet along the CO iceline, it was then determined when during chemical evolution each comet was most likely to have formed. Thereby a formation time classification for all 15 comets was proposed, with some degeneracy remaining between the early (<1 Myr) and late (>7.5 Myr) formation. With more samples of comets with sufficient molecular detections in the future, it will be possible to further test this chemical evolution classification scheme for formation histories of comets. It will be interesting to see if other comets support the idea of a chemical reset start, and if they too show best agreement with formation in the vicinity of the CO iceline.

\begin{acknowledgements}
This work was motivated by discussions at the International Space Science Institute in Bern during meetings of International Team 361, ``From Qualitative to Quantitative: Exploring the Early solar system by Connecting Comet Composition and Protoplanetary Disk Models'', led by Dr Boncho Bonev. The authors thank the anonymous referee for valuable comments and input that helped improve the manuscript, and for pointing out the published measurements of CO in comet 17P, which led to the inclusion of the CO abundance for this comet. C.E. also thanks Chunhua Qi for discussions and suggestions on which measured CO abundance value to use for comet 17P in this work, and Anita Cochran for an interesting discussion. Astrochemistry in Leiden is supported by the European Union A-ERC grant 291141 CHEMPLAN and the Netherlands Research School for Astronomy (NOVA). C.W. acknowledges support from the University of Leeds and the Science and Technology Facilities Council (grant No. ST/R000549/1). C.E. acknowledges the Virginia Initiative on Cosmic Origins (VICO) Fellowship for financial support.
\end{acknowledgements}


\bibliographystyle{aa} 
\bibliography{../../Downloads/paper_oxygen/bib_new}

\begin{thebibliography}{66}
\expandafter\ifx\csname natexlab\endcsname\relax\def\natexlab#1{#1}\fi

\bibitem[{{Adams}(2010)}]{adams2010}
{Adams}, F.~C. 2010, \araa, 48, 47

\bibitem[{{A'Hearn} {et~al.}(1995){A'Hearn}, {Millis}, {Schleicher}, {Osip}, \&
  {Birch}}]{ahearn1995}
{A'Hearn}, M.~F., {Millis}, R.~C., {Schleicher}, D.~O., {Osip}, D.~J., \&
  {Birch}, P.~V. 1995, \icarus, 118, 223

\bibitem[{{Aikawa} {et~al.}(1997){Aikawa}, {Umebayashi}, {Nakano}, \&
  {Miyama}}]{aikawa1997}
{Aikawa}, Y., {Umebayashi}, T., {Nakano}, T., \& {Miyama}, S.~M. 1997, \apjl,
  486, L51

\bibitem[{{Altwegg} {et~al.}(1994){Altwegg}, {Balsiger}, \&
  {Geiss}}]{altwegg1994}
{Altwegg}, K., {Balsiger}, H., \& {Geiss}, J. 1994, \aap, 290, 318

\bibitem[{{Biver} {et~al.}(2007){Biver}, {Bockel{\'e}e-Morvan}, {Boissier},
  {Crovisier}, {Colom}, {Lecacheux}, {Moreno}, {Paubert}, {Lis}, {Sumner},
  {Frisk}, {Hjalmarson}, {Olberg}, {Winnberg}, {Flor{\'e}n}, {Sandqvist}, \&
  {Kwok}}]{biver2007}
{Biver}, N., {Bockel{\'e}e-Morvan}, D., {Boissier}, J., {et~al.} 2007, \icarus,
  187, 253

\bibitem[{{Biver} {et~al.}(1999){Biver}, {Bockel{\'e}e-Morvan}, {Crovisier},
  {Davies}, {Matthews}, {Wink}, {Rauer}, {Colom}, {Dent}, {Despois}, {Moreno},
  {Paubert}, {Jewitt}, \& {Senay}}]{biver1999}
{Biver}, N., {Bockel{\'e}e-Morvan}, D., {Crovisier}, J., {et~al.} 1999, \aj,
  118, 1850

\bibitem[{{Biver} {et~al.}(2008){Biver}, {Bockel{\'e}e-Morvan}, {Crovisier},
  {Lecacheux}, {Lis}, {Boissier}, {Colom}, {Dello-Russo}, {Flor{\'e}n},
  {Frisk}, {Hjalmarson}, {Kwok}, {Menten}, {Moreno}, {Olberg}, {Parise},
  {Paubert}, {Sandqvist}, {Vervack}, {Weaver}, \& {Winnberg}}]{biver2008conf}
{Biver}, N., {Bockel{\'e}e-Morvan}, D., {Crovisier}, J., {et~al.} 2008, in
  Asteroids, Comets, Meteors 2008, Vol. 1405, 8149

\bibitem[{{Biver} {et~al.}(2006){Biver}, {Bockel{\'e}e-Morvan}, {Crovisier},
  {Lis}, {Moreno}, {Colom}, {Henry}, {Herpin}, {Paubert}, \&
  {Womack}}]{biver2006}
{Biver}, N., {Bockel{\'e}e-Morvan}, D., {Crovisier}, J., {et~al.} 2006, \aap,
  449, 1255

\bibitem[{{Biver} {et~al.}(2014){Biver}, {Bockel{\'e}e-Morvan}, {Debout},
  {Crovisier}, {Boissier}, {Lis}, {Dello Russo}, {Moreno}, {Colom}, {Paubert},
  {Vervack}, \& {Weaver}}]{biver2014}
{Biver}, N., {Bockel{\'e}e-Morvan}, D., {Debout}, V., {et~al.} 2014, \aap, 566,
  L5

\bibitem[{{Bockel{\'e}e-Morvan} {et~al.}(2014){Bockel{\'e}e-Morvan}, {Biver},
  {Crovisier}, {Lis}, {Hartogh}, {Moreno}, {de Val-Borro}, {Blake},
  {Szutowicz}, {Boissier}, {Cernicharo}, {Charnley}, {Combi}, {Cordiner}, {de
  Graauw}, {Encrenaz}, {Jarchow}, {Kidger}, {K{\"u}ppers}, {Milam},
  {M{\"u}ller}, {Phillips}, \& {Rengel}}]{bockelee-morvan2014}
{Bockel{\'e}e-Morvan}, D., {Biver}, N., {Crovisier}, J., {et~al.} 2014, \aap,
  562, A5

\bibitem[{{Bockel{\'e}e-Morvan} {et~al.}(1995){Bockel{\'e}e-Morvan}, {Brooke},
  \& {Crovisier}}]{bockelee-morvan1995}
{Bockel{\'e}e-Morvan}, D., {Brooke}, T.~Y., \& {Crovisier}, J. 1995, \icarus,
  116, 18

\bibitem[{{Bockel{\'e}e-Morvan} {et~al.}(2004){Bockel{\'e}e-Morvan},
  {Crovisier}, {Mumma}, \& {Weaver}}]{bockelee-morvan2004}
{Bockel{\'e}e-Morvan}, D., {Crovisier}, J., {Mumma}, M.~J., \& {Weaver}, H.~A.
  2004, {The composition of cometary volatiles}, 391--423

\bibitem[{{Bockel{\'e}e-Morvan} {et~al.}(2015){Bockel{\'e}e-Morvan}, {Debout},
  {Erard}, {Leyrat}, {Capaccioni}, {Filacchione}, {Fougere}, {Drossart},
  {Arnold}, {Combi}, {Schmitt}, {Crovisier}, {de Sanctis}, {Encrenaz},
  {K{\"u}hrt}, {Palomba}, {Taylor}, {Tosi}, {Piccioni}, {Fink}, {Tozzi},
  {Barucci}, {Biver}, {Capria}, {Combes}, {Ip}, {Blecka}, {Henry}, {Jacquinod},
  {Reess}, {Semery}, \& {Tiphene}}]{bockelee-morvan2015}
{Bockel{\'e}e-Morvan}, D., {Debout}, V., {Erard}, S., {et~al.} 2015, \aap, 583,
  A6

\bibitem[{{Bockel{\'e}e-Morvan} {et~al.}(2000){Bockel{\'e}e-Morvan}, {Lis},
  {Wink}, {Despois}, {Crovisier}, {Bachiller}, {Benford}, {Biver}, {Colom},
  {Davies}, {G{\'e}rard}, {Germain}, {Houde}, {Mehringer}, {Moreno}, {Paubert},
  {Phillips}, \& {Rauer}}]{bockelee-morvan2000}
{Bockel{\'e}e-Morvan}, D., {Lis}, D.~C., {Wink}, J.~E., {et~al.} 2000, \aap,
  353, 1101

\bibitem[{{Brooke} {et~al.}(1996){Brooke}, {Tokunaga}, {Weaver}, {Crovisier},
  {Bockel{\'e}e-Morvan}, \& {Crisp}}]{brooke1996}
{Brooke}, T.~Y., {Tokunaga}, A.~T., {Weaver}, H.~A., {et~al.} 1996, \nat, 383,
  606

\bibitem[{{Cleeves} {et~al.}(2013){Cleeves}, {Adams}, \&
  {Bergin}}]{cleeves13crex}
{Cleeves}, L.~I., {Adams}, F.~C., \& {Bergin}, E.~A. 2013, \apj, 772, 5

\bibitem[{{Cochran} {et~al.}(2012){Cochran}, {Barker}, \& {Gray}}]{cochran2012}
{Cochran}, A.~L., {Barker}, E.~S., \& {Gray}, C.~L. 2012, \icarus, 218, 144

\bibitem[{{Colangeli} {et~al.}(1999){Colangeli}, {Epifani}, {Brucato},
  {Bussoletti}, {De Sanctis}, {Fulle}, {Mennella}, {Palomba}, {Palumbo}, \&
  {Rotundi}}]{colangeli1999}
{Colangeli}, L., {Epifani}, E., {Brucato}, J.~R., {et~al.} 1999, \aap, 343, L87

\bibitem[{{Combes} {et~al.}(1988){Combes}, {Moroz}, {Crovisier}, {Encrenaz},
  {Bibring}, {Grigoriev}, {Sanko}, {Coron}, {Crifo}, {Gispert},
  {Bockel{\'e}e-Morvan}, {Nikolsky}, {Krasnopolsky}, {Owen}, {Emerich},
  {Lamarre}, \& {Rocard}}]{combes1988}
{Combes}, M., {Moroz}, V.~I., {Crovisier}, J., {et~al.} 1988, \icarus, 76, 404

\bibitem[{{Dello Russo} {et~al.}(2001){Dello Russo}, {Mumma}, {DiSanti},
  {Magee-Sauer}, \& {Novak}}]{dellorusso2001}
{Dello Russo}, N., {Mumma}, M.~J., {DiSanti}, M.~A., {Magee-Sauer}, K., \&
  {Novak}, R. 2001, \icarus, 153, 162

\bibitem[{{Dello Russo} {et~al.}(2008){Dello Russo}, R.~J.~Vervack, Weaver,
  Montgomery, Deshpande, Fernández, \& Martin}]{dellorusso2008}
{Dello Russo}, N., R.~J.~Vervack, J., Weaver, H.~A., {et~al.} 2008, The
  Astrophysical Journal, 680, 793

\bibitem[{{Dello Russo} {et~al.}(2011){Dello Russo}, {Vervack}, {Lisse},
  {Weaver}, {Kawakita}, {Kobayashi}, {Cochran}, {Harris}, {McKay}, {Biver},
  {Bockel{\'e}e-Morvan}, \& {Crovisier}}]{dellorusso2011}
{Dello Russo}, N., {Vervack}, R.~J., J., {Lisse}, C.~M., {et~al.} 2011, \apj,
  734, L8

\bibitem[{{Dello Russo} {et~al.}(2009){Dello Russo}, {Vervack}, {Weaver},
  {Kawakita}, {Kobayashi}, {Biver}, {Bockel{\'e}e-Morvan}, \&
  {Crovisier}}]{dellorusso2009}
{Dello Russo}, N., {Vervack}, R.~J., J., {Weaver}, H.~A., {et~al.} 2009, \apj,
  703, 187

\bibitem[{{Dello Russo} {et~al.}(2007){Dello Russo}, {Vervack}, {Weaver},
  {Biver}, {Bockel{\'e}e-Morvan}, {Crovisier}, \& {Lisse}}]{dellorusso2007}
{Dello Russo}, N., {Vervack}, R.~J., {Weaver}, H.~A., {et~al.} 2007, \nat, 448,
  172

\bibitem[{{Despois} {et~al.}(2005){Despois}, {Biver}, {Bockel{\'e}e-Morvan}, \&
  {Crovisier}}]{despois2005conf}
{Despois}, D., {Biver}, N., {Bockel{\'e}e-Morvan}, D., \& {Crovisier}, J. 2005,
  in IAU Symposium, Vol. 231, Astrochemistry: Recent Successes and Current
  Challenges, ed. D.~C. {Lis}, G.~A. {Blake}, \& E.~{Herbst}, 469--478

\bibitem[{{DiSanti} {et~al.}(2007{\natexlab{a}}){DiSanti}, {Anderson},
  {Villanueva}, {Bonev}, {Magee-Sauer}, {Gibb}, \&
  {Mumma}}]{disanti2007codepl73p}
{DiSanti}, M.~A., {Anderson}, W.~M., {Villanueva}, G.~L., {et~al.}
  2007{\natexlab{a}}, \apj, 661, L101

\bibitem[{{DiSanti} {et~al.}(2003){DiSanti}, {Mumma}, {Dello Russo},
  {Magee-Sauer}, \& {Griep}}]{disanti2003}
{DiSanti}, M.~A., {Mumma}, M.~J., {Dello Russo}, N., {Magee-Sauer}, K., \&
  {Griep}, D.~M. 2003, Journal of Geophysical Research (Planets), 108, 5061

\bibitem[{{DiSanti} {et~al.}(2007{\natexlab{b}}){DiSanti}, {Villanueva},
  {Bonev}, {Magee-Sauer}, {Lyke}, \& {Mumma}}]{disanti2007comet9p}
{DiSanti}, M.~A., {Villanueva}, G.~L., {Bonev}, B.~P., {et~al.}
  2007{\natexlab{b}}, \icarus, 187, 240

\bibitem[{{Eberhardt}(1999)}]{eberhardt1999}
{Eberhardt}, P. 1999, \ssr, 90, 45

\bibitem[{{Eberhardt} {et~al.}(1994){Eberhardt}, {Meier}, {Krankowsky}, \&
  {Hodges}}]{eberhardt1994}
{Eberhardt}, P., {Meier}, R., {Krankowsky}, D., \& {Hodges}, R.~R. 1994, \aap,
  288, 315

\bibitem[{{Eistrup} {et~al.}(2016){Eistrup}, {Walsh}, \& {van
  Dishoeck}}]{eistrup2016}
{Eistrup}, C., {Walsh}, C., \& {van Dishoeck}, E.~F. 2016, \aap, 595, A83

\bibitem[{{Eistrup} {et~al.}(2018){Eistrup}, {Walsh}, \& {van
  Dishoeck}}]{eistrup2018}
{Eistrup}, C., {Walsh}, C., \& {van Dishoeck}, E.~F. 2018, \aap, 613, A14

\bibitem[{{Fink}(2009)}]{fink2009}
{Fink}, U. 2009, \icarus, 201, 311

\bibitem[{{Gibb} {et~al.}(2007){Gibb}, {DiSanti}, {Magee-Sauer}, {Dello Russo},
  {Bonev}, \& {Mumma}}]{gibb2007}
{Gibb}, E.~L., {DiSanti}, M.~A., {Magee-Sauer}, K., {et~al.} 2007, \icarus,
  188, 224

\bibitem[{{Gibb} {et~al.}(2003){Gibb}, {Mumma}, {Dello Russo}, {DiSanti}, \&
  {Magee-Sauer}}]{gibb2003}
{Gibb}, E.~L., {Mumma}, M.~J., {Dello Russo}, N., {DiSanti}, M.~A., \&
  {Magee-Sauer}, K. 2003, \icarus, 165, 391

\bibitem[{{Grossman}(1972)}]{grossman1972}
{Grossman}, L. 1972, \gca, 36, 597

\bibitem[{{Hayashi}(1981)}]{hayashi1981}
{Hayashi}, C. 1981, Progress of Theoretical Physics Supplement, 70, 35

\bibitem[{{Kawakita} {et~al.}(2013){Kawakita}, {Kobayashi}, {Dello Russo},
  {Vervack}, {Hashimoto}, {Weaver}, {Lisse}, {Cochran}, {Harris},
  {Bockel{\'e}e-Morvan}, {Biver}, {Crovisier}, \& {McKay}}]{kawakita2013}
{Kawakita}, H., {Kobayashi}, H., {Dello Russo}, N., {et~al.} 2013, \icarus,
  222, 723

\bibitem[{{Krankowsky} {et~al.}(1986){Krankowsky}, {Lammerzahl}, {Herrwerth},
  {Woweries}, {Eberhardt}, {Dolder}, {Herrmann}, {Schulte}, {Berthelier},
  {Illiano}, {Hodges}, \& {Hoffman}}]{krankowski1986}
{Krankowsky}, D., {Lammerzahl}, P., {Herrwerth}, I., {et~al.} 1986, \nat, 321,
  326

\bibitem[{{Le Roy} {et~al.}(2015){Le Roy}, {Altwegg}, {Balsiger}, {Berthelier},
  {Bieler}, {Briois}, {Calmonte}, {Combi}, {De Keyser}, {Dhooghe}, {Fiethe},
  {Fuselier}, {Gasc}, {Gombosi}, {H{\"a}ssig}, {J{\"a}ckel}, {Rubin}, \&
  {Tzou}}]{leroy2015}
{Le Roy}, L., {Altwegg}, K., {Balsiger}, H., {et~al.} 2015, \aap, 583, A1

\bibitem[{{Lis} {et~al.}(1997){Lis}, {Keene}, {Young}, {Phillips},
  {Bockel{\'e}e-Morvan}, {Crovisier}, {Schilke}, {Goldsmith}, \&
  {Bergin}}]{lis1997}
{Lis}, D.~C., {Keene}, J., {Young}, K., {et~al.} 1997, \icarus, 130, 355

\bibitem[{{Magee-Sauer} {et~al.}(2008){Magee-Sauer}, {Mumma}, {DiSanti}, {Dello
  Russo}, {Gibb}, {Bonev}, \& {Villanueva}}]{magee-sauer2008}
{Magee-Sauer}, K., {Mumma}, M.~J., {DiSanti}, M.~A., {et~al.} 2008, \icarus,
  194, 347

\bibitem[{{McElroy} {et~al.}(2013){McElroy}, {Walsh}, {Markwick}, {Cordiner},
  {Smith}, \& {Millar}}]{mcelroy13}
{McElroy}, D., {Walsh}, C., {Markwick}, A.~J., {et~al.} 2013, \aap, 550, A36

\bibitem[{{McPhate} {et~al.}(1996){McPhate}, {Feldman}, {Weaver}, {A'Hearn},
  {Tozzi}, \& {Festou}}]{mcphate1996conf}
{McPhate}, J.~B., {Feldman}, P.~D., {Weaver}, H.~A., {et~al.} 1996, in
  AAS/Division for Planetary Sciences Meeting Abstracts \#28, AAS/Division for
  Planetary Sciences Meeting Abstracts, 09.29

\bibitem[{{Mumma} {et~al.}(2011){Mumma}, {Bonev}, {Villanueva}, {Paganini},
  {DiSanti}, {Gibb}, {Keane}, {Meech}, {Blake}, {Ellis}, {Lippi}, {Boehnhardt},
  \& {Magee-Sauer}}]{mumma2011}
{Mumma}, M.~J., {Bonev}, B.~P., {Villanueva}, G.~L., {et~al.} 2011, \apj, 734,
  L7

\bibitem[{{Mumma} \& {Charnley}(2011)}]{charnley11}
{Mumma}, M.~J. \& {Charnley}, S.~B. 2011, \araa, 49, 471

\bibitem[{{Mumma} {et~al.}(1996){Mumma}, {Disanti}, {dello Russo}, {Fomenkova},
  {Magee-Sauer}, {Kaminski}, \& {Xie}}]{mumma1996}
{Mumma}, M.~J., {Disanti}, M.~A., {dello Russo}, N., {et~al.} 1996, Science,
  272, 1310

\bibitem[{{Mumma} {et~al.}(2003){Mumma}, {DiSanti}, {Dello Russo},
  {Magee-Sauer}, {Gibb}, \& {Novak}}]{mumma2003}
{Mumma}, M.~J., {DiSanti}, M.~A., {Dello Russo}, N., {et~al.} 2003, Advances in
  Space Research, 31, 2563

\bibitem[{{Mumma} {et~al.}(2000){Mumma}, {DiSanti}, {Dello Russo},
  {Magee-Sauer}, \& {Rettig}}]{mumma2000}
{Mumma}, M.~J., {DiSanti}, M.~A., {Dello Russo}, N., {Magee-Sauer}, K., \&
  {Rettig}, T.~W. 2000, \apj, 531, L155

\bibitem[{{Mumma} {et~al.}(2005){Mumma}, {DiSanti}, {Magee-Sauer}, {Bonev},
  {Villanueva}, {Kawakita}, {Dello Russo}, {Gibb}, {Blake}, {Lyke}, {Campbell},
  {Aycock}, {Conrad}, \& {Hill}}]{mumma2005}
{Mumma}, M.~J., {DiSanti}, M.~A., {Magee-Sauer}, K., {et~al.} 2005, Science,
  310, 270

\bibitem[{{Ootsubo} {et~al.}(2012){Ootsubo}, {Kawakita}, {Hamada}, {Kobayashi},
  {Yamaguchi}, {Usui}, {Nakagawa}, {Ueno}, {Ishiguro}, {Sekiguchi}, {Watanabe},
  {Sakon}, {Shimonishi}, \& {Onaka}}]{ootsubo2012}
{Ootsubo}, T., {Kawakita}, H., {Hamada}, S., {et~al.} 2012, \apj, 752, 15

\bibitem[{{Padovani} {et~al.}(2018){Padovani}, {Ivlev}, {Galli}, \&
  {Caselli}}]{padovani2018}
{Padovani}, M., {Ivlev}, A.~V., {Galli}, D., \& {Caselli}, P. 2018, \aap, 614,
  A111

\bibitem[{{Paganini} {et~al.}(2014){Paganini}, {DiSanti}, {Mumma},
  {Villanueva}, {Bonev}, {Keane}, {Gibb}, {Boehnhardt}, \&
  {Meech}}]{paganini2014}
{Paganini}, L., {DiSanti}, M.~A., {Mumma}, M.~J., {et~al.} 2014, \aj, 147, 15

\bibitem[{{Qi} {et~al.}(2015){Qi}, {Hogerheijde}, {Jewitt}, {Gurwell}, \&
  {Wilner}}]{qi2015}
{Qi}, C., {Hogerheijde}, M.~R., {Jewitt}, D., {Gurwell}, M.~A., \& {Wilner},
  D.~J. 2015, \apj, 799, 110

\bibitem[{{Radeva} {et~al.}(2013){Radeva}, {Mumma}, {Villanueva}, {Bonev},
  {DiSanti}, {A'Hearn}, \& {Dello Russo}}]{radeva2013}
{Radeva}, Y.~L., {Mumma}, M.~J., {Villanueva}, G.~L., {et~al.} 2013, \icarus,
  223, 298

\bibitem[{{Rubin} {et~al.}(2011){Rubin}, {Tenishev}, {Combi}, {Hansen},
  {Gombosi}, {Altwegg}, \& {Balsiger}}]{rubin2011}
{Rubin}, M., {Tenishev}, V.~M., {Combi}, M.~R., {et~al.} 2011, \icarus, 213,
  655

\bibitem[{{Schwarz} \& {Bergin}(2014)}]{schwarz2014}
{Schwarz}, K.~R. \& {Bergin}, E.~A. 2014, \apj, 797, 113

\bibitem[{{Taquet} {et~al.}(2018){Taquet}, {van Dishoeck}, {Swayne}, {Harsono},
  {J{\o}rgensen}, {Maud}, {Ligterink}, {M{\"u}ller}, {Codella}, {Altwegg},
  {Bieler}, {Coutens}, {Drozdovskaya}, {Furuya}, {Persson}, {van't Hoff},
  {Walsh}, \& {Wampfler}}]{taquet2018}
{Taquet}, V., {van Dishoeck}, E.~F., {Swayne}, M., {et~al.} 2018, \aap, 618,
  A11

\bibitem[{{Trinquier} {et~al.}(2009){Trinquier}, {Elliott}, {Ulfbeck}, {Coath},
  {Krot}, \& {Bizzarro}}]{trinquier2009}
{Trinquier}, A., {Elliott}, T., {Ulfbeck}, D., {et~al.} 2009, Science, 324, 374

\bibitem[{{Villanueva} {et~al.}(2006){Villanueva}, {Bonev}, {Mumma},
  {Magee-Sauer}, {DiSanti}, {Salyk}, \& {Blake}}]{villanueva2006}
{Villanueva}, G.~L., {Bonev}, B.~P., {Mumma}, M.~J., {et~al.} 2006, \apj, 650,
  L87

\bibitem[{{Visser} {et~al.}(2009){Visser}, {van Dishoeck}, {Doty}, \&
  {Dullemond}}]{visser2009}
{Visser}, R., {van Dishoeck}, E.~F., {Doty}, S.~D., \& {Dullemond}, C.~P. 2009,
  \aap, 495, 881

\bibitem[{{Walsh} {et~al.}(2015){Walsh}, {Nomura}, \& {van Dishoeck}}]{walsh15}
{Walsh}, C., {Nomura}, H., \& {van Dishoeck}, E. 2015, \aap, 582, A88

\bibitem[{{Weaver} {et~al.}(1999){Weaver}, {Chin}, {Bockel{\'e}e-Morvan},
  {Crovisier}, {Brooke}, {Cruikshank}, {Geballe}, {Kim}, \&
  {Meier}}]{weaver1999}
{Weaver}, H.~A., {Chin}, G., {Bockel{\'e}e-Morvan}, D., {et~al.} 1999, \icarus,
  142, 482

\bibitem[{{Weaver} {et~al.}(2011){Weaver}, {Feldman}, {A'Hearn}, {Dello Russo},
  \& {Stern}}]{weaver2011}
{Weaver}, H.~A., {Feldman}, P.~D., {A'Hearn}, M.~F., {Dello Russo}, N., \&
  {Stern}, S.~A. 2011, \apj, 734, L5

\bibitem[{{Weaver} {et~al.}(1994){Weaver}, {Feldman}, {McPhate}, {A'Hearn},
  {Arpigny}, \& {Smith}}]{weaver1994}
{Weaver}, H.~A., {Feldman}, P.~D., {McPhate}, J.~B., {et~al.} 1994, \apj, 422,
  374

\bibitem[{{Woodney} {et~al.}(1997){Woodney}, {McMullin}, \&
  {A'Hearn}}]{woodney1997}
{Woodney}, L.~M., {McMullin}, J., \& {A'Hearn}, M.~F. 1997, Planetary and Space
  Science, 45, 717

\end{thebibliography}

\begin{appendix}
\section{Evolving modelled abundances and cometary abundances}
\label{abun_plots}

\begin{figure*}[!t]
\includegraphics[clip, trim = 2.5cm 0.4cm 2.5cm 0.75cm, width=0.33\textwidth]{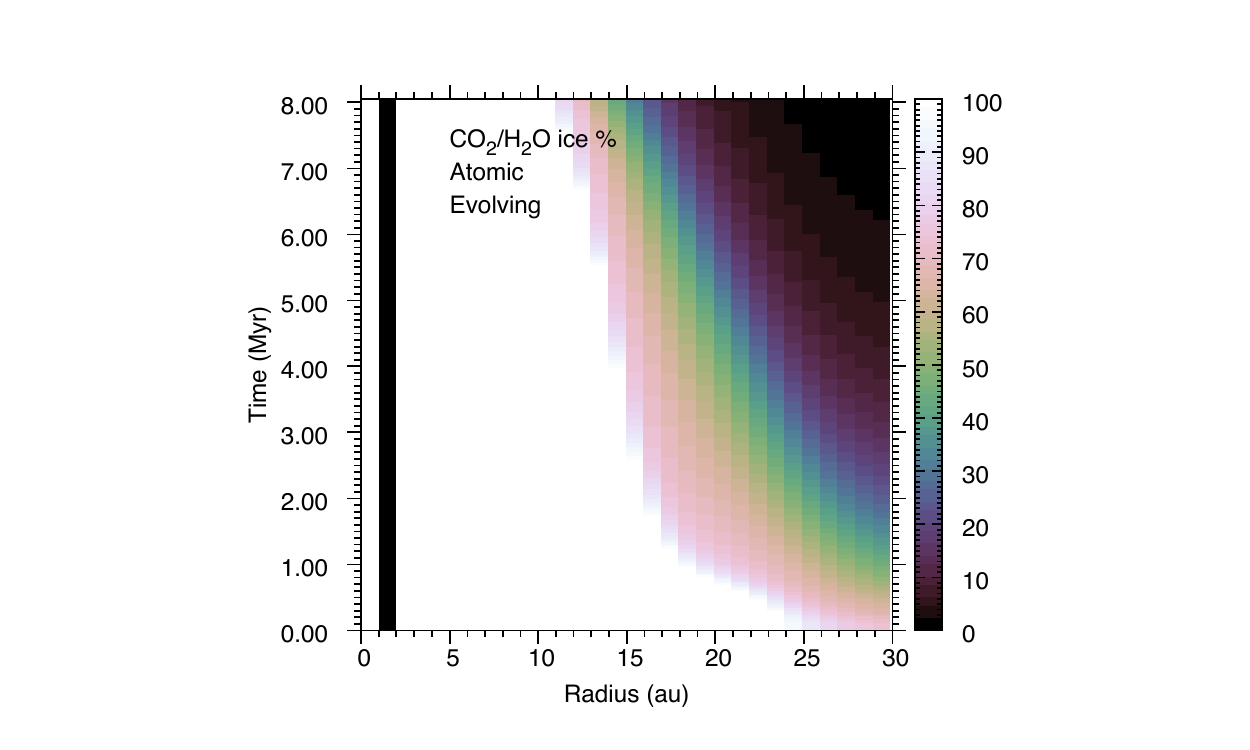}
\includegraphics[clip, trim = 2.5cm 0.4cm 2.5cm 0.75cm, width=0.33\textwidth]{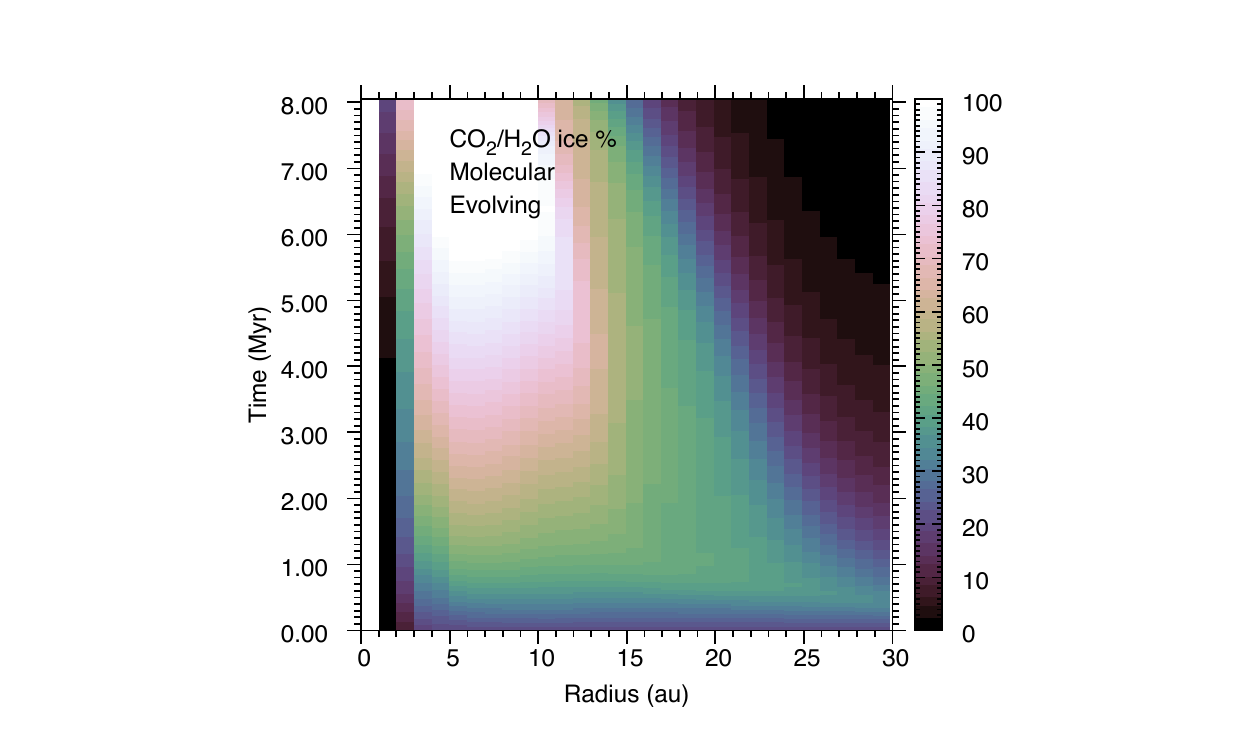}
\includegraphics[clip, trim = 2cm -0.5cm 2.25cm 0cm, width=0.33\textwidth]{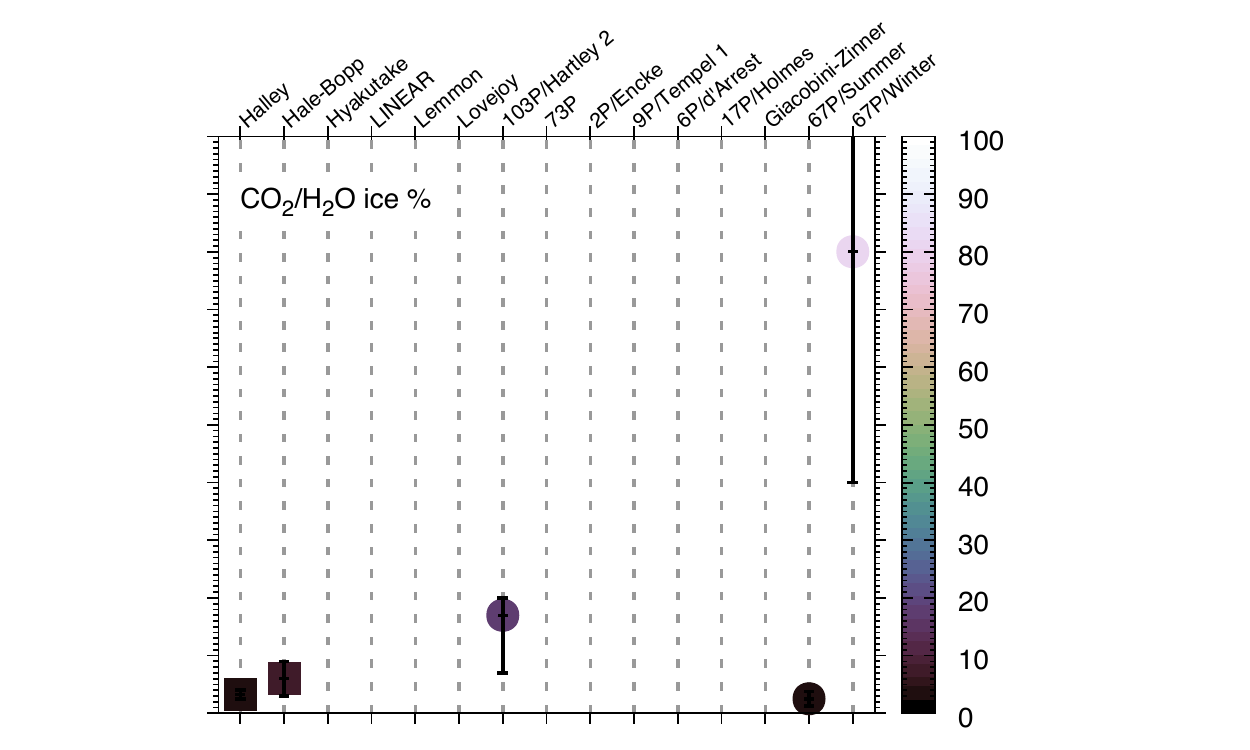}
\includegraphics[clip, trim = 2.5cm 0.4cm 2.5cm 0.75cm, width=0.33\textwidth]{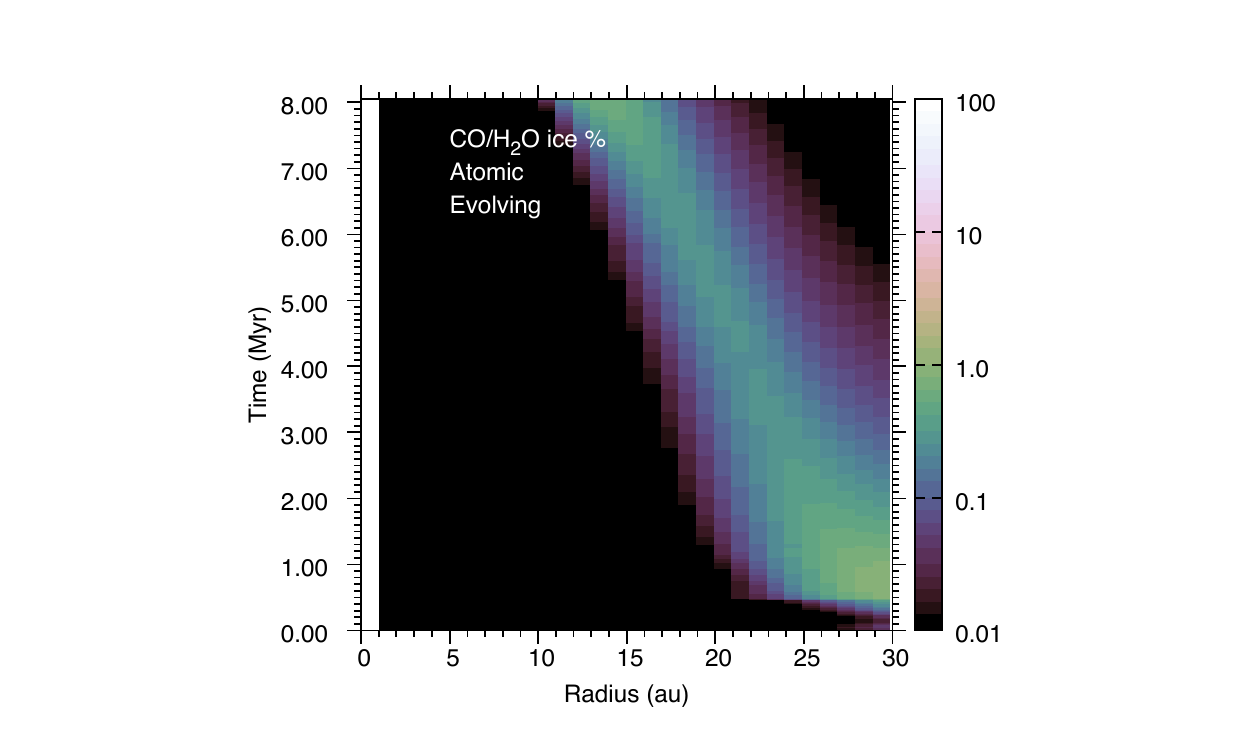}
\includegraphics[clip, trim = 2.5cm 0.4cm 2.5cm 0.75cm, width=0.33\textwidth]{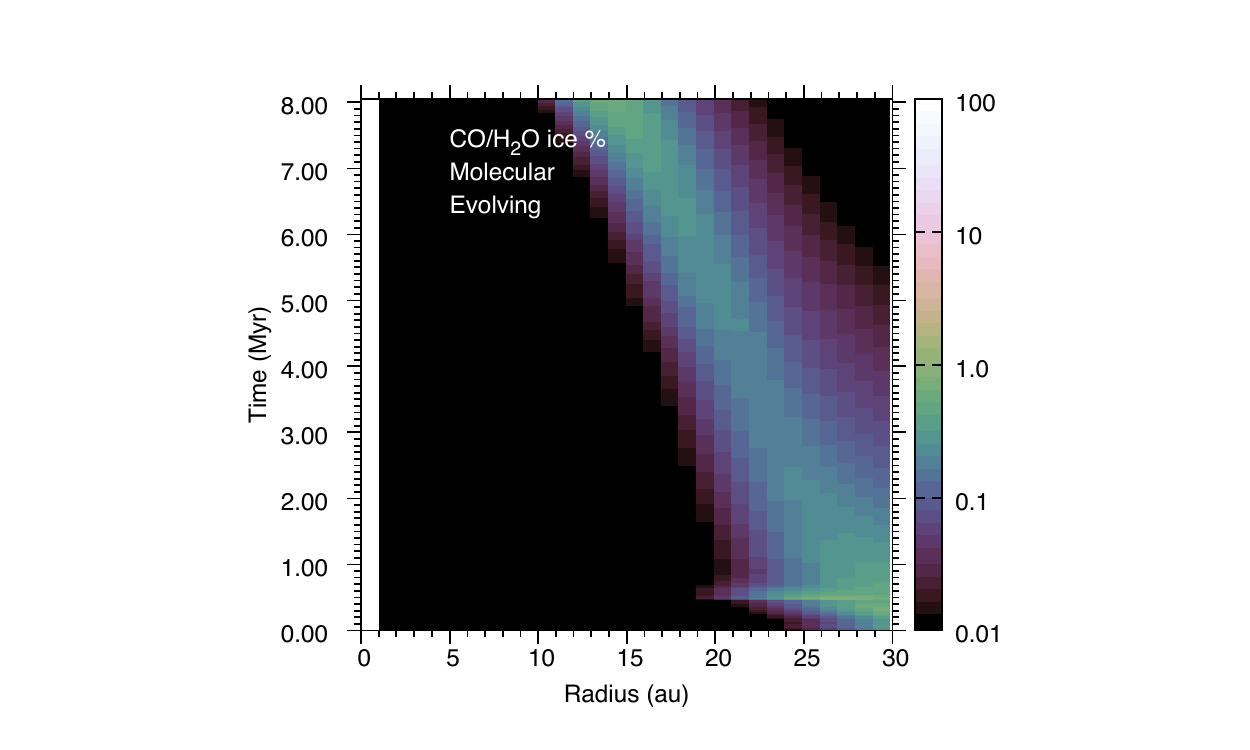}
\includegraphics[clip, trim = 2cm -0.5cm 2.25cm 0cm, width=0.33\textwidth]{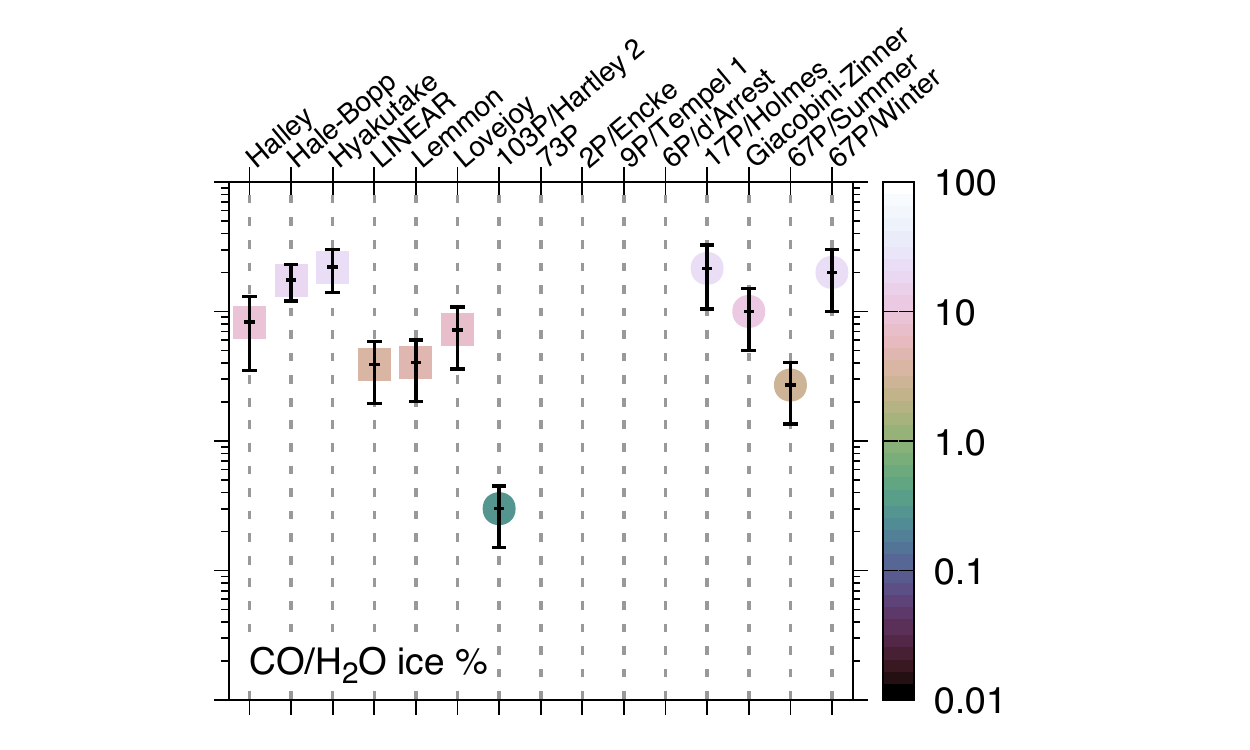}
\includegraphics[clip, trim = 2.5cm 0.4cm 2.5cm 0.75cm, width=0.33\textwidth]{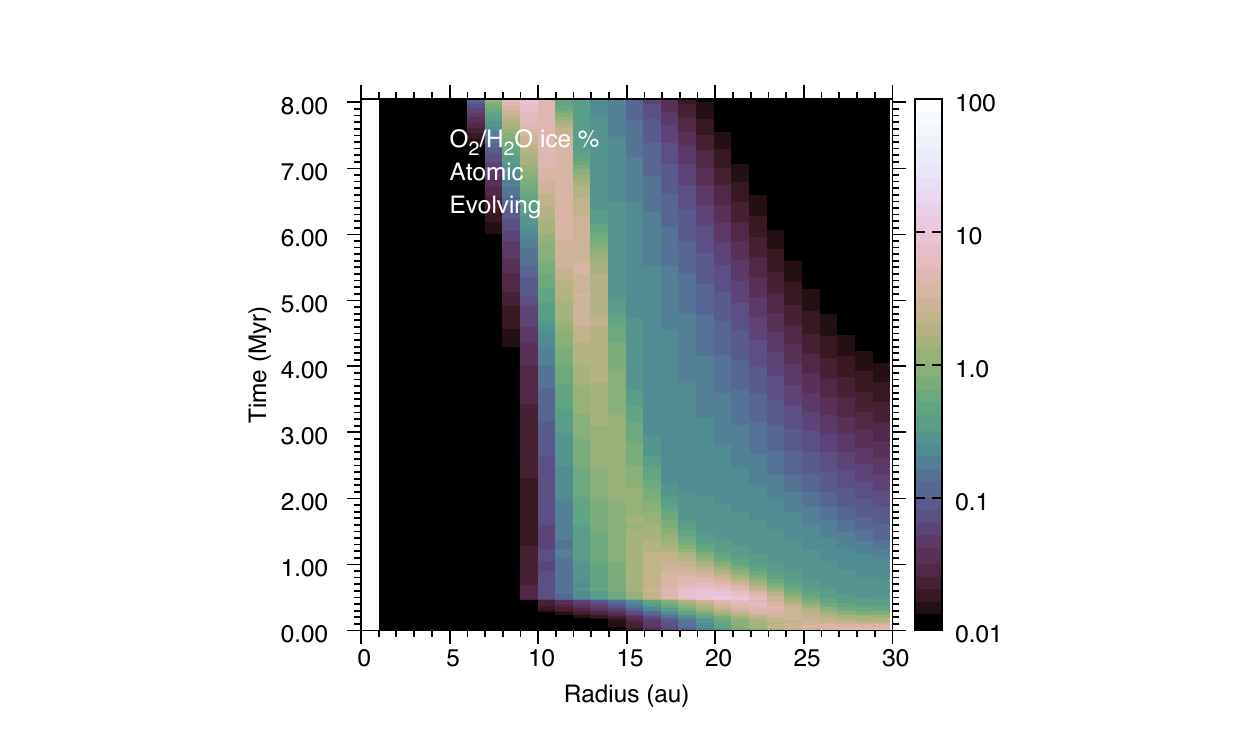}
\includegraphics[clip, trim = 2.5cm 0.4cm 2.5cm 0.75cm, width=0.33\textwidth]{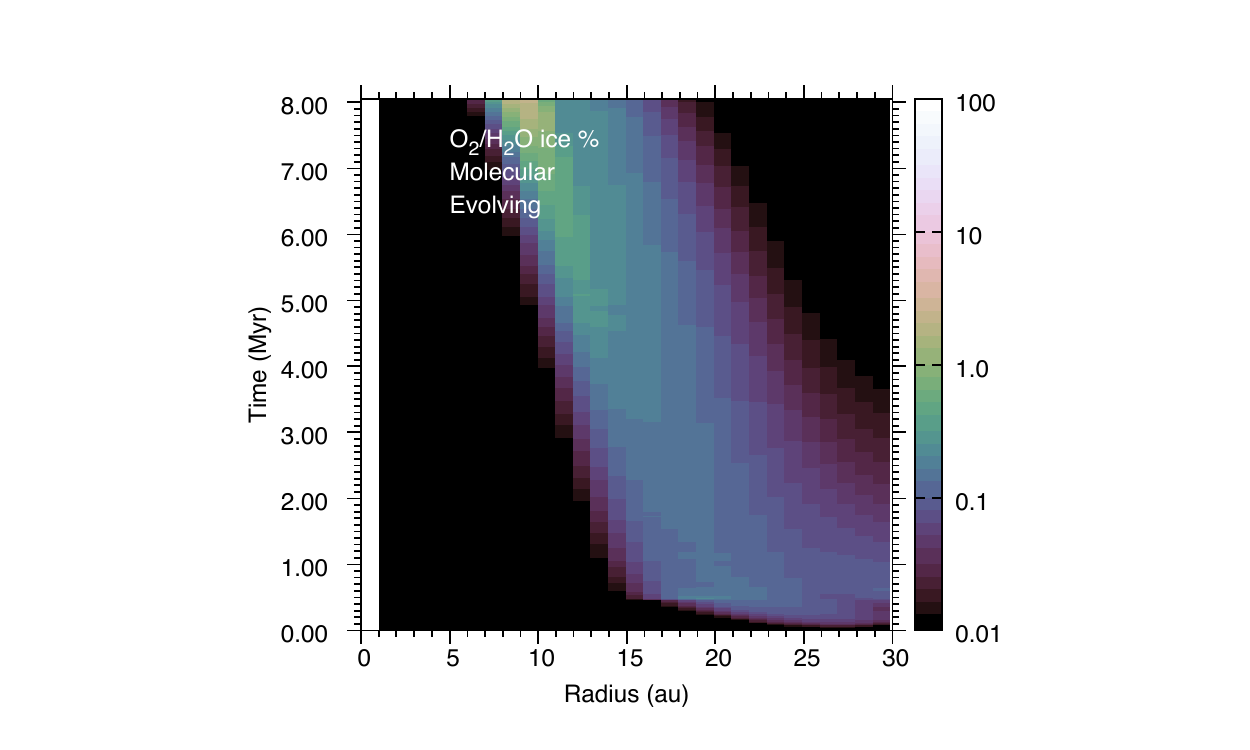}
\includegraphics[clip, trim = 2cm -0.5cm 2.25cm 0cm, width=0.33\textwidth]{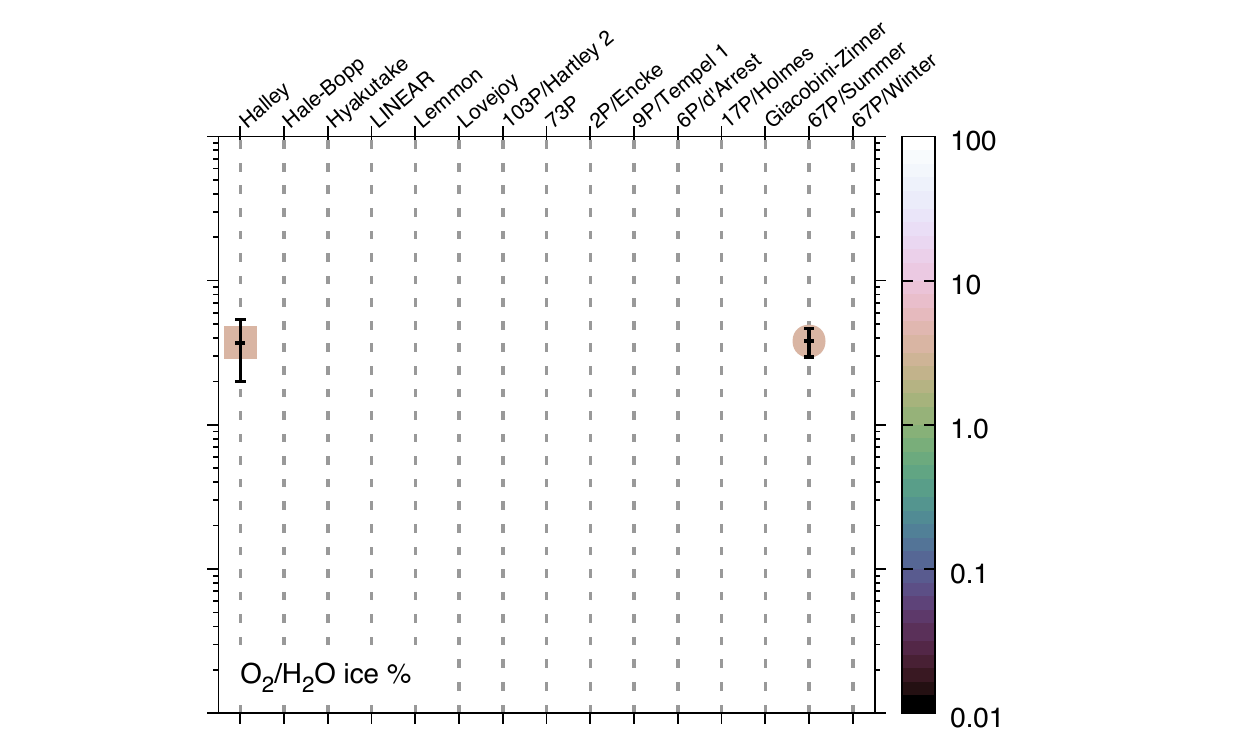}
\caption{Abundance of \ce{CO2}, \ce{CO}, and \ce{O2} ice relative to water ice as a function of radius and time for 
the protoplanetary disk model with evolving physical conditions and using fully atomic initial abundances (left) 
and fully molecular initial abundances (middle).  Note that the data for \ce{CO2} are shown on a linear scale 
as opposed to logarithmic because of the low dynamic range in the chemical model results.  
The right-hand column shows the corresponding values measured for each species in cometary comae.  
The vertical dashed lines are included solely to guide the eye. Oort cloud comets and Jupiter family comets 
are represented by the squares and circles, respectively. For comets without a stated observed error (or range), we have assumed a conservative error of 50\% of the observed ratio with respect to \ce{H2O} ice.}
\label{figure6}
\end{figure*}

\begin{figure*}[!t]
\includegraphics[clip, trim = 2.5cm 0.4cm 2.5cm 0.75cm, width=0.33\textwidth]{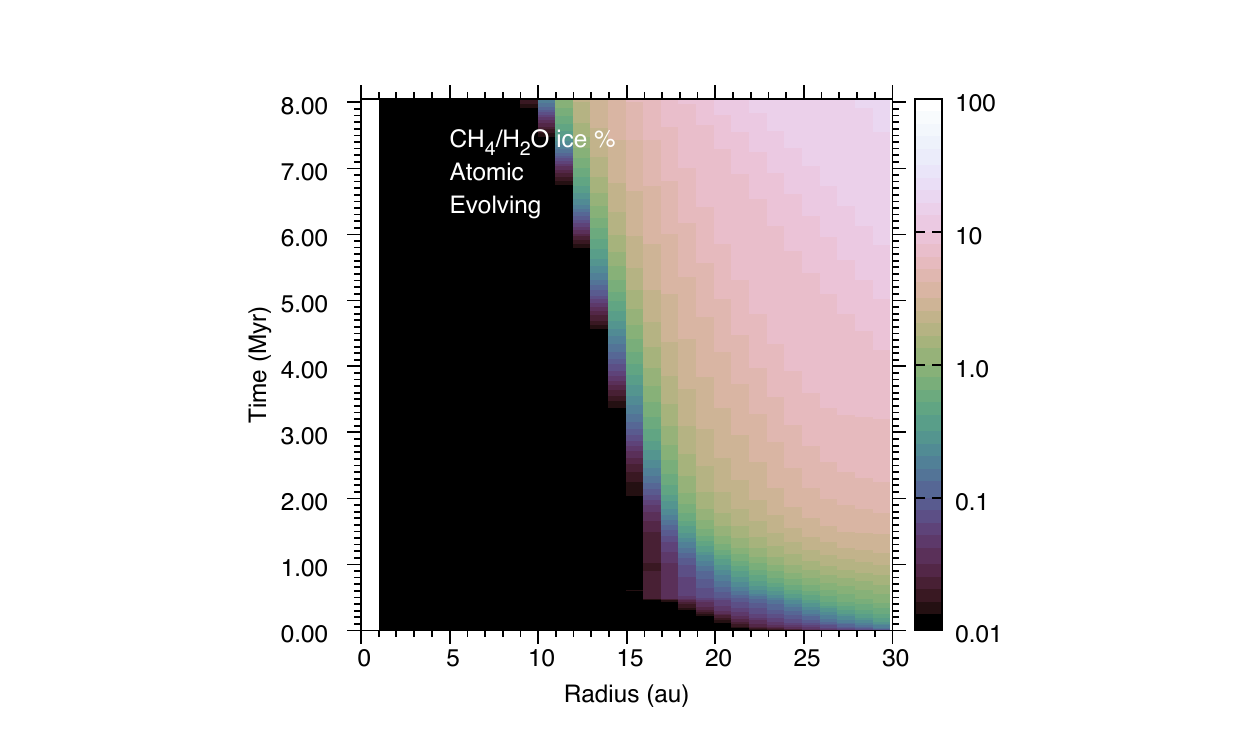}
\includegraphics[clip, trim = 2.5cm 0.4cm 2.5cm 0.75cm, width=0.33\textwidth]{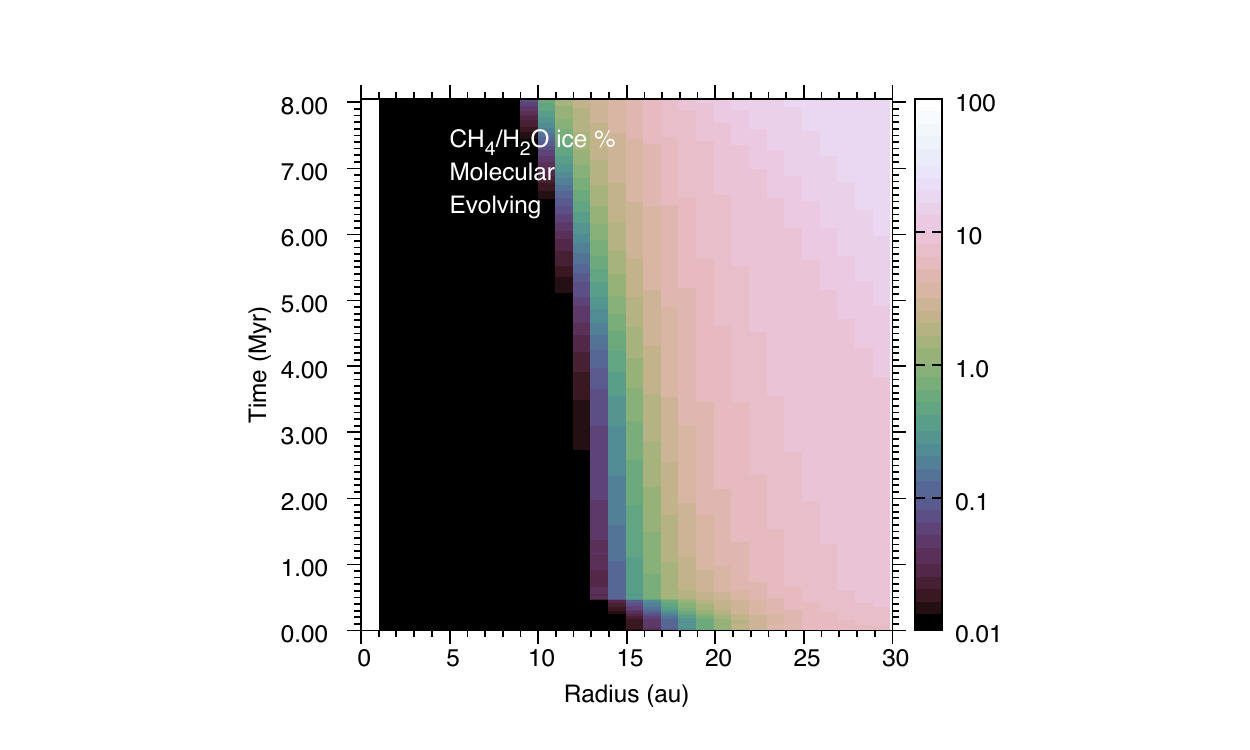}
\includegraphics[clip, trim = 2cm -0.5cm 2.25cm 0cm, width=0.33\textwidth]{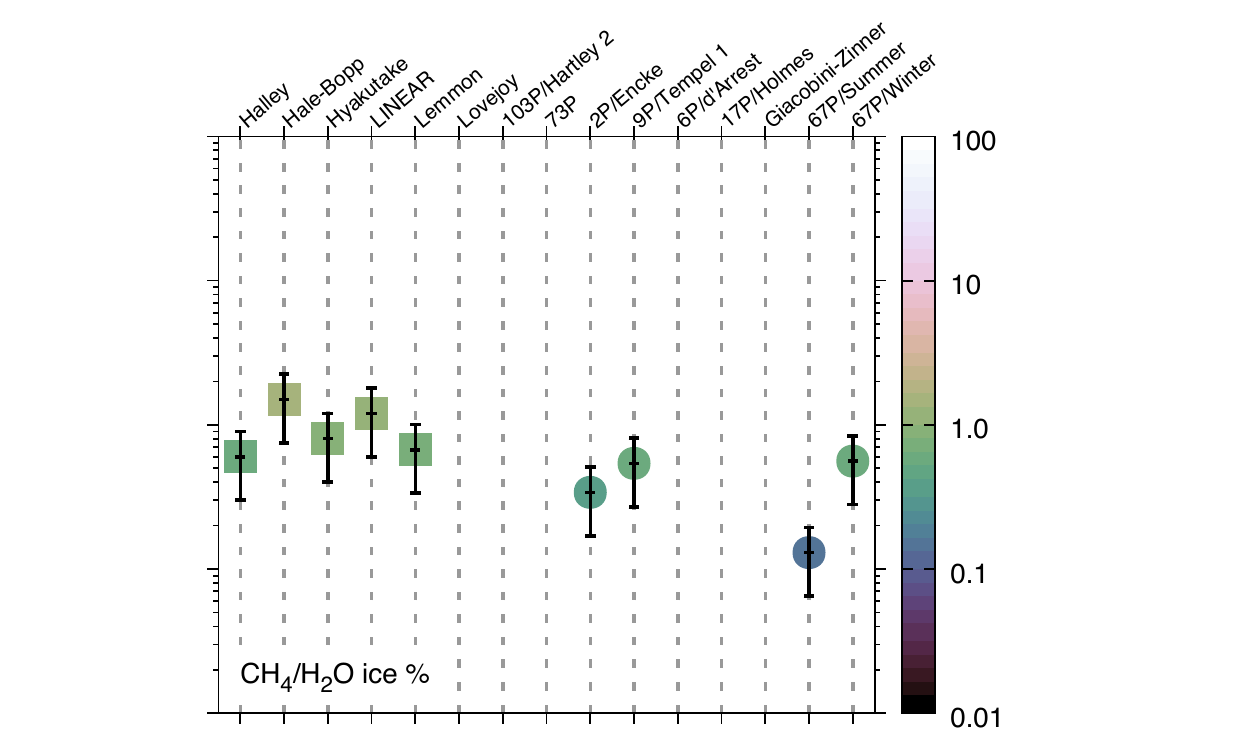}
\includegraphics[clip, trim = 2.5cm 0.4cm 2.5cm 0.75cm, width=0.33\textwidth]{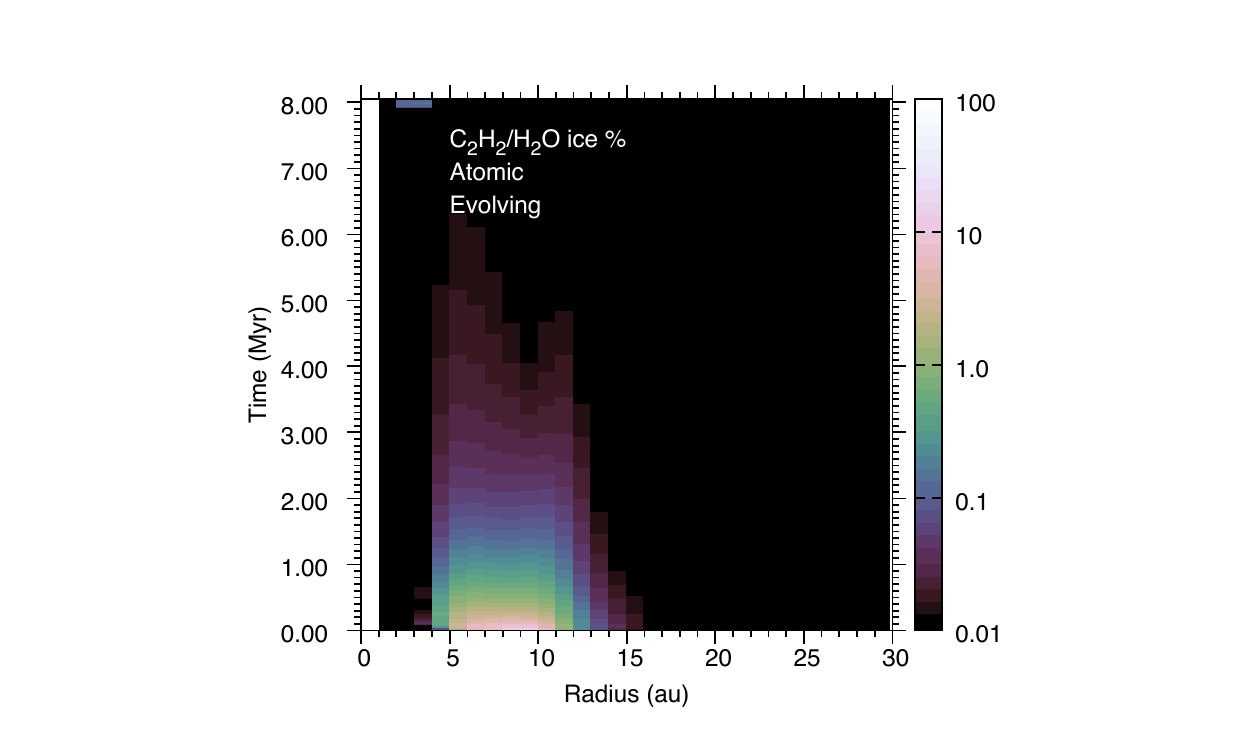}
\includegraphics[clip, trim = 2.5cm 0.4cm 2.5cm 0.75cm, width=0.33\textwidth]{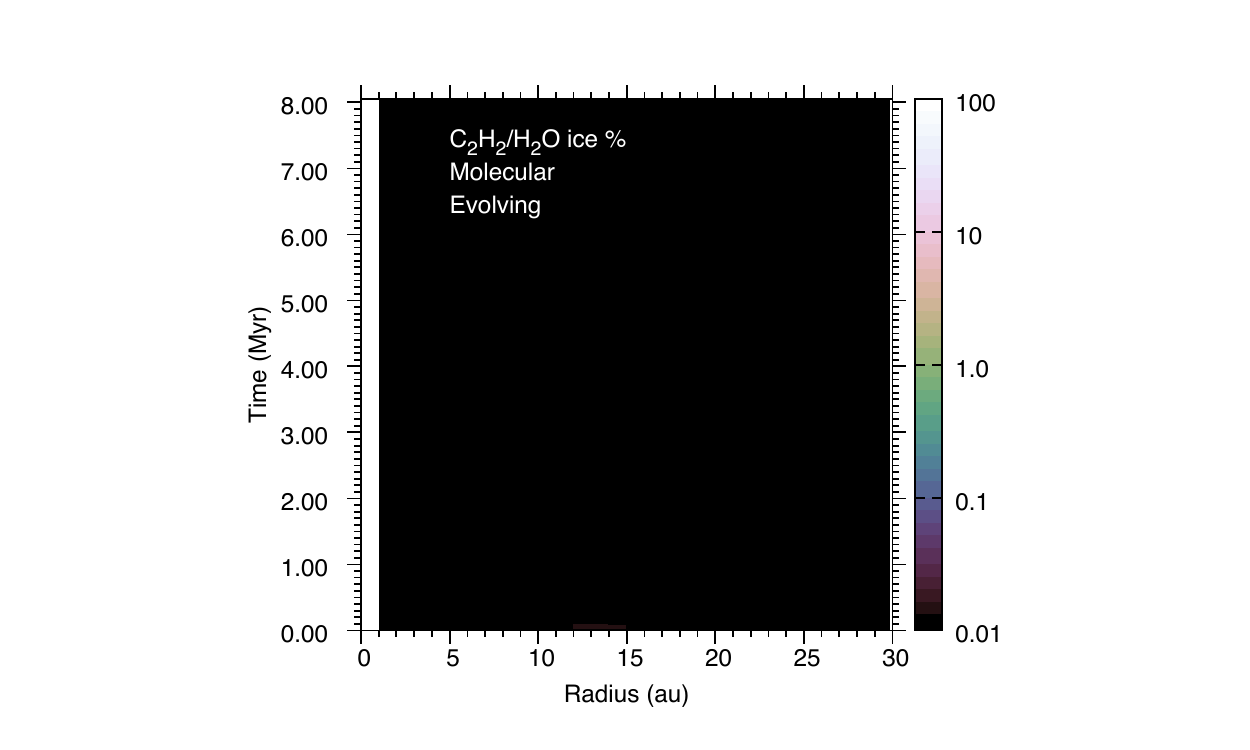}
\includegraphics[clip, trim = 2cm -0.5cm 2.25cm 0cm, width=0.33\textwidth]{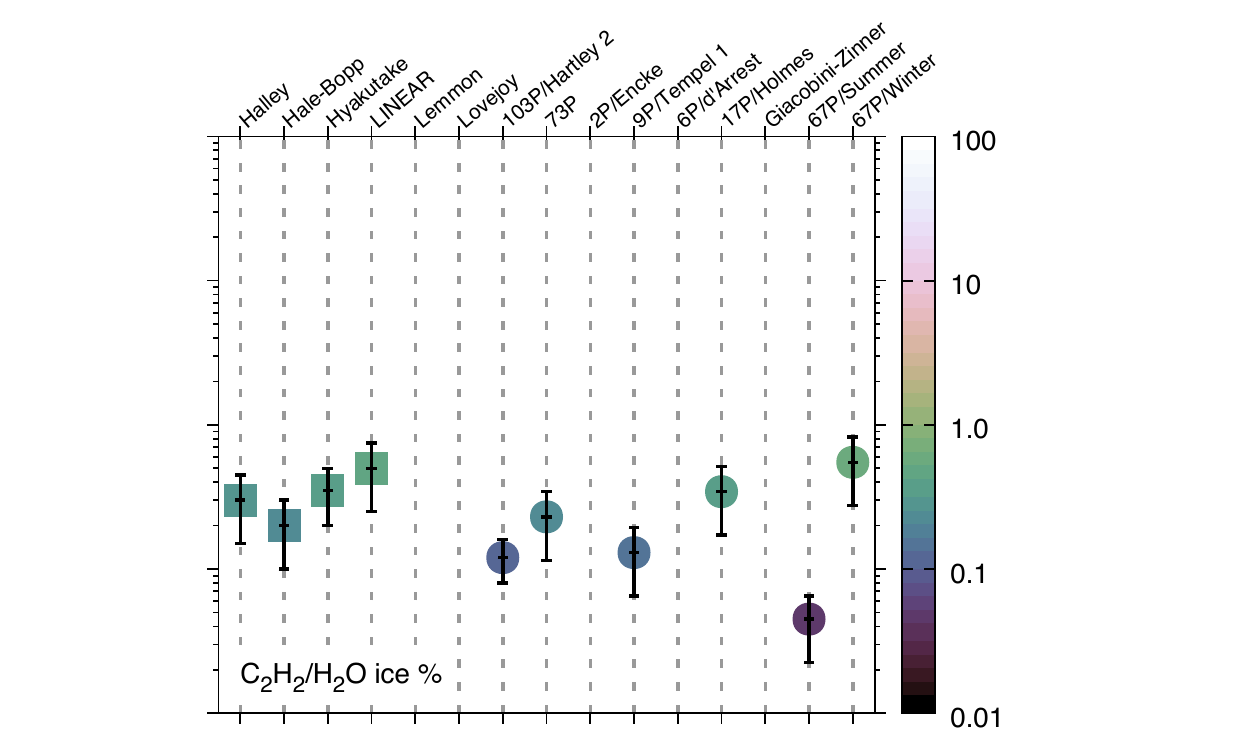}
\includegraphics[clip, trim = 2.5cm 0.4cm 2.5cm 0.75cm, width=0.33\textwidth]{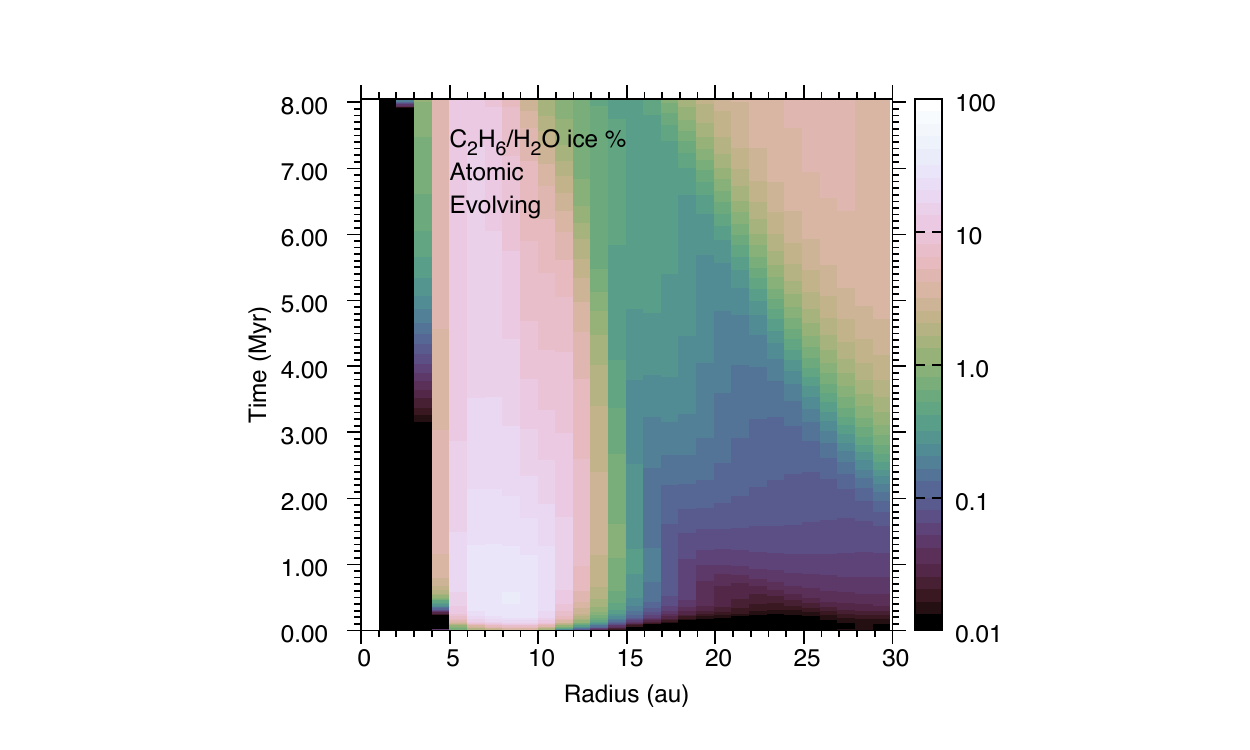}
\includegraphics[clip, trim = 2.5cm 0.4cm 2.5cm 0.75cm, width=0.33\textwidth]{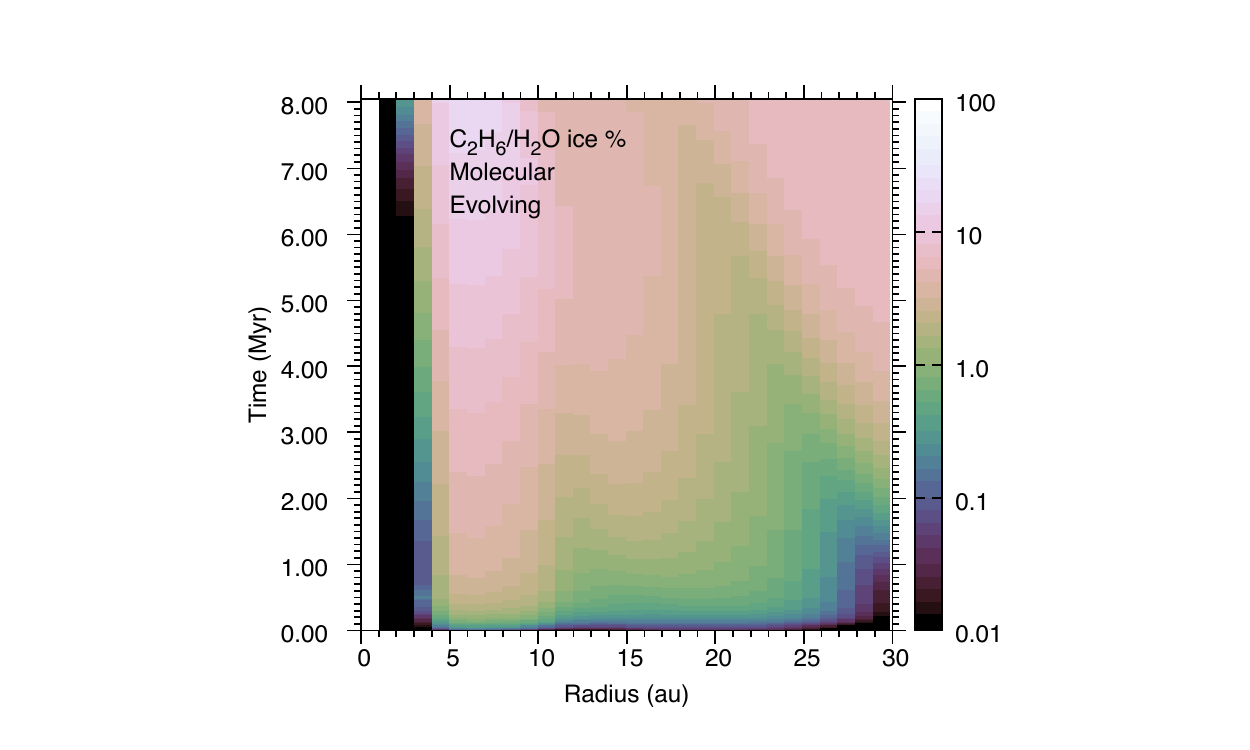}
\includegraphics[clip, trim = 2cm -0.5cm 2.25cm 0cm, width=0.33\textwidth]{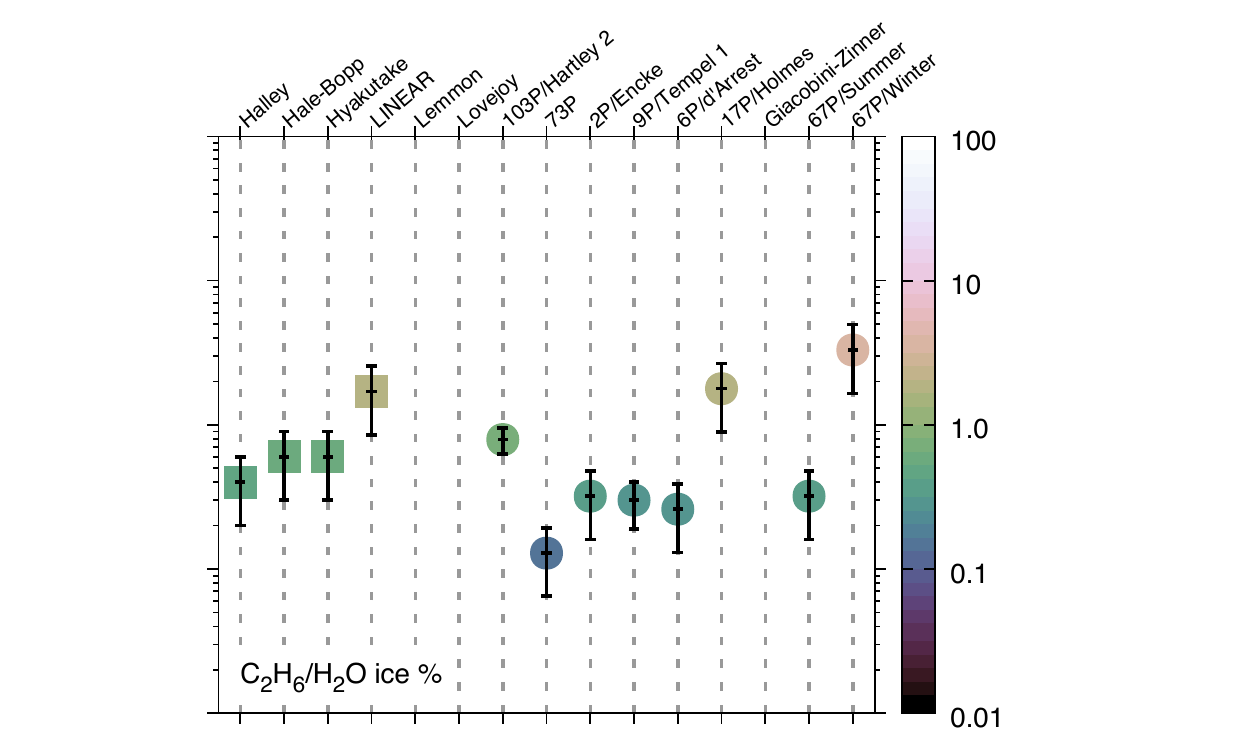}
\caption{Same as Figure~\ref{figure6} for \ce{CH4}, \ce{C2H2}, and \ce{C2H6} ice.}
\label{figure7}
\end{figure*}

\begin{figure*}[!t]
\includegraphics[clip, trim = 2.5cm 0.4cm 2.5cm 0.75cm, width=0.33\textwidth]{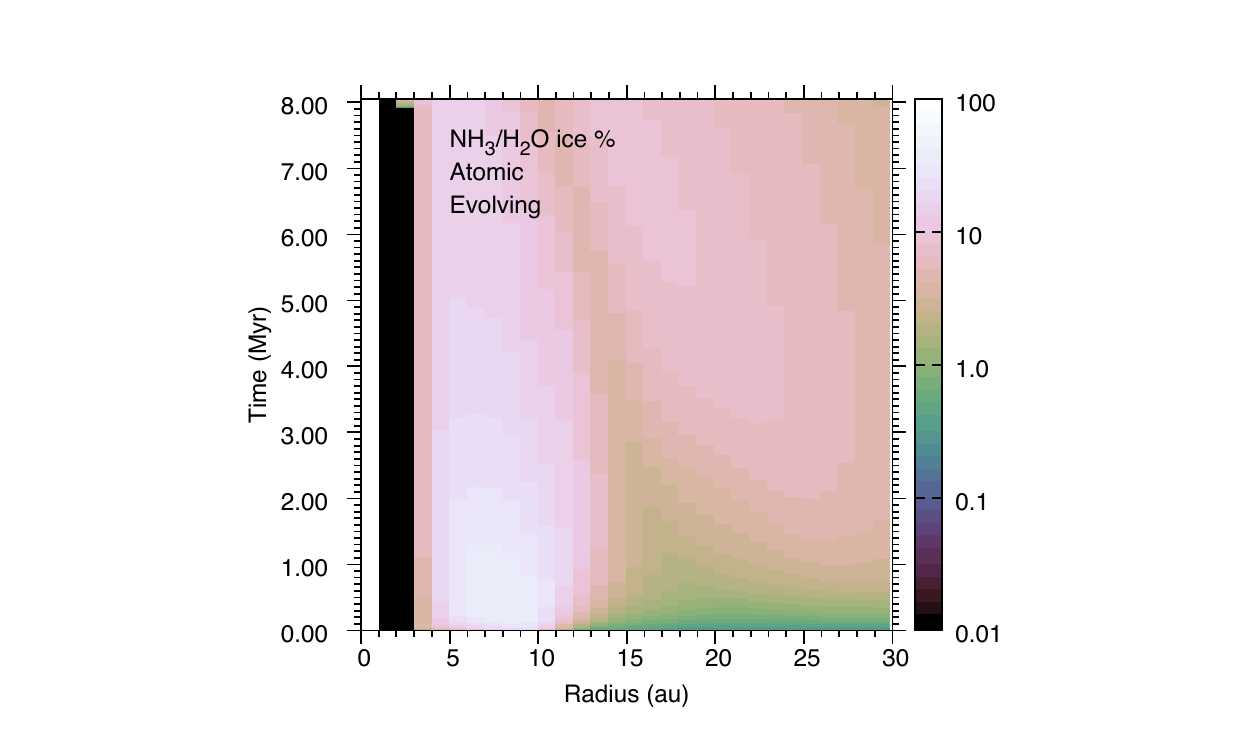}
\includegraphics[clip, trim = 2.5cm 0.4cm 2.5cm 0.75cm, width=0.33\textwidth]{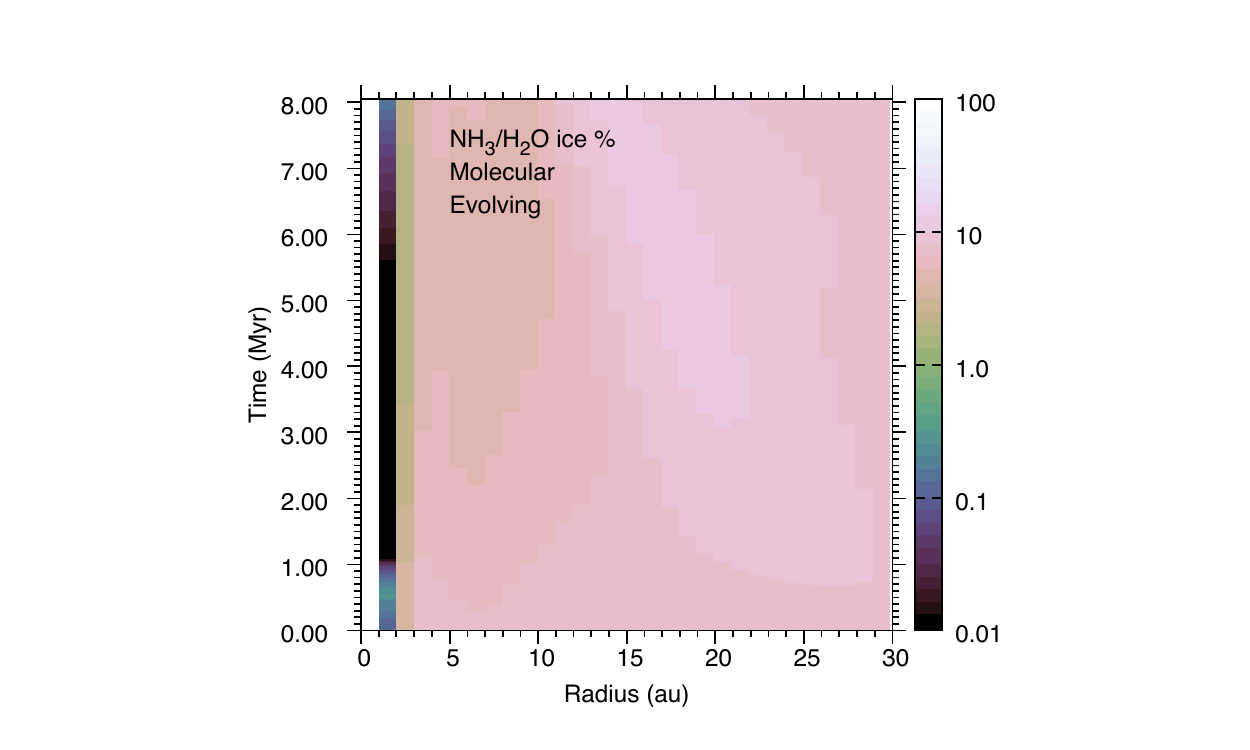}
\includegraphics[clip, trim = 2cm -0.5cm 2.25cm 0cm, width=0.33\textwidth]{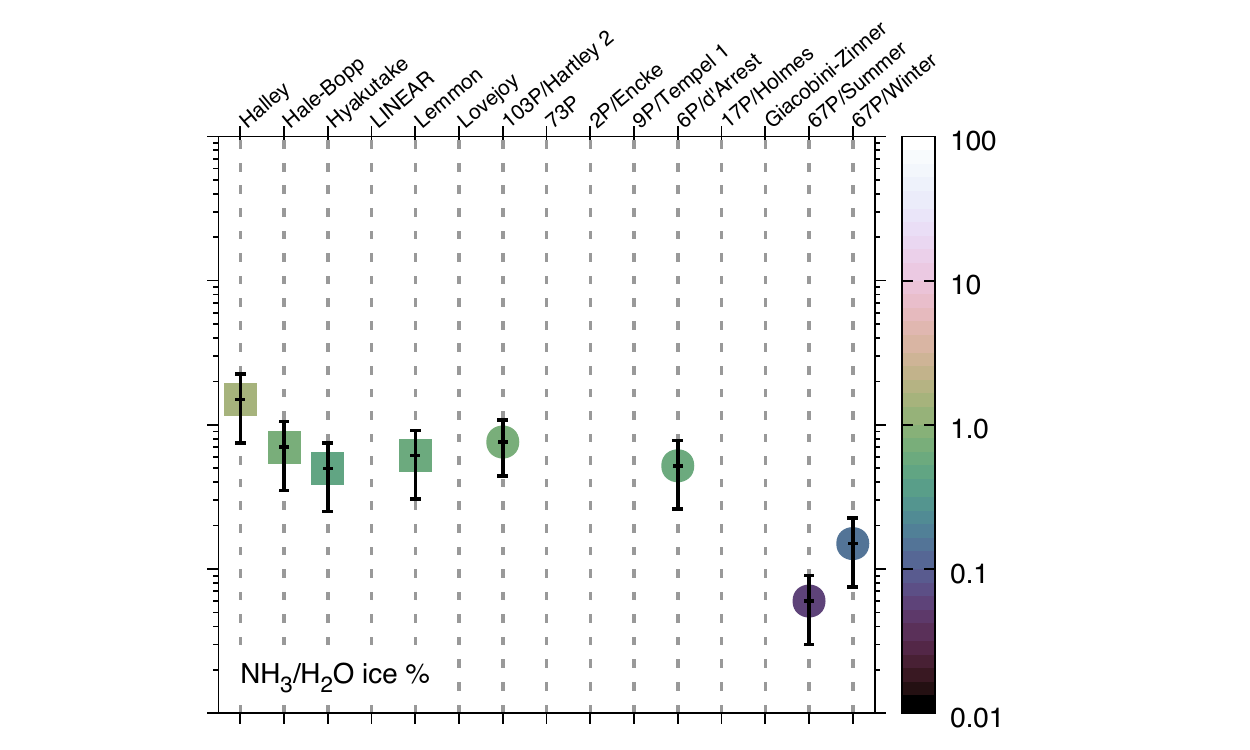}
\includegraphics[clip, trim = 2.5cm 0.4cm 2.5cm 0.75cm, width=0.33\textwidth]{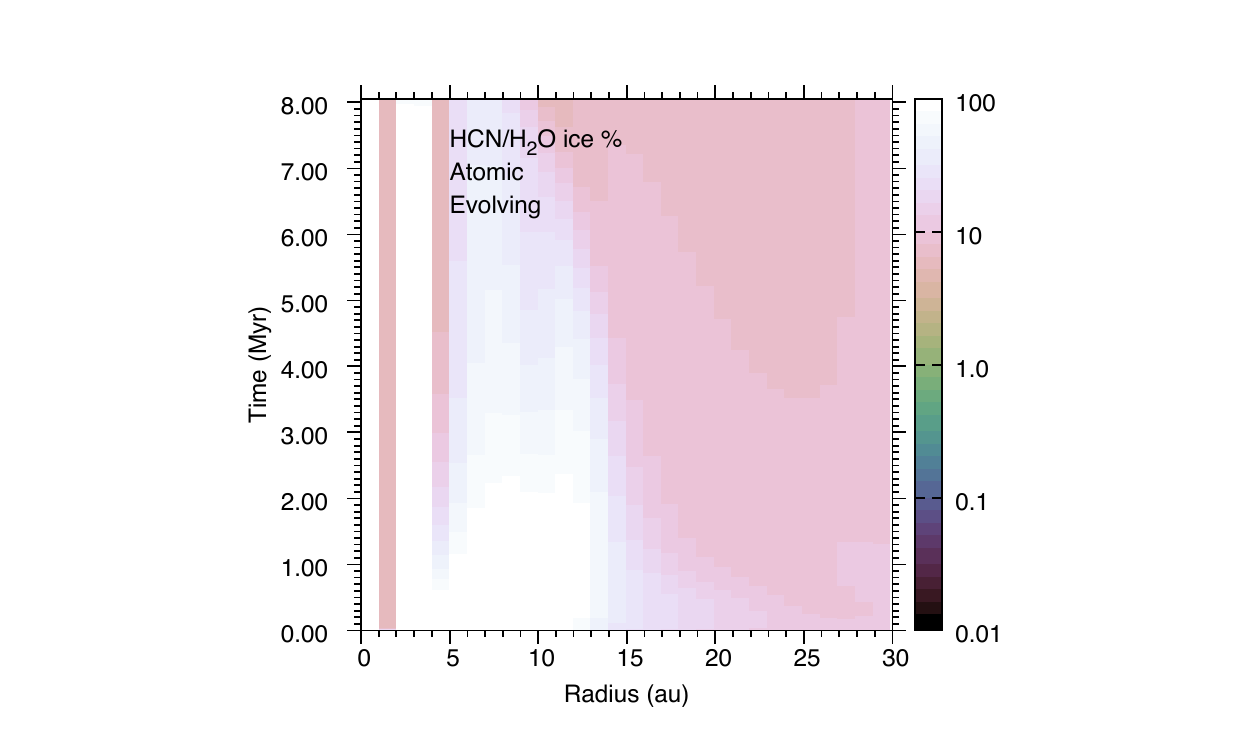}
\includegraphics[clip, trim = 2.5cm 0.4cm 2.5cm 0.75cm, width=0.33\textwidth]{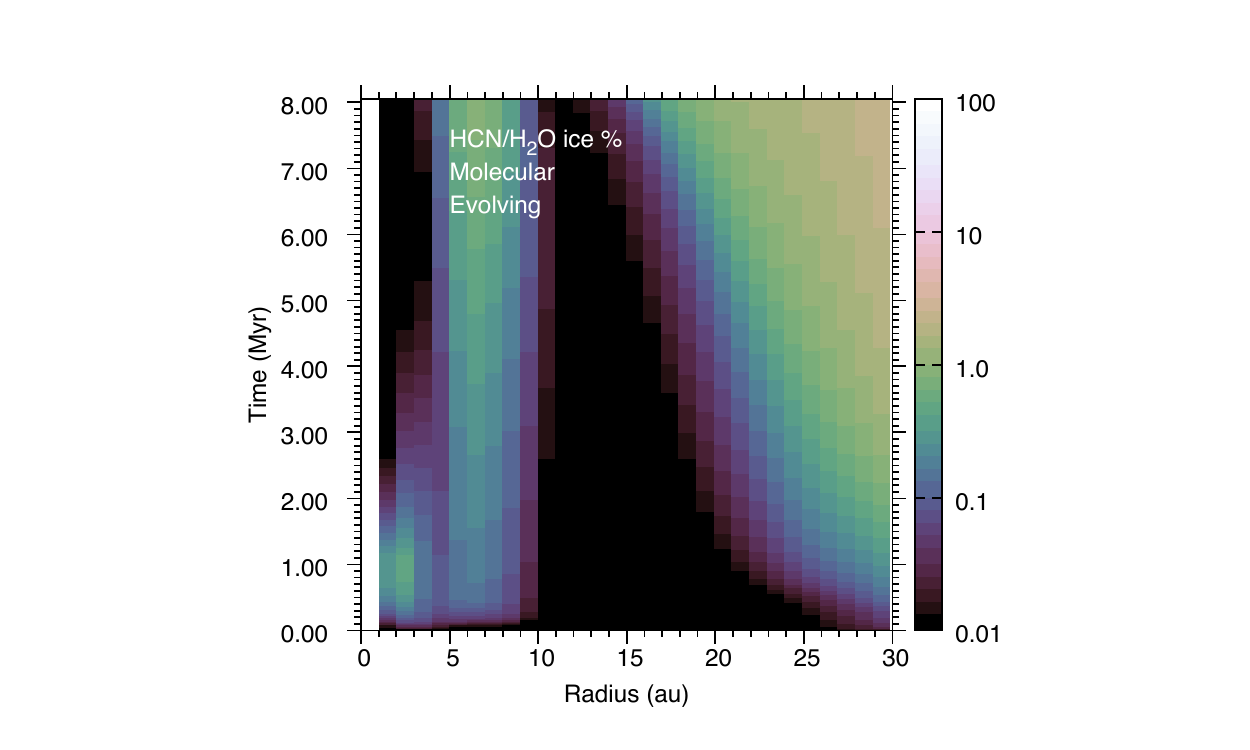}
\includegraphics[clip, trim = 2cm -0.5cm 2.25cm 0cm, width=0.33\textwidth]{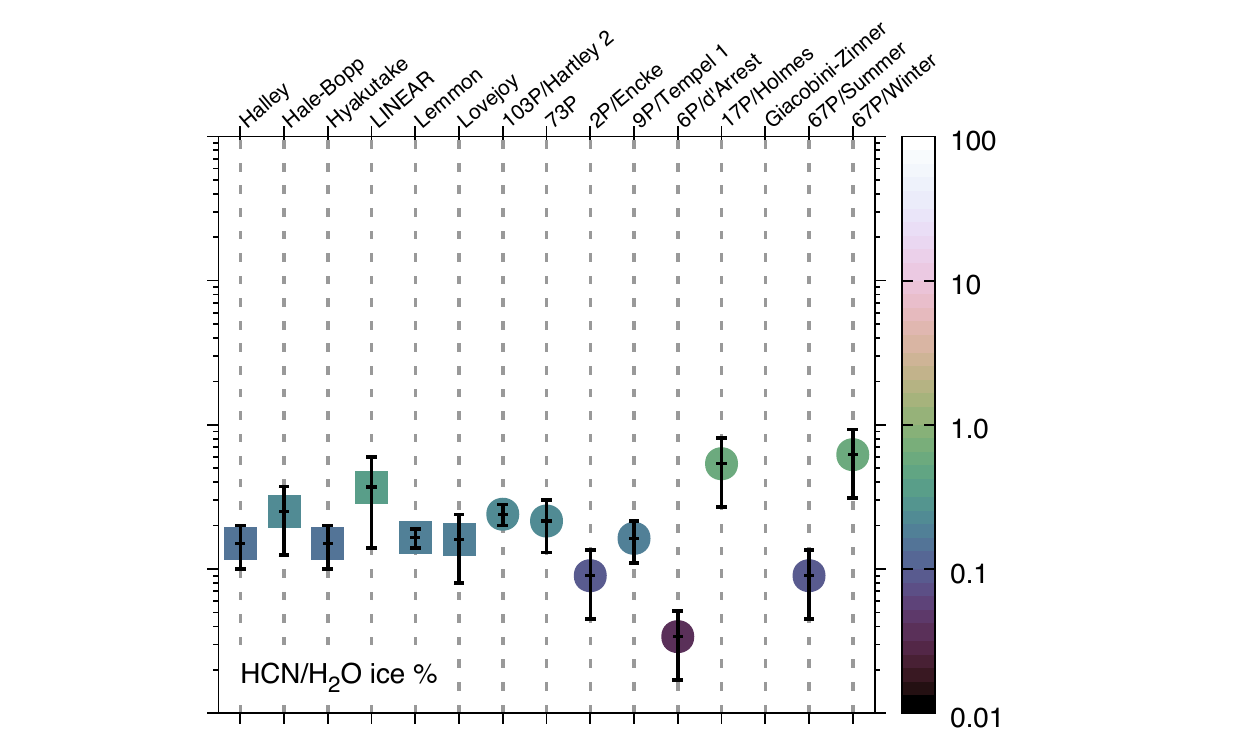}
\includegraphics[clip, trim = 2.5cm 0.4cm 2.5cm 0.75cm, width=0.33\textwidth]{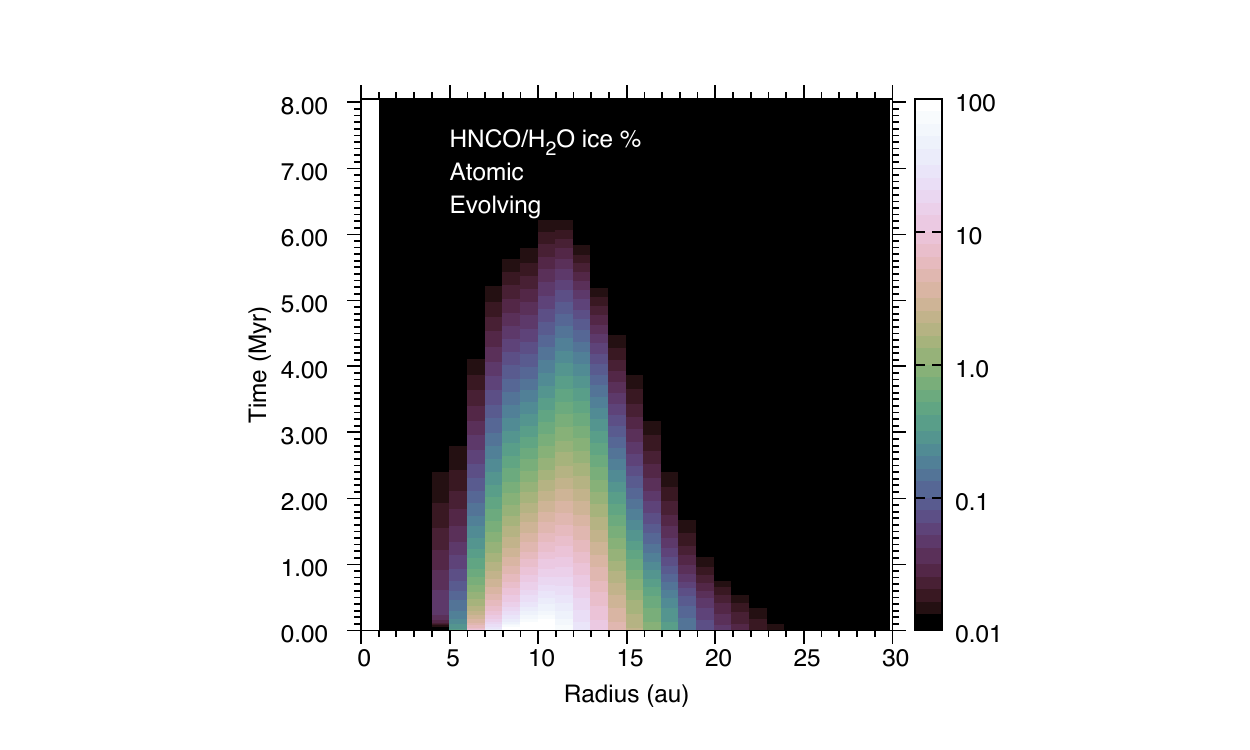}
\includegraphics[clip, trim = 2.5cm 0.4cm 2.5cm 0.75cm, width=0.33\textwidth]{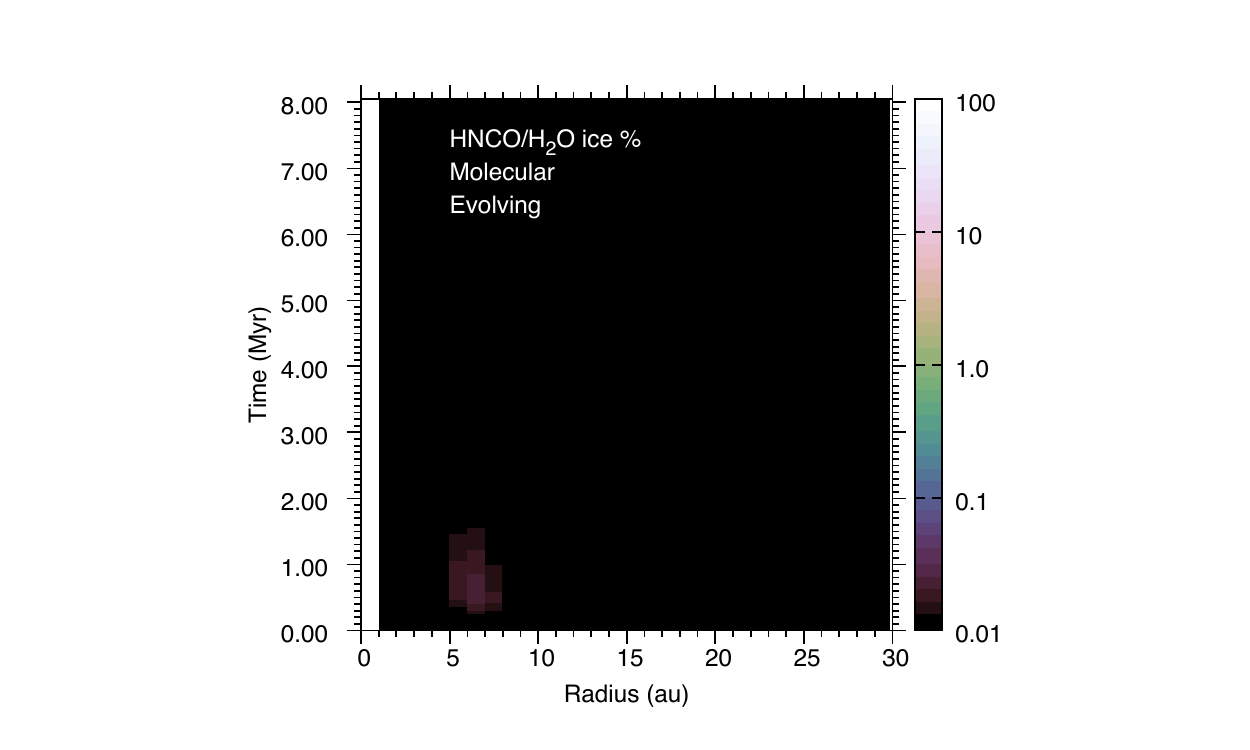}
\includegraphics[clip, trim = 2cm -0.5cm 2.25cm 0cm, width=0.33\textwidth]{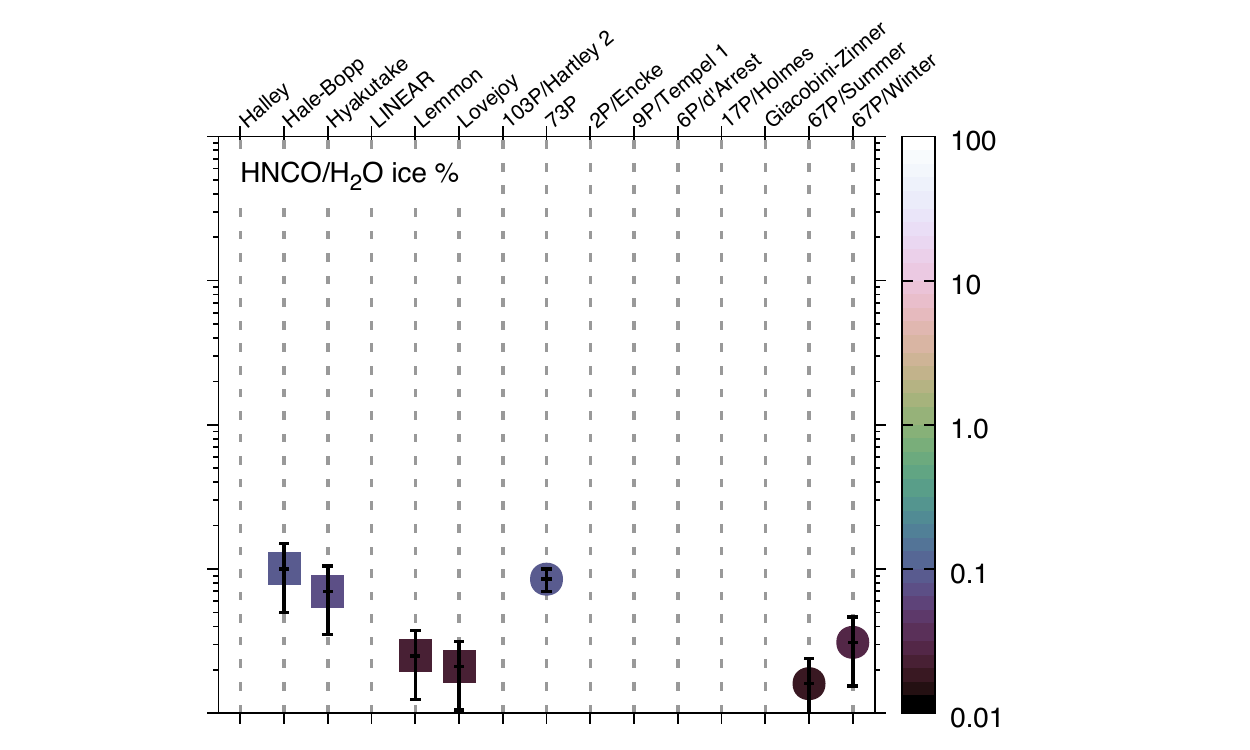}
\includegraphics[clip, trim = 2.5cm 0.4cm 2.5cm 0.75cm, width=0.33\textwidth]{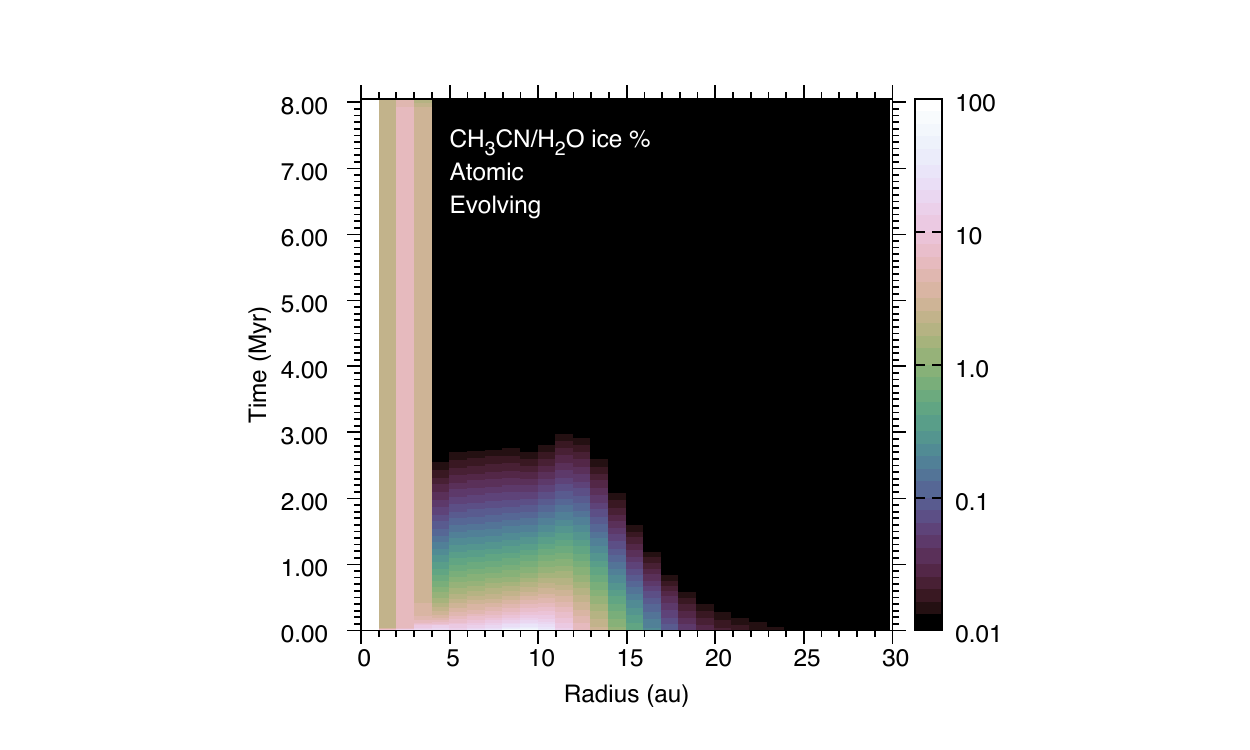}
\includegraphics[clip, trim = 2.5cm 0.4cm 2.5cm 0.75cm, width=0.33\textwidth]{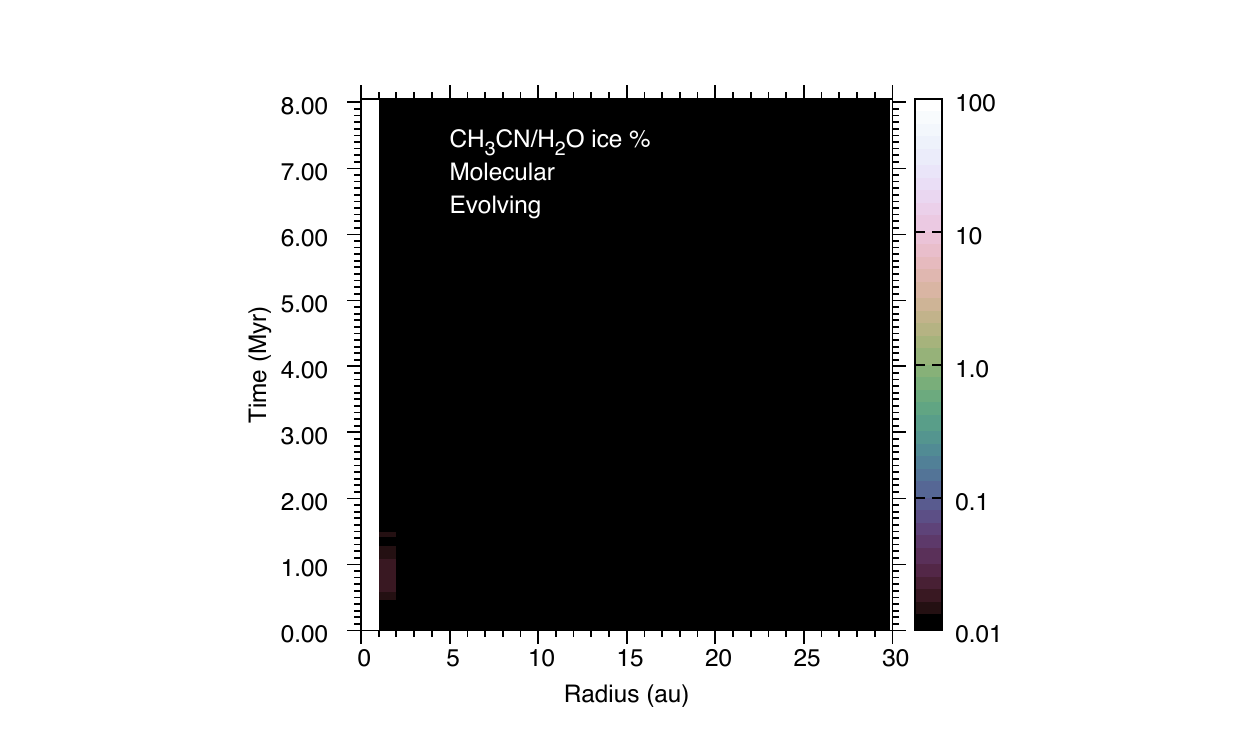}
\includegraphics[clip, trim = 2cm -0.5cm 2.25cm 0cm, width=0.33\textwidth]{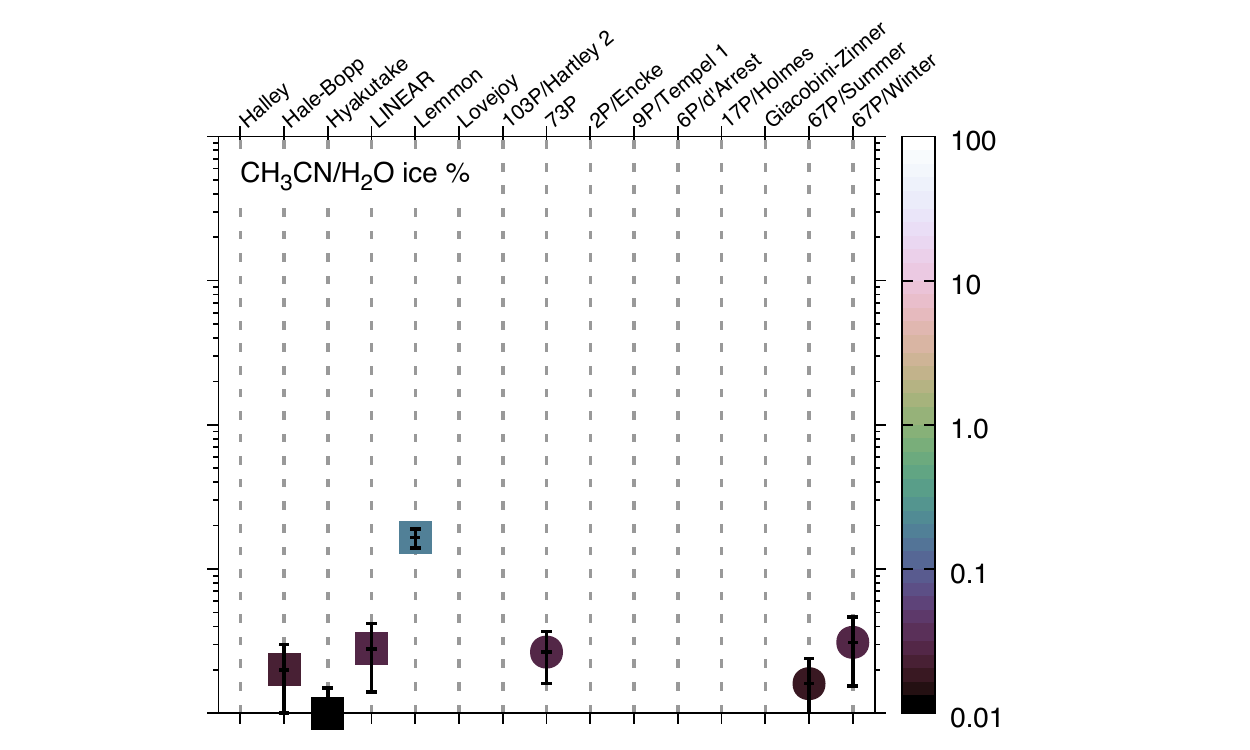}
\caption{Same as Figure~\ref{figure6} for \ce{NH3}, \ce{HCN}, \ce{HNCO} and \ce{CH3CN} ice.}
\label{figure8}
\end{figure*}

\begin{figure*}[!t]
\includegraphics[clip, trim = 2.5cm 0.4cm 2.5cm 0.75cm, width=0.33\textwidth]{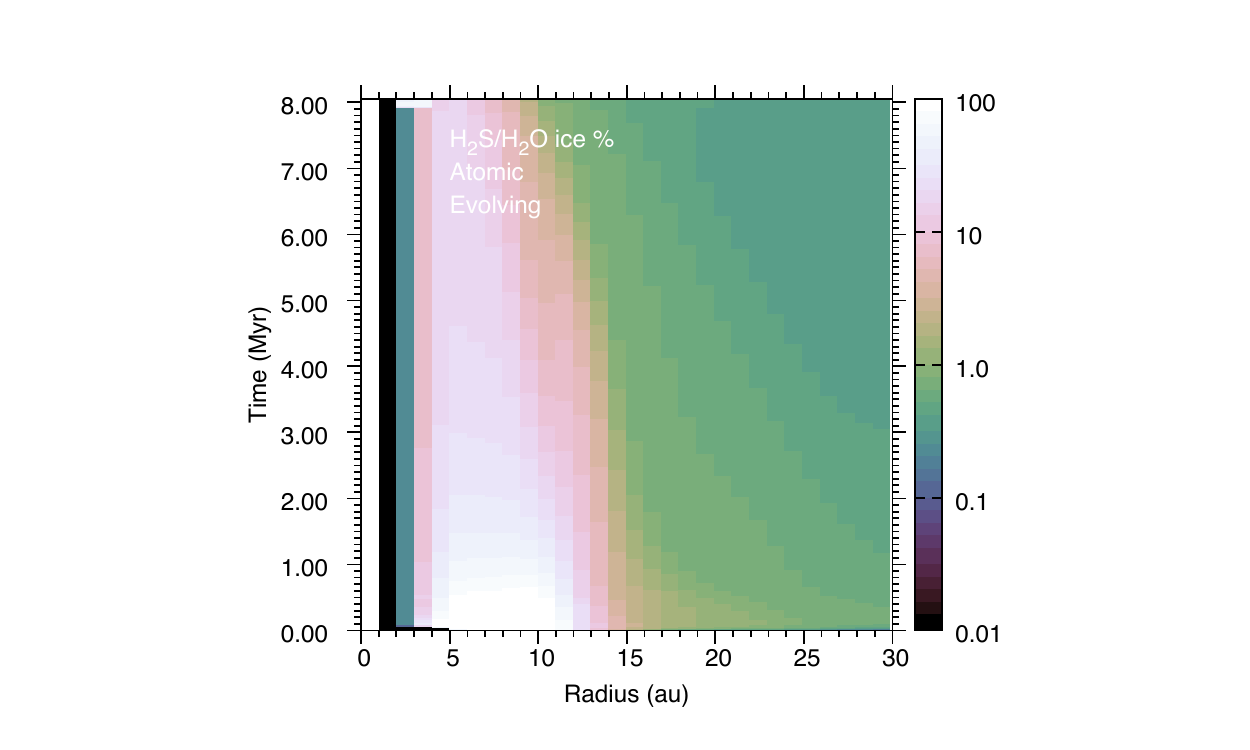}
\includegraphics[clip, trim = 2.5cm 0.4cm 2.5cm 0.75cm, width=0.33\textwidth]{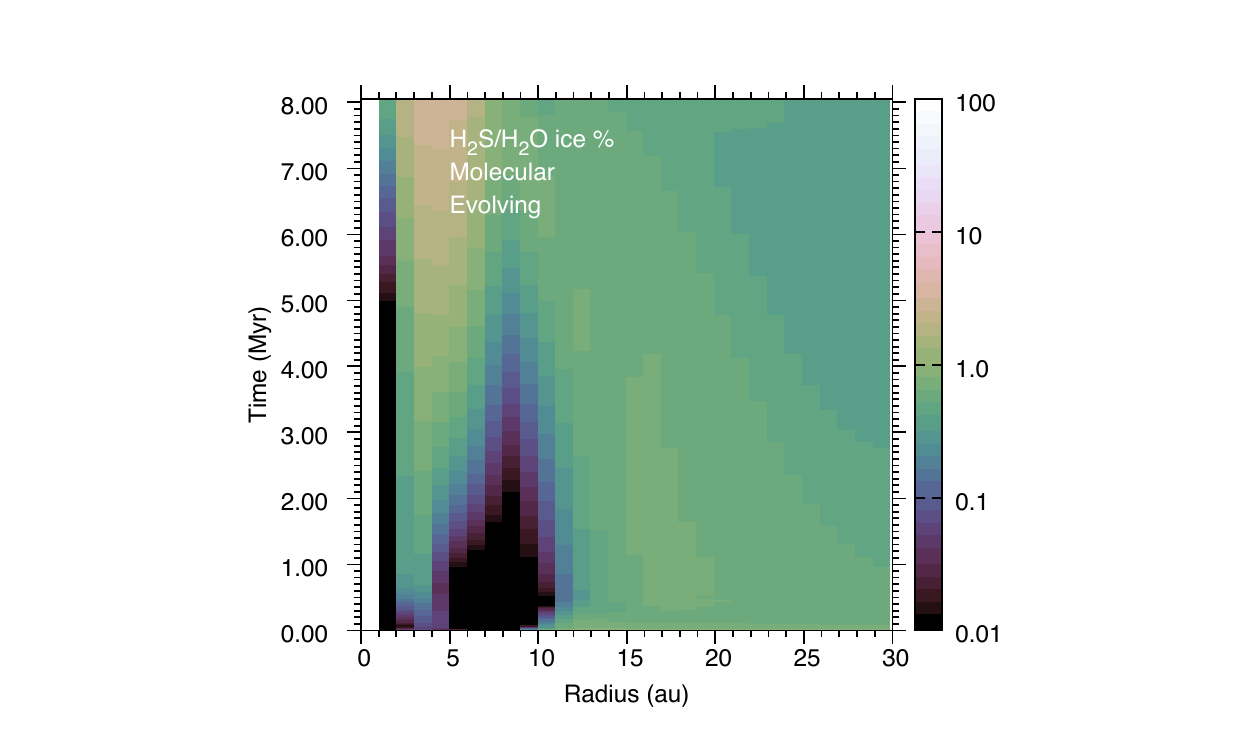}
\includegraphics[clip, trim = 2cm -0.5cm 2.25cm 0cm, width=0.33\textwidth]{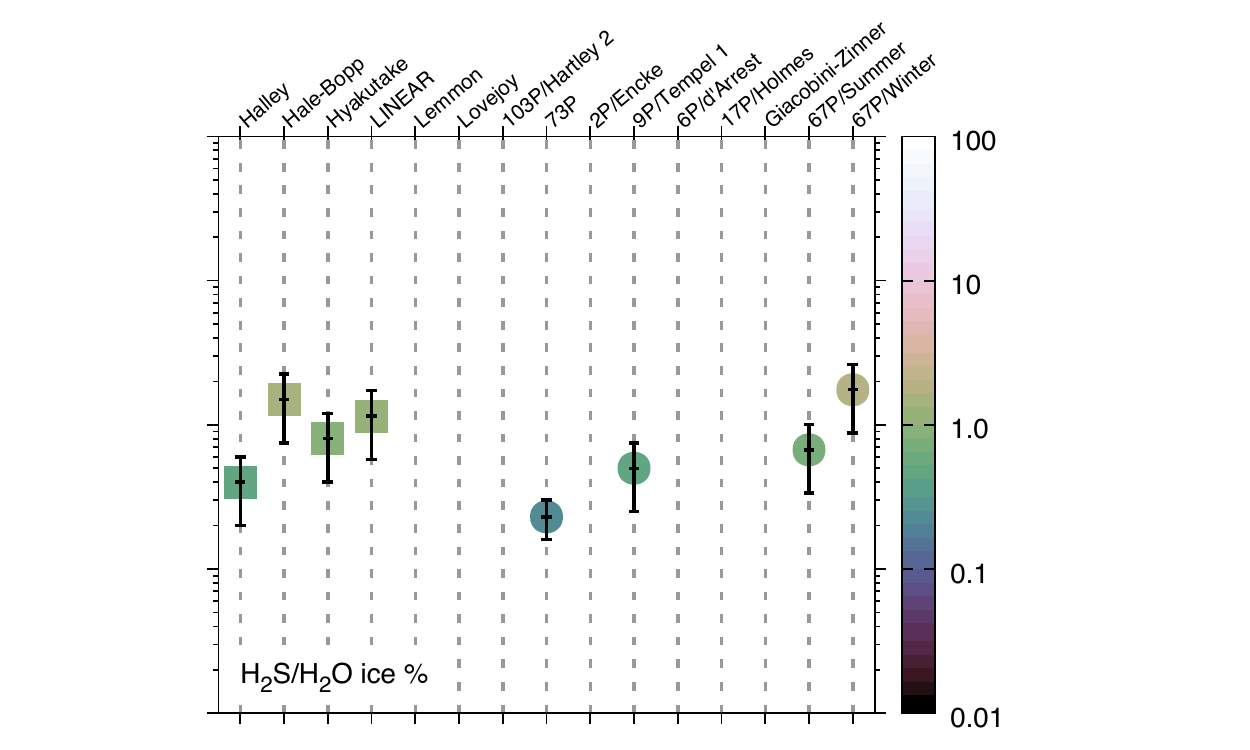}
\includegraphics[clip, trim = 2.5cm 0.4cm 2.5cm 0.75cm, width=0.33\textwidth]{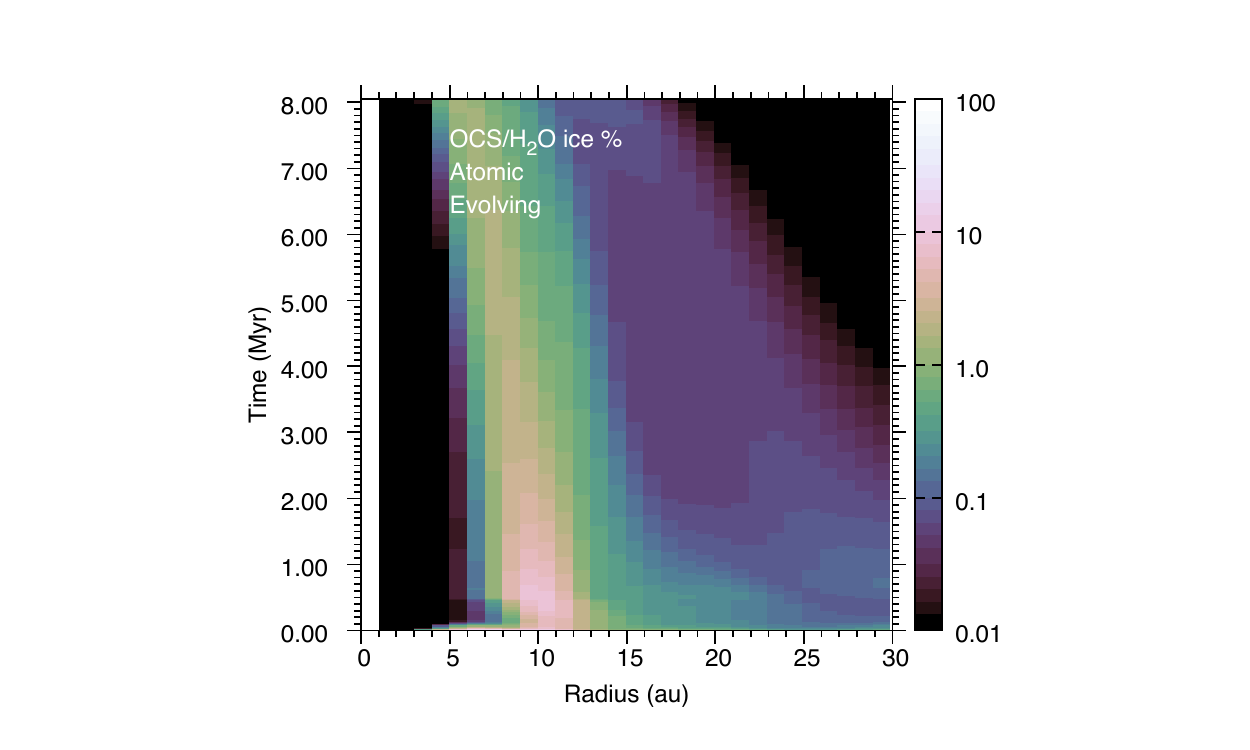}
\includegraphics[clip, trim = 2.5cm 0.4cm 2.5cm 0.75cm, width=0.33\textwidth]{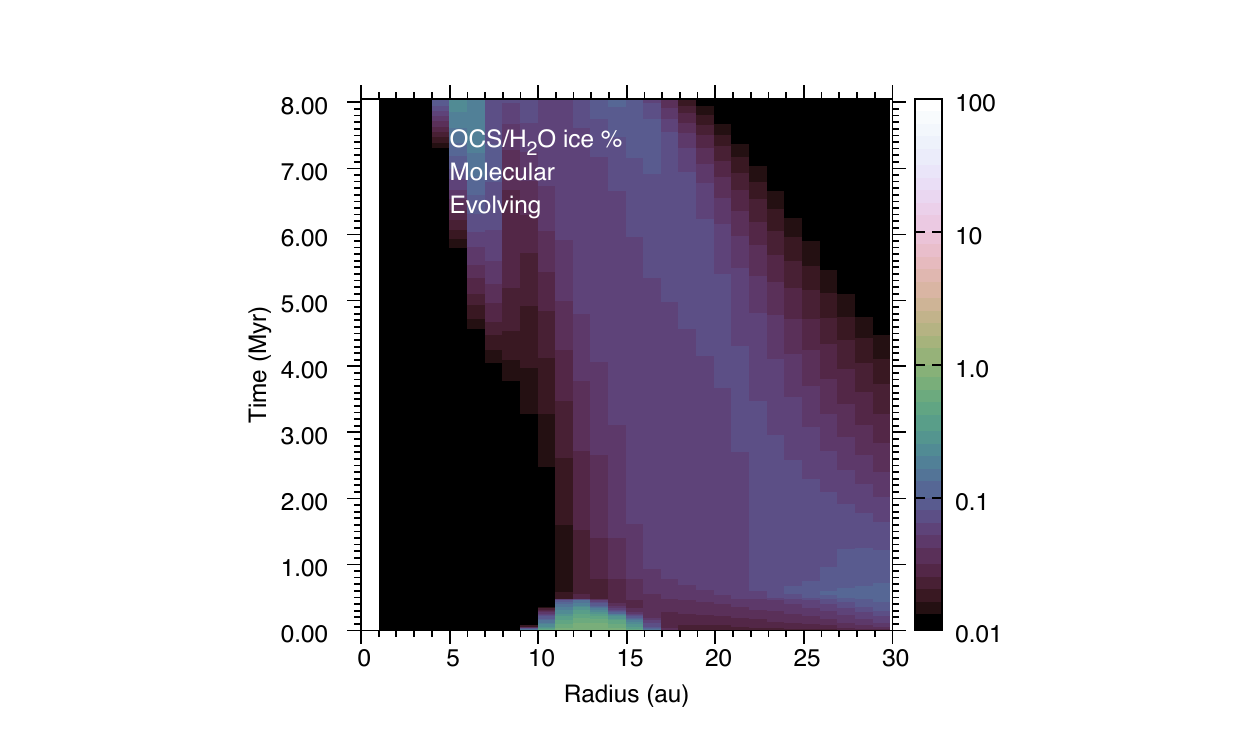}
\includegraphics[clip, trim = 2cm -0.5cm 2.25cm 0cm, width=0.33\textwidth]{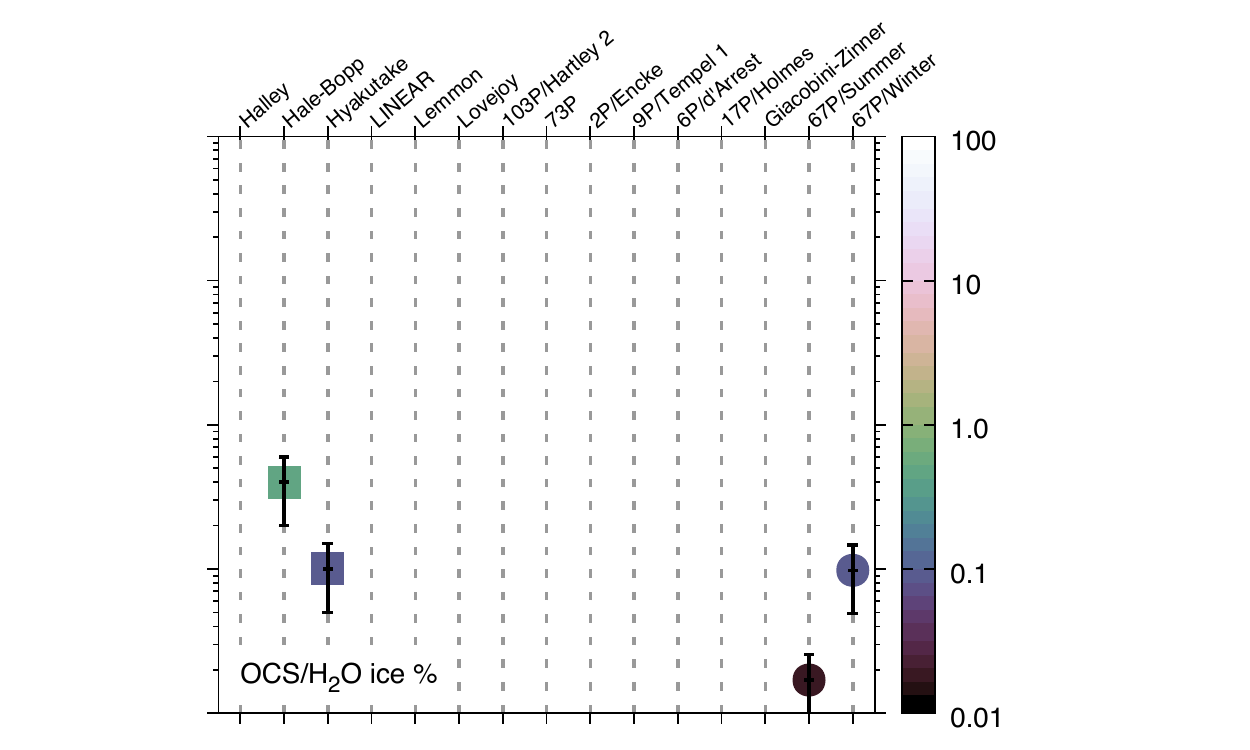}
\includegraphics[clip, trim = 2.5cm 0.4cm 2.5cm 0.75cm, width=0.33\textwidth]{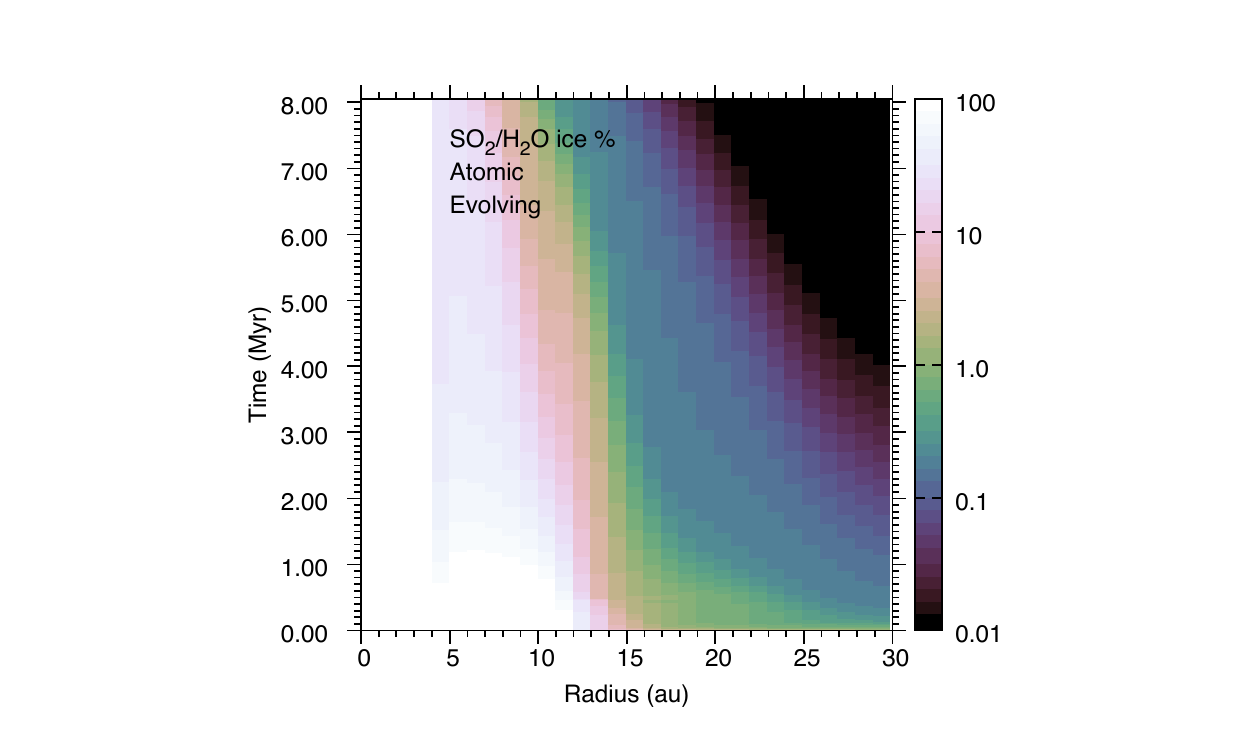}
\includegraphics[clip, trim = 2.5cm 0.4cm 2.5cm 0.75cm, width=0.33\textwidth]{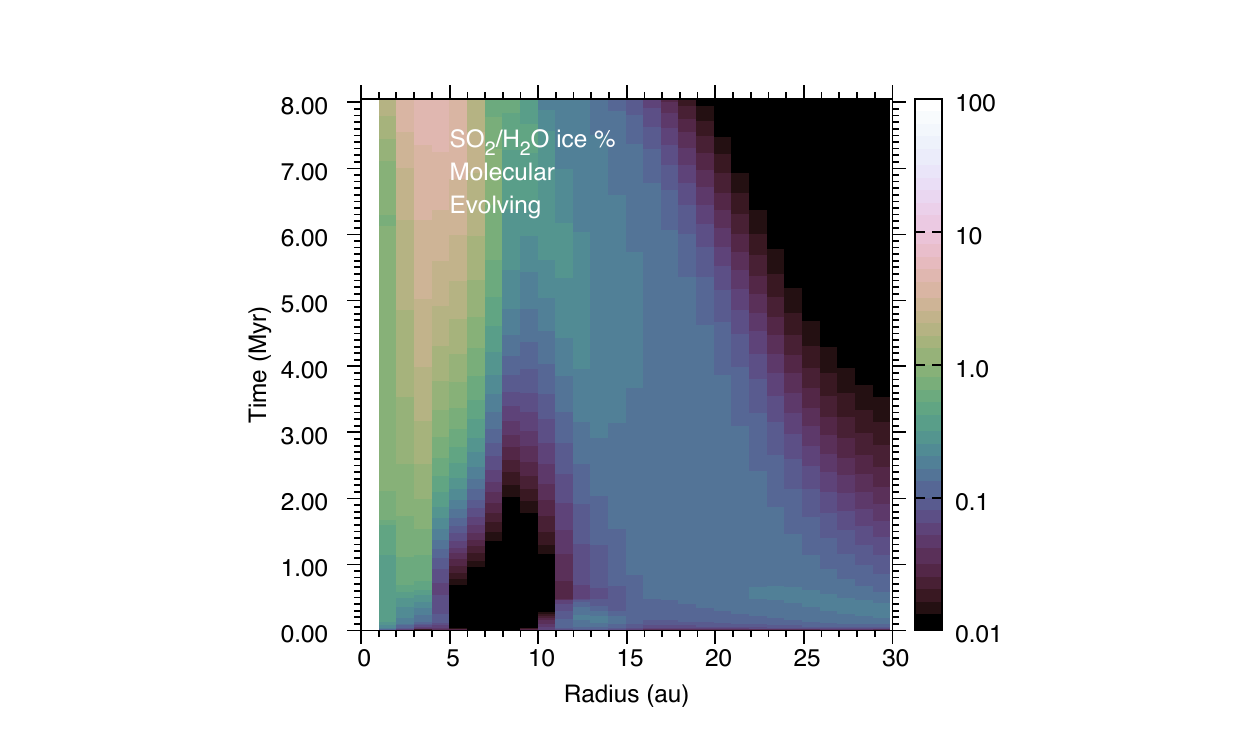}
\includegraphics[clip, trim = 2cm -0.5cm 2.25cm 0cm, width=0.33\textwidth]{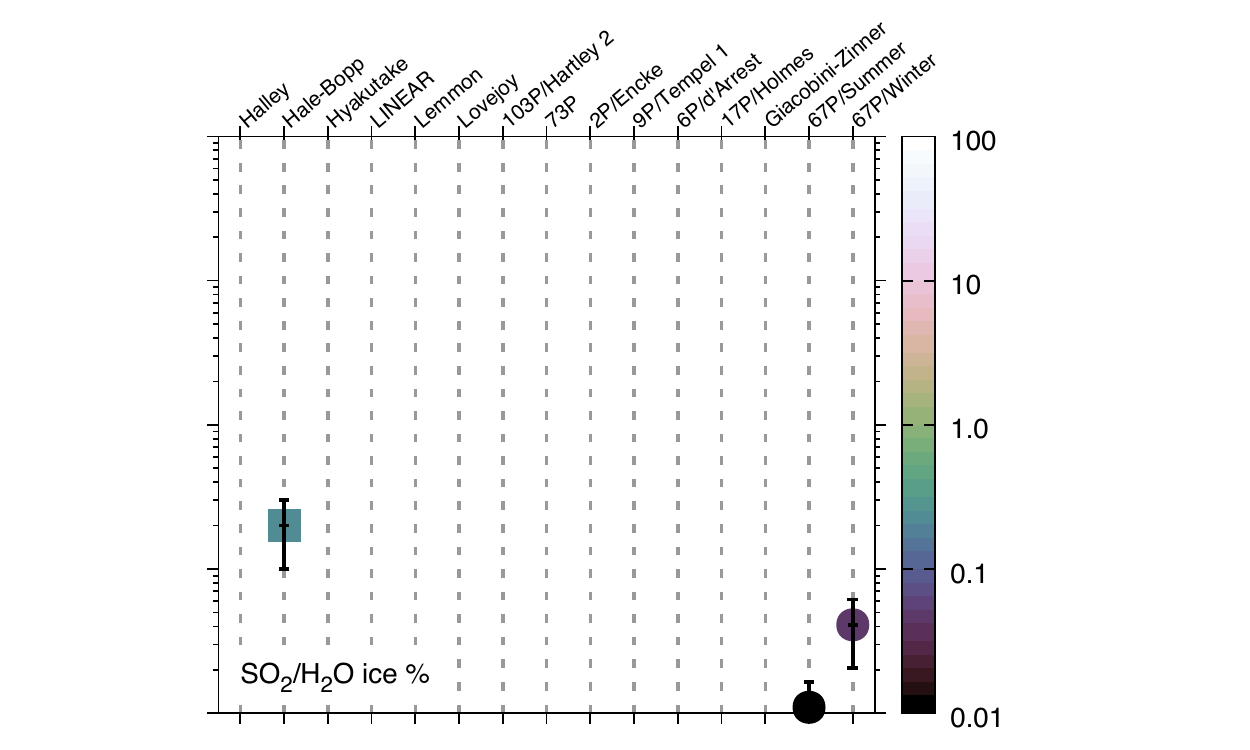}
\caption{Same as Figure~\ref{figure6} for \ce{H2S}, \ce{OCS}, and \ce{SO2} ice.}
\label{figure9}
\end{figure*}

\begin{figure*}[!t]
\includegraphics[clip, trim = 2.5cm 0.4cm 2.5cm 0.75cm, width=0.33\textwidth]{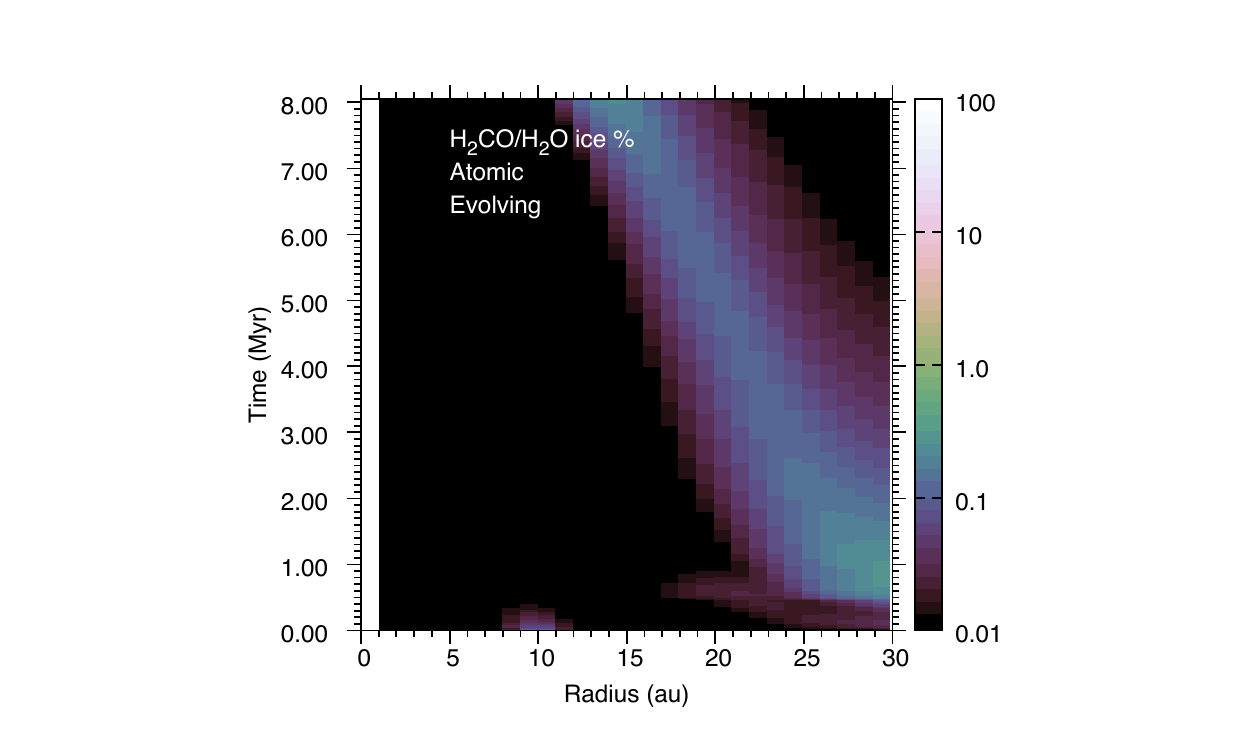}
\includegraphics[clip, trim = 2.5cm 0.4cm 2.5cm 0.75cm, width=0.33\textwidth]{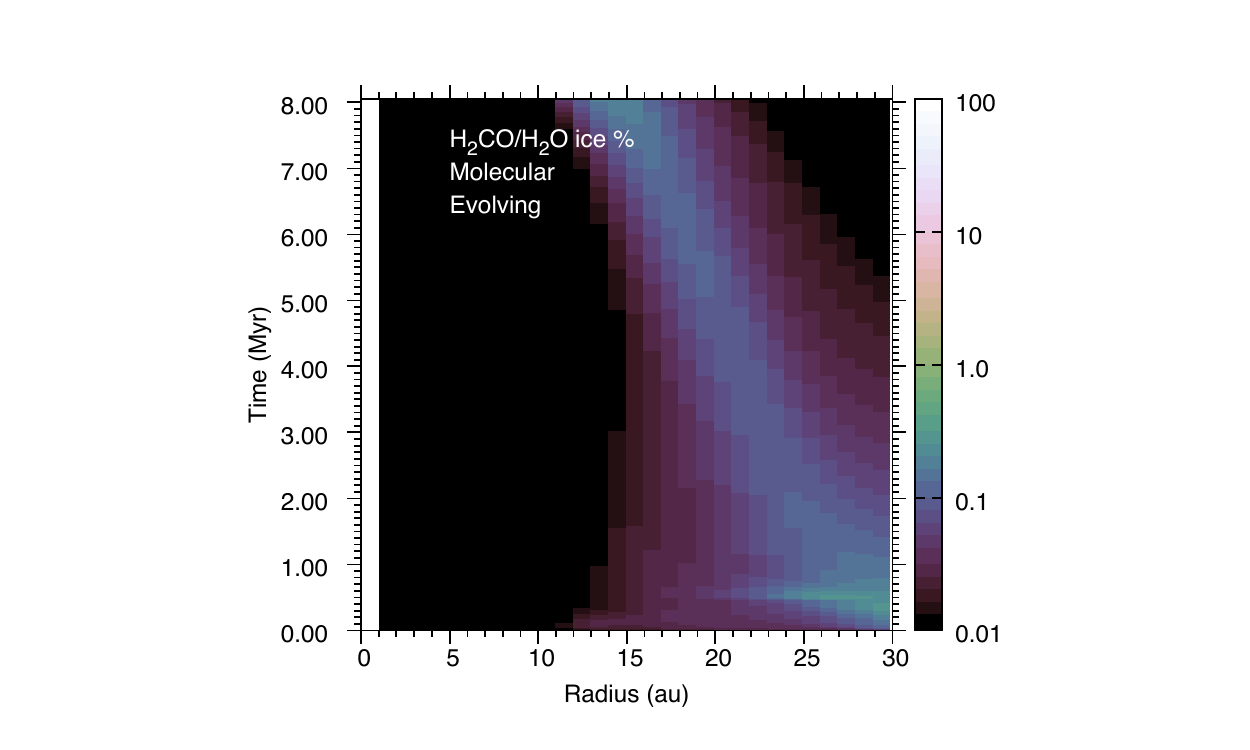}
\includegraphics[clip, trim = 2cm -0.5cm 2.25cm 0cm, width=0.33\textwidth]{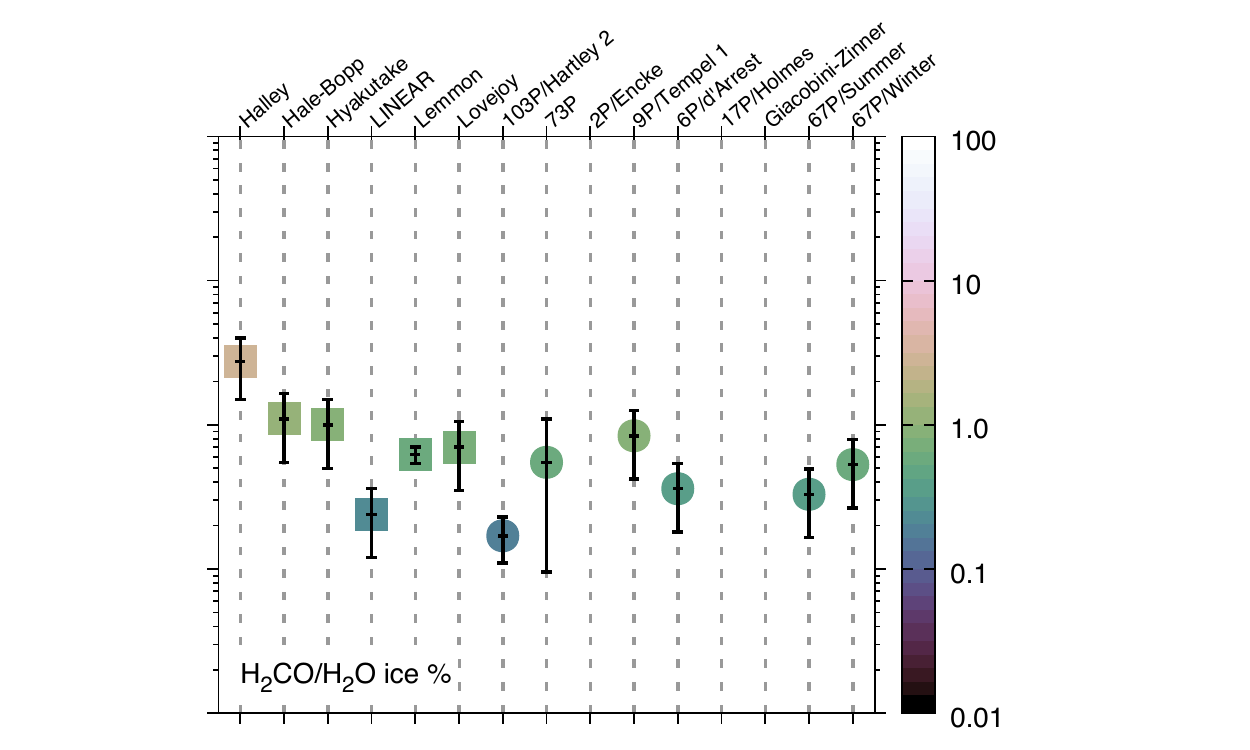}
\includegraphics[clip, trim = 2.5cm 0.4cm 2.5cm 0.75cm, width=0.33\textwidth]{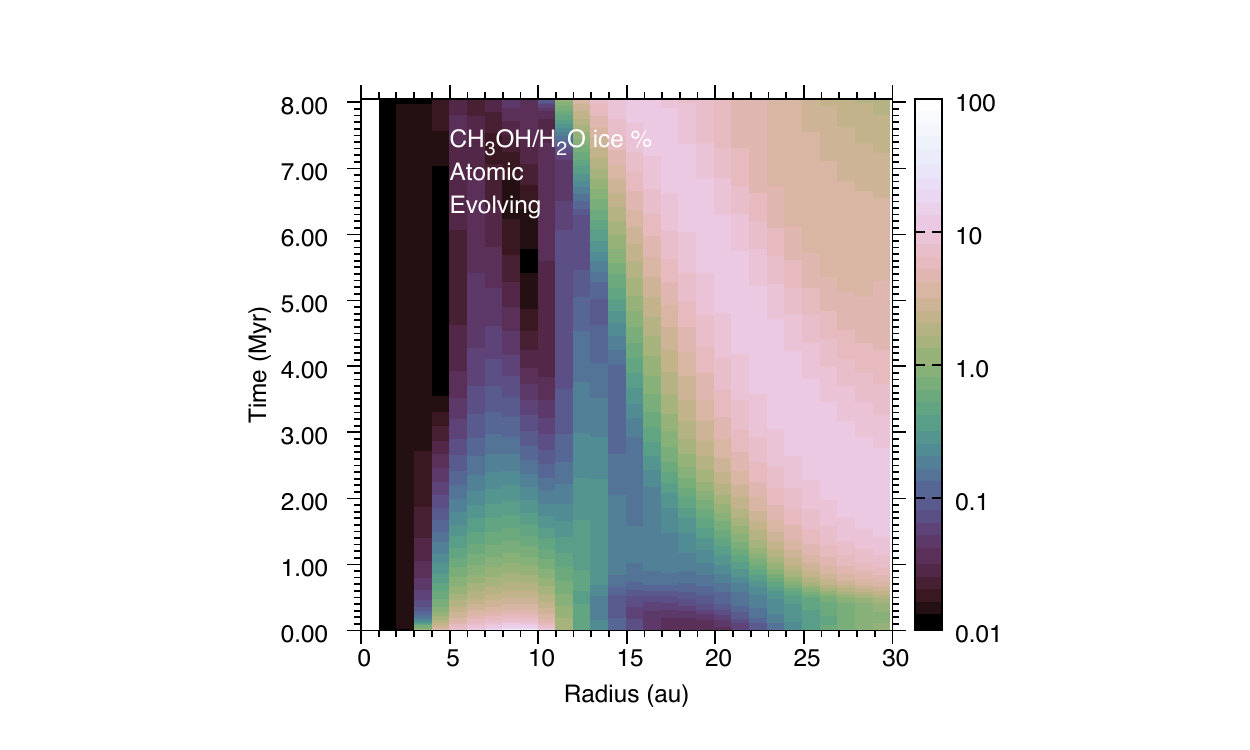}
\includegraphics[clip, trim = 2.5cm 0.4cm 2.5cm 0.75cm, width=0.33\textwidth]{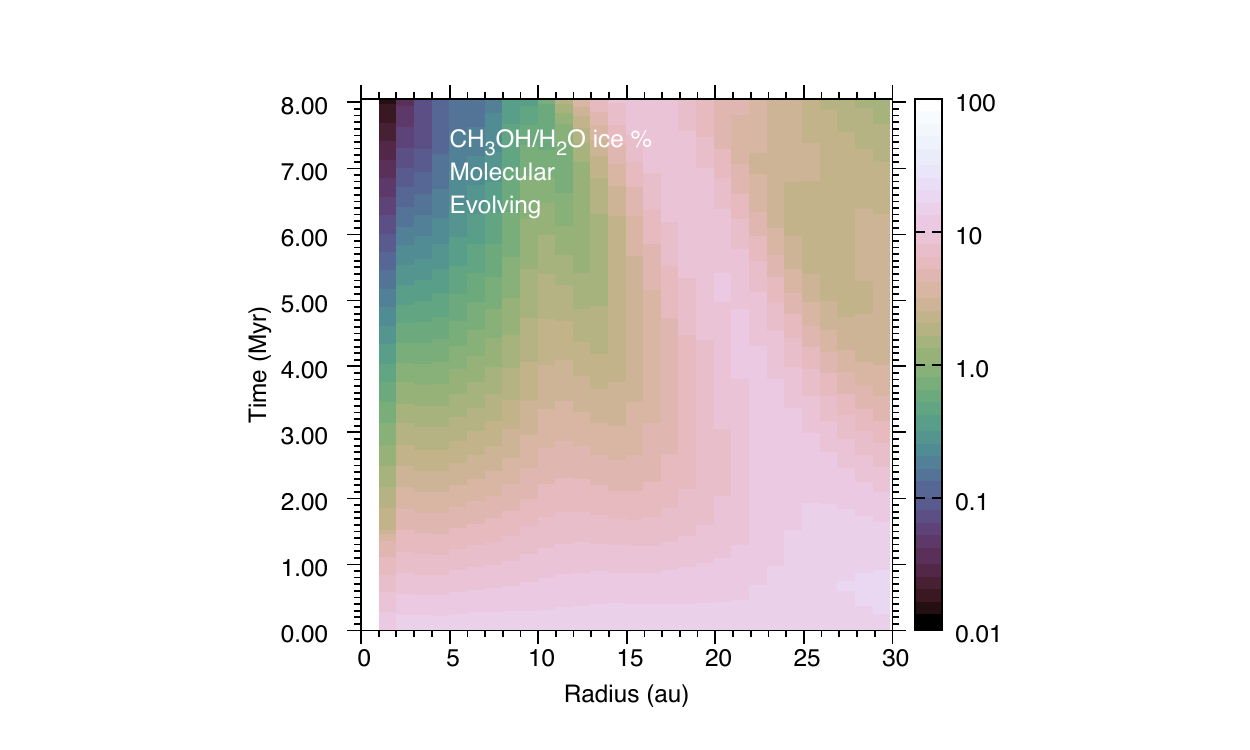}
\includegraphics[clip, trim = 2cm -0.5cm 2.25cm 0cm, width=0.33\textwidth]{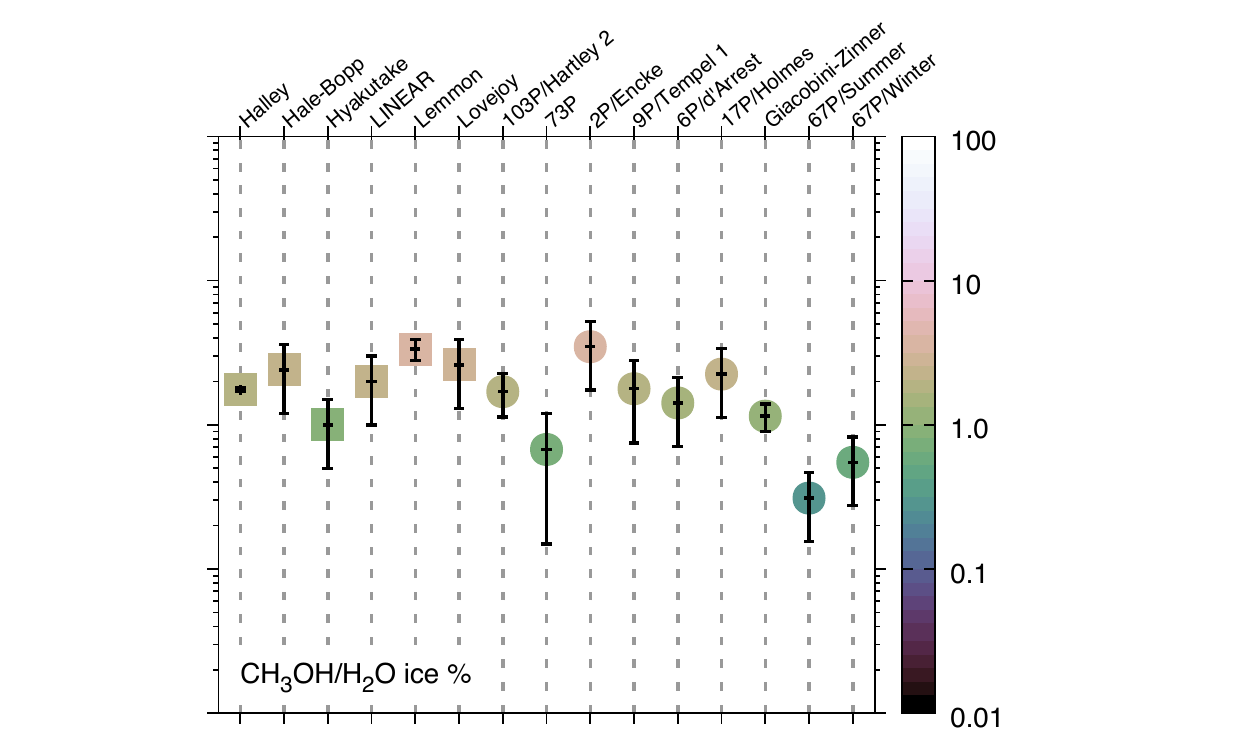}
\caption{Same as Figure~\ref{figure6} for \ce{H2CO} and \ce{CH3OH} ice.}
\label{figure10}
\end{figure*}

This appendix features abundance ratio maps (Figs. \ref{figure6}-\ref{figure10}) in radius and time for the molecular species considered in this work, as well as HCN, \ce{NH3}, HNCO, \ce{CH3CN} and \ce{C2H2}. These abundances are taken from \citet{eistrup2018}. Each row is for one molecular ice species. The left columns of the figures are evolving abundances for the reset scenario (``Atomic''), and the middle columns for inheritance scenario (``Molecular'') from the 0.1 MMSN evolving disk from \citet{eistrup2018}. The right column features the observed abundances (with given errors, or, if no error was given, assuming a conservative 50\% error relative to the observed species abundance with respect to \ce{H2O} ice) of a given ice species, in those of the 15 comets, where the molecule has been observed. 

The colourbar next to the right column indicates that high abundance with respect to \ce{H2O} ice gives a light colour, and low abundance gives a darker colour. The cometary abundances can also be read off vertically on the $y$-axis. The colour abundances of the comets are intended to enable easy visual comparison between the modelled (left and middle columns) and observed (right column) abundances.
 
\end{appendix}

\end{document}